\documentclass[]{aa}
\usepackage[varg]{txfonts} % times font in math formulae
\usepackage{graphicx}
\usepackage{color}
\usepackage[breaklinks,bookmarks,bookmarksnumbered]{hyperref}
\usepackage{mathabx}
\usepackage{amsmath}

\usepackage{natbib}

\title{A multi-wavelength analysis for interferometric (sub-)mm observations of protoplanetary disks}
\subtitle{Radial constraints on the dust properties and the disk structure}
\titlerunning{A multi-wavelength analysis for interferometric (sub-)mm observations of protoplanetary disks}

\author{
M.~Tazzari\inst{\ref{ESO}, \ref{ExCluster}}, 
L.~Testi\inst{\ref{ESO}, \ref{ExCluster}, \ref{Arcetri}}, 
B.~Ercolano\inst{\ref{ExCluster}, \ref{USM}}, 
A.~Natta\inst{\ref{Arcetri}, \ref{DIAS}}, 
A.~Isella\inst{\ref{Rice}}, 
C.~J.~Chandler\inst{\ref{NRAO}}, 
L.~M.~Pérez\inst{\ref{NRAO}}, 
S.~Andrews\inst{\ref{CfA}}, 
D.~J.~Wilner\inst{\ref{CfA}},  
L.~Ricci\inst{\ref{CfA}},
T.~Henning\inst{\ref{MPIA}}, 
H.~Linz\inst{\ref{MPIA}},
W.~Kwon\inst{\ref{KOREA}},
S.~A.~Corder\inst{\ref{JAO}}, 
C.~P.~Dullemond\inst{\ref{ITA}}, 
J.~M.~Carpenter\inst{\ref{Caltech}},  
A.~I.~Sargent\inst{\ref{Caltech}}, 
L.~Mundy\inst{\ref{Maryland}}, 
S.~Storm\inst{\ref{Maryland}}, 
N.~Calvet\inst{\ref{Michigan}},  
J.~A.~Greaves\inst{\ref{SAndrews}}, 
J.~Lazio\inst{\ref{JPL}}, 
A.~T.~Deller\inst{\ref{ASTRON}}
}
\authorrunning{Tazzari, M. et al.}

\institute{European Southern Observatory, Karl-Schwarzschild-Str. 2, D-85748 Garching, Germany \email{mtazzari@eso.org} \label{ESO}
\and
Excellence Cluster Universe, Boltzmannstr. 2, D-85748 Garching, Germany \label{ExCluster}
\and
INAF-Osservatorio Astrofisico di Arcetri, Largo E. Fermi 5, I-50125 Firenze, Italy \label{Arcetri}
\and
Universitats-Sternwarte M\"unchen, Scheinerstra{\ss}e 1, D-81679 M\"unchen, Germany \label{USM}
\and
Dublin Institute for Advanced Studies, School of Cosmic Physics, 31 Fitzwilliam Place, Dublin 2, Ireland \label{DIAS}
\and
Department of Physics and Astronomy, Rice University, 6100 Main Street, Houston, TX, 77005, USA \label{Rice}
\and
National Radio Astronomy Observatory, P.O. Box O, Socorro, NM 87801, USA \label{NRAO}
\and
Harvard-Smithsonian Center for Astrophysics, 60 Garden Street, Cambridge, MA 02138, USA \label{CfA}
\and
Max Planck Institute for Astronomy, K\"onigstuhl 17, D-69117 Heidelberg, Germany \label{MPIA}
\and
Korea Astronomy and Space Science Institute, 776 Daedeok-daero, Yuseong-gu, Daejeon 34055, Republic of Korea\label{KOREA}
\and
Joint ALMA Observatory (JAO), Av. Alonso de Cordova 3107, Vitacura, Santiago, Chile \label{JAO}
\and
Institute for Theoretical Astrophysics, Heidelberg University, Albert-Ueberle-Strasse 2, D-69120 Heidelberg, Germany \label{ITA}
\and
Department of Astronomy, California Institute of Technology, MC 249-17, Pasadena, CA 91125, USA \label{Caltech}
\and
Department of Astronomy, University of Maryland, College Park, MD 20742 \label{Maryland}
\and
Department of Astronomy, University of Michigan, 830 Dennison Building, 500 Church Street, Ann Arbor, MI 48109, USA \label{Michigan}
\and
School of Physics and Astronomy, University of St Andrews, North Haugh, St Andrews,  KY16 9SS Scotland, UK \label{SAndrews}
\and
Jet Propulsion Laboratory, California Institute of Technology Pasadena, CA 91109, USA \label{JPL}
\and
The Netherlands Institute for Radio Astronomy (ASTRON), Dwingeloo, The Netherlands \label{ASTRON}
}

\usepackage{color}

\newcommand{\sur}{\mathrm{sur}}
\renewcommand{\mid}{\mathrm{mid}}
\newcommand{\klambda}{\mbox{k$\lambda$}}

\definecolor{linkcolor}{rgb}{0,0,0.6}
\newcommand{\paper}{Article}
\newcommand{\emcee}{\textit{emcee}}

\newcommand{\twolayer}{two-layer}

\newcommand{\Msun}{M_\odot}
\newcommand{\Rbar}{\overline{R}}

\abstract
{The growth of dust grains from sub-$\mu$m to mm and cm size is the first step towards the formation of planetesimals.  
Theoretical models of grain growth predict dust properties to change as a function of disk radius, mass, age and other physical conditions.
High angular resolution observations at several (sub-)mm wavelengths constitute the ideal tool to directly probe the bulk of dust grains and to investigate the radial distribution of their properties.}
{We lay down the methodology for a multi-wavelength analysis of (sub-)mm and cm continuum interferometric observations to constrain self-consistently the disk structure and the radial variation of the dust properties. The computational architecture is massively parallel and highly modular.}
{The analysis is based on the simultaneous fit in the uv-plane of observations at several wavelengths with a model for the disk thermal emission and for the dust opacity. The observed flux density at the different wavelengths is fitted by posing constraints on the disk structure and on the radial variation of the grain size distribution.}
{We apply the analysis to {observations of three protoplanetary disks (AS 209, FT Tau, DR Tau) for which a combination of spatially resolved observations in the range $\sim$0.88mm to $\sim$10mm is available (from SMA, CARMA, and VLA). In these disks} we find evidence of a decrease in the maximum dust grain size, $a_{\max}$, with radius. We derive large $a_{\max}$ values up to 1\,cm in the inner disk $15\,\mathrm{AU}\leq R\leq 30\,$AU and smaller grains with $a_{\max}\sim 1\,$mm in the outer disk ($R\gtrsim 80\,$AU). Our analysis of the AS 209 protoplanetary disk confirms previous literature results showing $a_{\max}$ decreasing with radius.}
{Theoretical studies of planetary formation through grain growth are plagued by the lack of direct information on the radial distribution of the dust grain size. In this paper we develop a multi-wavelength analysis that will allow this missing quantity to be constrained for statistically relevant samples of disks and to investigate possible correlations with disk or stellar parameters.}


\begin{document}

\maketitle

\section{Introduction}
Planetary systems form in the dusty gaseous circumstellar disks that orbit young pre-main sequence (PMS) stars. According to the core accretion scenario, the formation of planets proceeds with the formation of rocky cores through the growth of dust aggregates from mm/cm to km sizes and the subsequent accretion of gas onto these cores to form the planetary atmosphere (\citealt{safronov:1972, Wetherill:1980lr} and references therein). The initial population of sub-micron sized dust particles coming from the ancestral ISM cloud \citep{Mathis:1977yq} is processed by micro-physical interactions that determine the evolution of their shape, size and structure (\citealt{Testi:2014kx} and references therein).

Determining the size distribution of dust grains in protoplanetary disks is of paramount importance in order to  understand the initial conditions where planet formation takes place. 
In the recent theoretical studies by \citet{Brauer:2008kq} and \citet{Birnstiel:2010jk} the evolution of dust grain populations have been modeled taking into account dust growth processes (coagulation and fragmentation) as well as dynamical mechanisms responsible for the transport of dust grains (radial drift, vertical settling, turbulent mixing). {According to these studies, radial drift effectively depletes the large grain population in disks within 1\,Myr (see also \citealt{Laibe:2014qy}) unless it is halted by the occurrence of dust traps \citep{Whipple:1972tk,Weidenschilling:1977lr, Klahr:1997fk, Pinilla:2012xy} or a different aerodynamic behaviour in the case the grains are extremely fluffy aggregates \citep{Okuzumi:2012qy, Kataoka:2013uq}.}

The major limitation of theoretical grain growth studies is the lack of direct information on the actual size of dust grains occurring in the different regions of the protoplanetary disks. Observing a protoplanetary disk at different wavelengths allows different parts of the disk to be investigated. Optical, near- and mid-infrared observations show evidence for micron-sized dust grains in protoplanetary disks \citep{Bouwman:2001qv, van-Boekel:2003nr, Juhasz:2010fj, Miotello:2012fv}, however they effectively trace the dust content only of the inner disk and its surface layer (both directly heated by the impinging stellar radiation). In order to study the disk midplane, where the bulk of the dust mass resides and planet formation is thought to occur, millimeter and sub-millimeter continuum observations are needed. At these longer wavelengths most of the disk becomes optically thin to its own thermal radiation and therefore directly probes the entire dust emitting volume. Moreover, approximating the  (sub-)mm dust opacity  with a power law $\kappa_\nu\propto\nu^\beta$, changes in the spectral index $\beta$ can be linked to changes in the dust properties \citep{Stognienko:1995yu, Henning:2010qf}. In particular, \citet{Draine:2006ty} and \citet{Natta:2004yu} showed that $\beta$ can be interpreted in terms of grain growth: large $\beta$-values are produced by small grains ($\mu$m to mm size), whereas small values $\beta\lesssim 1$ are a signature of dust grains larger than 1 mm. 

From multi-wavelength observations one can gain insight into the dust opacity spectral index, and therefore on the level of dust processing occurring in a protoplanetary disk \citep{Wilner:2000qd, Testi:2001rf, Testi:2003kq, Rodmann:2006hl}. Past mm and sub-mm photometry studies allowed a disk-averaged $\beta$ value to be inferred for several disks in different star forming regions, suggesting: first, that grain growth processes are particularly efficient, being able to produce large grains (size $\sim 1\,$mm) within relatively short timescales of $1-3\,$Myr \citep{Ricci:2010fv, Ricci:2010eu}; and second, local or global dust grain retention mechanisms must be occurring in disks in order to account for the observed presence of large grains in evolved disks \citep{Testi:2014kx}.

In recent years, the improved observational capabilities offered by mm and sub-mm interferometers allowed us to use spatially resolved observations to measure for the first time the dust grain properties as a function of the disk radius and to compare them with the theoretical predictions of grain growth models. 
Initial attempts using 1.3 and 3mm observations of a sample of disks showed hints of possible radial variation of the dust properties \citep{2010ApJ...714.1746I, 2011A&amp;A...529A.105G} {and} recent studies have successfully extended the wavelength range to cm wavelengths using the VLA \citep{Banzatti:2011ff, 2013A&amp;A...558A..64T, 2012ApJ...760L..17P, Menu:2014dz, Andrews:2014pb, Perez:2015fk, Guidi:2015}. {Azimuthal variations in the dust distribution have  been recently discovered with ALMA early science observations in several disks and interpreted as dust traps, with the large grains being more concentrated 
than the small grains and the molecular gas \citep{van-der-Marel:2013jk, van-der-Marel:2015lr, Fukagawa:2013qv, Perez:2014nr, Casassus:2015lr, Marino:2015fk}.}

\cite{2012ApJ...760L..17P} used interferometric observations between 0.88 and 9.8 mm to constrain the radial profile of the $\beta$ index in the AS 209 protoplanetary disk (and recently extended to CY Tau and DoAr 25 by \citealt{Perez:2015fk}). They fitted separately each wavelength assuming a dust opacity constant with radius and derived, for each wavelength, a disk temperature profile and an optical depth profile $\tau_\nu \,\propto \,\Sigma\, \kappa_\nu$, where $\Sigma$ is the dust surface density and $\kappa_\nu$ is the dust opacity.  Then, given the fact that the surface density must be wavelength independent, they ascribed the changes in $\tau_\nu$ to variations of $\kappa_\nu$ (and therefore of the dust properties), thus deriving the $\beta(R)$ profile. They found $\beta\sim 0.5$ (mm- and cm-sized grains) in the inner disk region ($R\lesssim 50\,AU$) and $\beta \gtrsim \beta_{\mathrm{ISM}}=1.7$ (sub-mm-sized grains) in the outer disk.  

Improving on the results of \cite{Banzatti:2011ff},  \cite{2013A&amp;A...558A..64T} carried out the first attempt of a self-consistent modeling of the disk structure and the dust properties by fitting interferometric observations of the CQ Tau protoplanetary disk using a dust opacity model that allowed radial variations in the dust grain size. With such modeling, they were able to fit simultaneously observations at 1.3, 2.6 and 7mm, finding evidence of dust grains up to a few cm in the inner disk (< 40 AU) to a few mm at 80 AU.
{The work by \cite{2013A&amp;A...558A..64T} laid down the basis of the study we present here, where we have developed further such multi-wavelength fit technique.}

{In this paper we present a data analysis procedure that exploits the wealth of information carried by multi-wavelength (sub-)mm observations in order to characterize the disk structure and the level of grain growth in the disk in a self-consistent way. Previous studies fitted observations at different wavelengths separately, inferring for each of them a different temperature and surface density profile (note that the discrepancy in the temperature profile inferred from fitting two different wavelengths can be as large as a factor of 2, especially in the outer disk, see Figure~\ref{fig.comparison.perez.structure} in Appendix~\ref{app:code.benchmark}). Then, making assumptions on the average temperature and surface density profile that should characterize the disk, these studies reconciled the wavelength-dependent discrepancies between the models by deriving a ${\beta(R)}$ profile.
Adopting a more typical forward-modeling technique, the multi-wavelength analysis we present here derives a self-consistent disk model defined by unique temperature, 
%${T(R)}$, 
surface density 
%${\Sigma(R)}$ 
and 
%maximum 
grain size 
%${a_{\mathrm{max}}(R)}$ 
profile that make it capable of reproducing the observed disk emission at all the wavelengths simultaneously. Indeed the multi-wavelength nature of the analysis enables us to break the degeneracy between the disk temperature, the dust opacity and the dust surface density and thus provides us with a self-consistent physical description of the disk structure.}

With ALMA reaching full science and the major upgrade of the VLA, the improving quality of the (sub-)mm observational datasets in terms of angular resolution and sensitivity make such automated multi-wavelength analysis an ideal tool to investigate the disk structure and the dust properties for a large number of disks.
 
The \paper\ is organized as follows. In Section 2 we describe in detail the method, by introducing the architecture of the {analysis technique}, describing the details of disk and of the dust models and clarifying the Bayesian approach adopted for the analysis. In Section 3 the details of the observations are reported. In Section 4 we present the fit results and in Section 5 we compare and discuss the results obtained for the different disks. In Section 6 we draw our conclusions. {In Appendix~\ref{app:mcmc.details} we describe the implementation of the Bayesian analysis in detail, in Appendix~\ref{app:code.benchmark} we report the results produced during the bench-marking of our {analysis} against previous results by \citet{2012ApJ...760L..17P} and in Appendix~\ref{app:fits.results} we present the fit results for each disk.}

\section{Analysis technique} 
We have developed a method to constrain the dust properties and the disk structure through a self-consistent modeling of sub-mm and mm observations. The method is based on the simultaneous $(u,v)-$plane fit of several interferometric observations at different wavelengths. The strength of the method lies in the fact that it allows the derivation of a unique disk structure and dust size distribution capable of reproducing the observed flux at all the fitted wavelengths (forward-fit).

The method adopts a Bayesian approach and performs an affine-invariant Markov Chain Monte Carlo (MCMC) developed by \citet{Goodman:2010} to explore parameter space in an efficient way. 
The results of the fit provide probability distributions for the value of each  free parameter, estimates of their correlations, and a best-fit model from which residual maps can be computed.

Furthermore, we designed the method to have a modular and flexible architecture. It is \textit{modular} in the sense it allows the disk and the dust models to be changed independently of each other, making it suitable for studying disks with particular morphologies (e.g., with holes or asymmetries) or for testing different dust opacity models. 
It is \textit{flexible} as it is designed to fit independently each observed $(u,v)-$point rather than binned values of deprojected visibilities (which would require an \textit{a priori} knowledge of the disk center, inclination and position angle). This makes the method ready to be applied to disks with non-axisymmetric surface brightness distributions and to be expanded in future to fit the disk inclination and position angle self-consistently with the disk structure.

It is worth noting that the major requirement of this method is the availability of a fast and efficient disk and dust model. Since the MCMC usually requires one to two millions model evaluations in order to converge, the speed and the efficiency of the model become extremely important and determine the overall computation time. For this, a huge effort was put into the optimization of the disk and the dust models as well as into the computation of synthetic visibilities. The average time required for one posterior evaluation\footnote{A posterior evaluation consists of: an execution of the disk and the dust models; the computation of the synthetic visibilities through 4 Fast Fourier Transforms (FFT) (for the test case, we fit 4 wavelengths and the matrix sizes were 1024x1024, 1024x1024, 4096x4096, 4096x4096, which are defined by the range of (u,v) distances sampled by the observed visibilities); the sampling of the synthetic visibilities at the location of the antennas (for the test case we have approximately 2.5 millions of uv-points).} is approximately 30 seconds on an Intel\textregistered\ Xeon\textregistered\ CPU E5-2680 at 2.70GHz, thus implying that 1 million evaluations would require more than 1 year to be computed. However, since at each step half of the model evaluations required by the affine invariant MCMC can be computed independently of each other, the overall computation time can be  shortened to 2 days by parallelizing the code and running it on hundreds of cores (we achieved a extremely good scalability of the code performance up to 200 cores, using the implementation and parameters discussed in Sect.\ref{sec:fitting.method}).

\subsection{Disk model}
\label{sec:disk.model}
We compute the disk structure and its thermal emission adopting the \twolayer\  disk model of \citet{1997ApJ...490..368C} with the refinements by \citet{2001ApJ...560..957D}. According to the \twolayer\  approximation, the disk is modeled as a \textit{surface} layer directly heated by the radiation of the central star and an internal layer - hereafter called \textit{midplane} - heated by the radiation reprocessed by the surface layer. Accretion, if present, is another process that would contribute to the heating of the midplane. This internal heating is most efficient in the very inner regions of the disk and only if the accretion rate is very high \citep[see e.g.][and references therein]{Dullemond:2007kb} however for this particular study we neglect it. Assuming that the disk is vertically isothermal (separately in each layer) and in hydrostatic equilibrium under the gas pressure and the gravitational field of the star, the \twolayer\  approximation allows us to compute the structure of the disk by solving the vertical radiative transfer equation at each radius. The disk model is computed over a logarithmic radial grid between an inner and an outer radius, respectively ${R_{\mathrm{in}}=0.1\,}$AU and ${R_{\mathrm{out}}\geq300\,}$AU (the exact value of ${R_{\mathrm{out}}}$ is chosen to be much larger than the continuum emission observed for the particular object that is being fitted). 

Once the surface and the midplane temperature profiles, respectively ${T_{\sur}(R)}$ and ${T_{\mid}(R)}$ are computed, the radial profile of the disk thermal emission (assumed to be optically thin) at a given wavelength ${\nu}$ is given by ${I_{\nu}^{\mathrm{tot}}(R)= I_{\nu}^{\sur}(R) + I_{\nu}^{\mid}(R) }$
where:
\begin{align}
\label{eq:Isur}
I_{\nu}^{\sur}(R) &= \frac{1}{d^{2}}
\left\{1+\exp{\left[-\frac{\Sigma_{\mathrm{d}}(R)\kappa_{\nu}^{\mid}(R)}{\cos{i}}\right]}\right\}B_{\nu}[T_{\sur}(R)]\,\Delta \Sigma(R) \kappa_{\nu}^{\sur}\,,\\
\label{eq:Imid}
I_{\nu}^{\mid}(R) &= \frac{\cos{i}}{d^{2}}
\left\{1-\exp{\left[-\frac{\Sigma_{\mathrm{d}}(R)\kappa_{\nu}^{\mid}(R)}{\cos{i}}\right]}\right\}B_{\nu}[T_{\mid}(R)]\,,
\end{align}
where ${\Sigma_{\mathrm{d}}(R)}$ is the dust surface density, ${\kappa_{\nu}^{\sur}}$ and ${\kappa_{\nu}^{\mid}(R)}$ are respectively the surface and the midplane dust opacities (see the next Section for details), ${B_{\nu}(T)}$ is the blackbody brightness, ${i}$ is the disk inclination (${i=0^{\circ}}$ for a face-on disk), ${\Delta \Sigma}$ is the surface density of the surface layer and  ${d}$ is the distance to the disk. These equations are derived for a geometrically thin disk ($H/R \ll 1$) and for low disk inclinations (if the disk is seen at $i>70^\circ$ a proper ray tracing is needed to account for the optical depth induced by the vertical structure of the disk). We refer the reader to \citet{2001ApJ...560..957D} for a complete derivation of the expressions above.

To complete the definition of the model, a radial profile for the gas and the dust surface density must be specified. In continuity with previous studies \citep{Andrews:2009zr, 2009ApJ...701..260I}, we parametrize the surface density adopting a self-similar solution for an accretion disk (solution derived assuming viscosity to be constant in time: \citealt{1974MNRAS.168..603L,1998ApJ...495..385H}):\begin{equation}
\label{eq:sigma.parametrization}
\Sigma_{\mathrm{g}}(R) = 
\Sigma_{0}\left(\frac{R}{R_{0}} \right)^{-\gamma}
\exp{\left[-\left(\frac{R}{R_{\mathrm{c}}} \right)^{2-\gamma} \right]}\,,
\end{equation}
where ${\Sigma_{0}}$ is a constant, ${R_{0}}$ is a scale radius that we keep fixed to ${R_{0}=40\,}$AU and ${R_{c}}$ is the spatial scale of the exponential cutoff. Assuming a constant dust-to-gas mass ratio ${\zeta=0.01}$ throughout the disk, the dust surface density will be ${\Sigma_{\mathrm{d}}=\zeta\, 
\Sigma_{\mathrm{g}}}$.
{A fixed $\zeta$ across the disk is a commonly used simplifying assumption, which we also adopt in our models as we cannot constrain independently the gas and dust surface density profiles with our observations. It is expected \citep[eg.,][]{Birnstiel:2014kx} that the gas and dust disk surface densities evolve differently over time. This is also confirmed by some observations that show extended, dust depleted outer gaseous disks \citep[eg.,][]{de-Gregorio-Monsalvo:2013fj}}. 
In addition, as discussed above, we assume that viscous evolution has been shaping the surface density profile, but that it is not important for the disk heating balance. This is a common approximation that generally describes well the observations; high spatial resolution observations reveal that these smooth surface densities are an approximation to the real dust distribution (e.g., ALMA observations of the HL~Tauri disk, \citealt{2041-8205-808-1-L3}). In this paper, considering the limited angular resolution of the observations we are analyzing, there is no need to adopt a more detailed radial profile. 

We now discuss two fundamental modifications we implemented in the \twolayer\  disk model for the present study.
First, the two-layer model is applicable only if, at every radius, the disk absorbs all the impinging stellar radiation. However, due to the exponential tapering of the dust surface density, the outermost disk region will eventually become optically thin to the stellar radiation. 
In this outer region, instead of adopting the two-layer model (which becomes numerically unstable), we assume that the disk temperature decreases radially as a power law $T_{\mid}(R)\propto R^{-k}$, where $k$ is obtained by fitting the $T_{\mid}$ profile in the optically thick disk region. Hereafter we will call $R\geq\Rbar$ the region where we apply this power law assumption.  
Secondly, following \citet{2009ApJ...701..260I} we impose a lower limit on the midplane temperature, namely the equilibrium temperature with the interstellar radiation field. We model this by adding an extra radiative flux impinging on the midplane ${\sigma_{\mathrm{SB}}T^{4}_{\mathrm{ext}}}$, where the temperature of the external radiation field ${T_{\mathrm{ext}}=10}$\,K, and $\sigma_{\mathrm{SB}}$ is the Stephan-Boltzmann constant. As a result, at each radius the effective midplane temperature is given by ${[T_{\mid}^{4}(R)+10^{4}]^{1/4}}$, where ${T_{\mid}(R)}$ is the temperature computed by the \twolayer\  model for ${R \leq \Rbar}$, and ${T_{\mid}(R)\propto R^{-k}}$ for ${R>\Rbar}$. This additional flux contribution is negligible in the inner region of the disk and starts to be relevant only in the outer parts, where ${T_{\mid}}$ becomes comparable to 10\,K.

The geometry of the disk on the sky is defined by specifying the disk inclination ${i}$ (the angle between the disk rotation axis and the line of sight, where $i=90^\circ$ corresponds to an edge-on view) and position angle ${PA}$ (the angle between the disk semi-major axis as it appears on sky and the North direction, measured East of North). In the present study ${i}$ and ${PA}$ are fixed parameters (see Table \ref{table:objects.properties} for detailed references).

\begin{table}
\centering
\caption{Stellar and disk properties.}
\resizebox{\hsize}{!}{
\begin{tabular}{lc c c c c cc}
\hline
\hline 
Object & SpT & ${L_{\star}}$ & ${M_{\star}}$ & ${T_{\star}}$ & \textit{i} & P.A.& Ref.\\
& & (${L_{\odot}}$) & (${M_{\odot}}$) & (K) & (${\,^{\circ}\,}$) & (${\,^{\circ}\,}$) & \\
\hline
AS 209  	& K5 	&	1.5\phantom{0} 	& 0.9\phantom{0} & 4250 & 38\phantom{\tablefootmark{a}}	& 86\phantom{\tablefootmark{a}} & 1\\
DR Tau 	& K7		&	1.09	& 0.8\phantom{0} & 4060	& 25\tablefootmark{a}	& 75\tablefootmark{a} & 2 \\
FT Tau 	& K6-M3.5\tablefootmark{b}	&	0.38	& 0.85 & 5000	& 23\phantom{\tablefootmark{a}}	& 29\phantom{\tablefootmark{a}} & 3 \\
\hline
\end{tabular}
}
\label{table:objects.properties}
\tablefoot{For each disk we report the spectral type of the central star, its luminosity, mass and surface temperature. We also report the parameters defining the disk geometry, namely the inclination and the position angle (measured East of North). \tablefoottext{a}{From \citet{Eisner:2014lr}}; \tablefoottext{b}{From \citet{Luhman:2010qy,Luhman:2010fk}}.}
\tablebib{
(1)~\citet{Andrews:2009zr};  
(2)~\citet{Ricci:2010eu}; 
(3)~\citet{2011A&amp;A...529A.105G, Kenyon:1995th}.}
\end{table}

Second, in our model we account for the possible contamination by emission from ionized gas, which is mostly relevant at longer wavelengths and is caused by thermal or non-thermal emission processes (e.g., free-free or gyro-synchrotron emission) from free electrons in the densest parts of a wind or the stellar corona \citep{Rodmann:2006hl, Ubach:2012ul}. We estimate and subtract this contamination by assuming that this emission is unresolved by our observations and dominates the long-wavelength emission at and beyond 3.6~cm. 
We estimate the maximum possible contamination extrapolating to millimetre wavelengths the flux density observed with the VLA at 6~cm (assuming that at these wavelengths there is no contribution from dust emission) and using an optically thick spherical wind approximation \citep[F$_\nu\sim\nu^{0.6}$:][]{Panagia:1975mz}. This assumption is conservative in the sense that provides the maximum possible contamination at shorter wavelengths.

\subsection{Dust model}
\label{sec:dust.model}

In order to account for the settling of large grains toward the disk midplane, as predicted by the models and observed by the lack of large grains in the disk atmosphere (based, e.g., on studies of silicate features: \citealt{Natta:2007ye} and references therein), we adopt two different dust size distributions for the surface and for the midplane which result in two different opacities, ${\kappa_{\nu}^{\sur}}$ and ${\kappa_{\nu}^{\mid}}$ respectively.

The choice of the dust grain size distribution aims at simply modeling a protoplanetary disk with a surface layer mainly composed of small grains and a midplane layer of larger pebbles \citep{Dullemond:2004ly,Tanaka:2005fj}. For both the surface and the midplane we adopt a power law distribution ${n(a) \propto\, a^{-q}}$ for ${a_{\mathrm{min}}<a<a_{\mathrm{max}}}$, where ${a}$ is the dust grain radius and ${q>0}$ is the power law index, choosing ${a_{\mathrm{min}}^{\sur}=a_{\mathrm{min}}^{\mid}=10\,}$nm, kept fixed throughout the disk. The particular value chosen for ${a_{\min}}$ does not affect our results as long as ${a_{\min}}\ll 1\,\mu m$ \citep{Miyake:1993lr}. For the surface we assume ${a_{\mathrm{max}}^{\sur}=1\,\mu}$m constant throughout the disk, whereas for the midplane we allow a radial variation of ${a_{\mathrm{max}}^{\mid}}$ modeled as follows:
\begin{equation}
\label{eq:amax.parametrization}
a_{\mathrm{max}}^{\mid}(R)= a_{\mathrm{max0}}\left(\frac{R}{R_{0}}\right)^{b_{\mathrm{max}}}\,,
\end{equation}
where $R_0=40\,$AU and ${a_{\mathrm{max0}}}$ is a normalization constant. 
In the disk surface we choose ${q=3.5}$ which describes well the size distribution of interstellar dust grains \citep{Mathis:1977yq,Draine:1984kx} out of which protoplanetary disks form, whereas for the disk midplane 
we assume ${q=3}$ that accounts for an enhanced population of large grains. 
\begin{figure}
\centering
\resizebox{0.9\hsize}{!}{\includegraphics{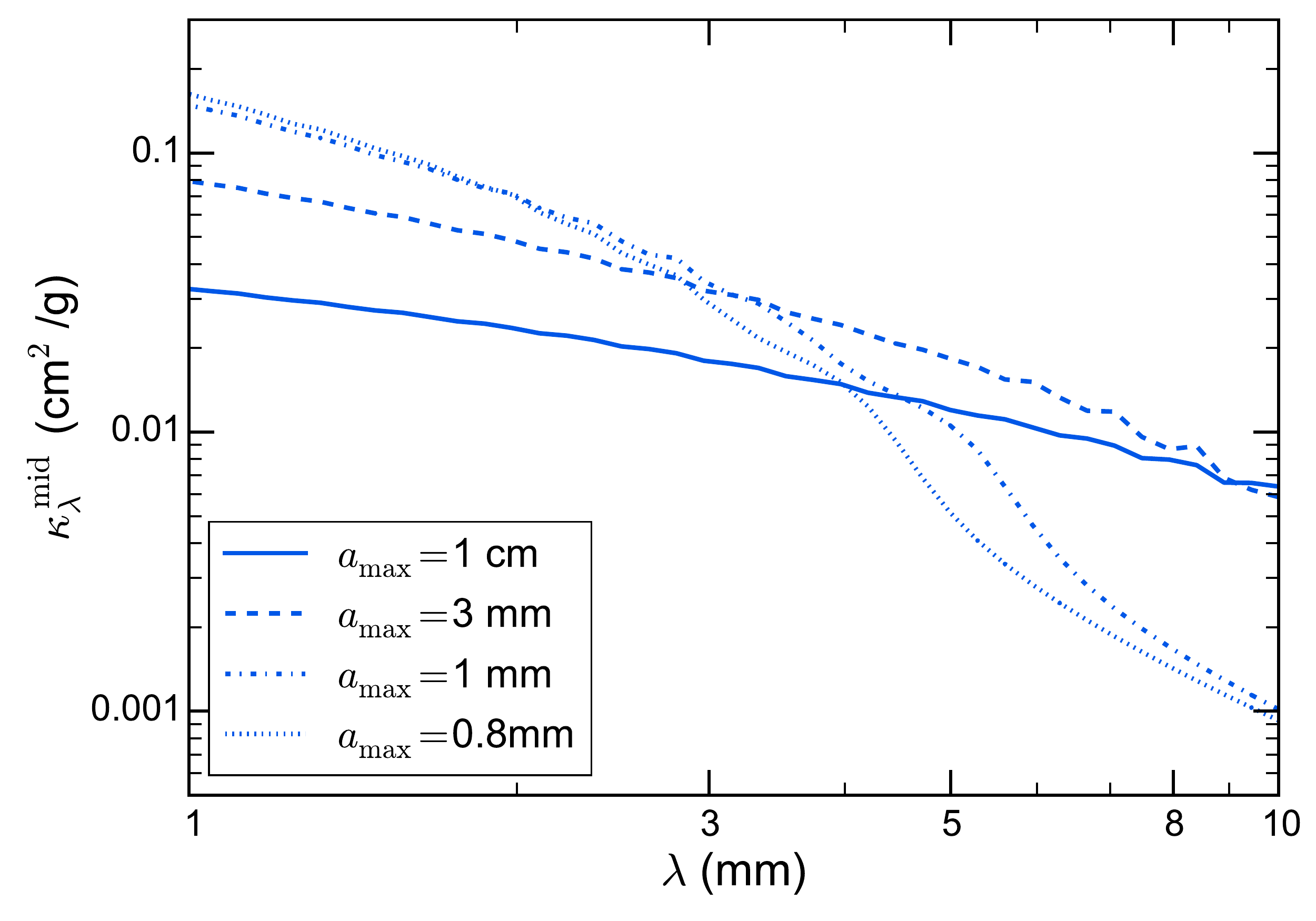}}
\caption{Dust opacity (per gram of dust) for ${a_{\mathrm{max}}=0.8,\ 1,\ 3\,}$mm and 1\,cm, computed for the midplane population of grains assuming the composition of 5.4\% astronomical silicates, 20.6\% carbonaceaous material, 44\% water ice and 30\% vacuum and a size distribution ${n(a) \propto\, a^{-q}}$ for ${a_{\mathrm{min}}<a<a_{\mathrm{max}}}$ with ${q=3.0}$ and ${a_{\mathrm{min}}=10\,}$nm.}
\label{fig:opacity.midplane.representative}
\end{figure}
{In Figure~\ref{fig:opacity.midplane.representative} we show the midplane dust opacity as a function of wavelength, computed for some ${a_{\mathrm{max}}}$ values ranging from 0.8mm to 1cm (which are representative values of the grain sizes we find in the analyzed disks, see below Section~\ref{sec:results}). The impact of the choice of ${q}$ on the resulting ${\beta}$ value can be seen in Figure~4 of \citet{Testi:2014kx}.}

Both in the surface and in the midplane we assume the same dust composition. Similarly to \citet{Banzatti:2011ff} and \cite{2013A&amp;A...558A..64T}, we assume the following simplified volume fractional abundances from \citet{Pollack:1994vn}: 5.4\% astronomical silicates, 20.6\% carbonaceaous material, 44\% water ice and 30\% vacuum, thus implying an average dust grain density of 0.9\,g/cm${^{3}}$. We choose the fractional abundances reported above for continuity with previous studies on grain growth, however we note that more recent estimates based on analysis of data from the Spitzer Space Telescope  can be found in the literature (e.g.,  \citealt{Juhasz:2010fj}).

Given the dust composition and the grain size distribution, we compute the dust opacity using the Bruggeman effective medium theory \citep{Bruggeman:1935ys} to calculate the dielectric function of the composite spherical grain and Mie theory to derive the dust absorption coefficients.
The complex optical constants used to compute the dielectric function are taken from \citet{Weingartner:2001fr} for silicates, \citet{Zubko:1996zr} for carbonacaeous material and \citet{Warren:1984mz} for water ice. 

\subsection{Modeling methodology}
\label{sec:fitting.method}
We adopt a Bayesian approach, which provides  probability distribution functions (PDFs) for the free parameters of the model using a variant of the Markov Chain Monte Carlo (MCMC) algorithms family. \citet{2009ApJ...701..260I} provide a general description of the use of this methodology for modeling protoplanetary disks. 

Our disk model is defined by the following free parameters: ${\Sigma_{0}}$, ${R_{\mathrm{c}}}$, ${\gamma}$ (to define the disk structure) and ${a_{\mathrm{max0}}}$, ${b_{\mathrm{max}}}$ (to define the dust distribution). In the present study the following parameters are kept fixed: the dust properties (e.g. composition, shape, porosity), the disk inclination ${i}$, the disk position angle ${PA}$, the contaminating free-free flux density at each wavelength $F_{\mathrm{ff}}( \nu)$. We note, however, that in a more general approach these parameters could be added to the Bayesian analysis and derived consistently with the disk structure and the grain size distribution.

Given a set of values for these parameters, we compute the synthetic disk images (one per fitted wavelength), and then we sample their Fourier Transform at the same ${(u,v)-}$plane locations of the observed visibilities. We finally compute the total $\chi^2$ as the sum of the several partial $\chi^2_\lambda$ computed for each wavelength, namely:
\begin{equation}
\chi^{2}=\sum_\lambda \chi^2_\lambda = \sum_{\lambda}\sum_{j=1}^{N_{\lambda}} \left|V^{\mathrm{obs}}_{\lambda,j}-V^{\mathrm{mod}}_{\lambda,j} \right|^{2}\cdot w_{\lambda,j}\,,
\end{equation}
where $w_{\lambda,j}$ is the visibility weight\footnote{The visibility weights $w_{\lambda,j}$ are computed theoretically as described by \citet{Wrobel:1999gf} and then re-scaled in order to assign the same weight to each wavelength.}, $V^{\mathrm{obs}}_{\lambda,j}$ and $V^{\mathrm{mod}}_{\lambda,j}$ are respectively the observed and the synthetic visibilities and $N_\lambda$ is the number of visibility points at the wavelength $\lambda$.

The posterior PDF is computed as the product of the likelihood of the observations given the model, namely $\exp(-\chi^2/2)$, by the prior PDF $p(\boldsymbol \theta)$ (where $\boldsymbol \theta$ is a point in the 5D space of parameters). We assume a uniform prior for all the parameters, therefore $p(\boldsymbol \theta)=1$ for $\boldsymbol \theta \in \Theta$ and $p(\boldsymbol \theta)=0$ otherwise, where $\Theta$ is a 5D domain in the space of parameters {defined by the ranges reported in Table~\ref{table:space.parameters}.  }
\begin{table}[h!]
\caption{Ranges defining the space of parameters explored by the Markov Chain.}
\begin{center}
\begin{tabular}{cccc}
\hline
\hline
Parameter		&	Min	&	Max 	& Unit\\
\hline
${\gamma}$		&	-1	&	4 		& --\\
${\Sigma_{0}}$	&	0.1	&	200		& g/cm${^{2}}$\\
${R_{\mathrm{c}}}$ & 5	&	300		& AU\\
${a_{\mathrm{max0}}}$&0.001	&	100		& cm\\
${b_{\mathrm{max}}}$& -5	&	2	& --\\
\hline
\end{tabular}
\end{center}
\label{table:space.parameters}
\end{table}%
In principle, the domain $\Theta$ can be changed from disk to disk (e.g., the $R_c$ interval can be changed depending on the disk size determined from observations), however in this study we kept it fixed to the intervals reported in {Table~\ref{table:space.parameters}}, which have been chosen to be large enough to be suitable for different disks.

In addition to the disk structure and the dust size distribution, we also derive the precise position of the disk centroid by adding two \textit{nuisance} parameters for each wavelength, $\Delta \alpha_0$  and $\Delta \delta_0$, that measure the angular offset between the disk centroid with respect to its nominal position $(\alpha_0,\ \delta_0)$. This method allows the derivation of the correct centre position for the model, even without any prior information on the star proper motion or other systematic position offsets between the different wavelength observations. 

The fit is performed using a variant of the Markov Chain Monte Carlo (MCMC) algorithm \citep{Goodman:2010} which allows the parameters space to be explored  by several Markov chains at the same time (an \textit{ensemble} of hundreds to thousands \textit{walkers}\footnote{In this affine-invariant MCMC, the chains are also called \textit{walkers}.}) which interact with each other in order to converge to the maximum of the posterior. The advantage of using several chains simultaneously is twofold: on the one hand, it allows a more complete exploration of the parameters space (each chain starts from a different initial location); on the other hand, it allows the computation to be massively parallelized. We perform the MCMC using the implementation provided in the Python package \emcee\ (\citealt{2013PASP..125..306F}) which offers the possibility to run the computation in parallel on several cores and has been used for an increasing number of astrophysical problems in recent years. In this study, we perform MCMC with 1000 chains: they are initialized with uniform random distribution in the parameters space ${\Theta}$, let evolve for a \textit{burn-in} phase of some hundreds steps, and finally let sample the posterior to get a sufficient number of independent posterior samples (for a detailed explanation of the criteria we used to assess the convergence of the chain, see Appendix~\ref{app:mcmc.details}).

The outcome of the MCMC is a collection (a \textit{chain}) of posterior samples out of which we can estimate the 1D and 2D marginal distributions of the free  parameters (marginalization means that all but one (two) parameters of the posterior are integrated over). From the 1D marginal distributions, we do a point estimate of each parameter using the median and we estimate the uncertainty as the central interval; i.e., from the 16\% to the 84\% percentile. These estimators give a good representation of the marginalized distribution, and reduce to the usual central credibility interval in the Gaussian case. 
For each fit, we present a \textit{staircase} plot with the 1D and 2D marginalized distributions of the interesting physical parameters.

For this study, the optimization of the disk model and of the imaging routines allowed us to execute the fits efficiently on hundreds of cores (hosted at the Computational Center for Particle and Astrophysics, C2PAP) and thus to reduce the  time needed to perform each multi-wavelength fit to one or two days. For further details on the implementation, see Appendix~\ref{app:mcmc.details}.

\section{Observations}

In this study we apply our method to three protoplanetary disks for which multi-wavelength (sub-)mm observations are available: AS~209, DR~Tau and FT~Tau. The AS~209 observations have already been presented by \cite{2012ApJ...760L..17P} and we refer the reader to that paper for their details. DR~Tau and FT~Tau observations from CARMA and VLA are now described in turn. A summary can also be found in {Table~\ref{table:observations}}.

\begin{table*}
\begin{center}
\caption{Details of the observations used for the fits.}
\begin{tabular}{l c c c c c c c c c}
\hline
\hline
Object & $\alpha$ & $\delta$ & Telescope & $\lambda$ & $F_\lambda$ & \multicolumn{2}{c}{Beam properties} & $F_\mathrm{ff}\,$ & Ref. \\
& (J2000.0) & (J2000.0) & & (mm) & (mJy) & FWHM (\,$^\second$\,) & P.A. (${\,^{\circ}\,}$) & ($\mu$Jy) & \\ 
\hline
AS 209 & 16 49 15.30 & $-$14 22 08.7 & SMA 	& 0.88 & 580\phantom{.00}$\pm$60\phantom{.0} & 0.61$\times$0.45 & \phantom{+}32.4 	& -- & 1\\
	    &			&			& CARMA	& 2.80 & \phantom{0}40\phantom{.00}$\pm$60\phantom{.0} & 0.93$\times$0.63	& $-$21.9	& -- & 1\\
	    &			&			& VLA 	& 8.00 & \phantom{00}1.1\phantom{0}$\pm$\phantom{0}0.1 & 0.25$\times$0.18	& $-$77.0	& \phantom{0}80 & 1\\
	    &			&			& VLA 	& 9.83 & \phantom{00}0.48$\pm$\phantom{0}0.1 	& 0.30$\times$0.20	& $-$79.0	& 100	& 1\\
DR Tau & 04 47 06.20	& +16 58 42.8 	& CARMA	& 1.30 & 137\phantom{.00}$\pm$20\phantom{.0} 	& 0.25$\times$0.22	& $-$63.8	& -- & 3\\
	    &			&			& VLA 	& 7.05 & \phantom{00}1.4\phantom{0}$\pm$\phantom{0}0.1 	& 0.43$\times$0.33	& $-$54.2	& <122 & 3 \\
             &			&			& VLA 	& 7.22 & \phantom{00}1.4\phantom{0}$\pm$\phantom{0}0.1	& 0.45$\times$0.34	& $-$54.1	& <120 & 3 \\
FT Tau  & 04 23 39.19 & +24 56 14.1 	& CARMA	& 1.30 & 107\phantom{.00}$\pm$10\phantom{.0} 	& 0.52$\times$0.39	& $-$60.0	& --	& 2\\
             &			&			& CARMA	& 2.60 & \phantom{0}24\phantom{.00}$\pm$\phantom{0}2\phantom{.0}  & 1.13$\times$1.02	& $-$77.3	& --	& 2\\
             &			&			& VLA 	& 8.00 & \phantom{00}1.1\phantom{0}$\pm$\phantom{0}0.1 	& 0.48$\times$0.32 & \phantom{+}71.3 & <220 & 3 \\
             &			&			& VLA 	& 9.83 & \phantom{00}0.7\phantom{0}$\pm$\phantom{0}0.1	& 0.57$\times$0.36 & \phantom{+}70.9 & <190 & 3 \\
\hline
\end{tabular}
\label{table:observations}
\end{center}
\tablefoot{For each disk we report the coordinates and the properties of the observations: $\lambda $ is the wavelength of the combined continuum data, $F_\lambda$ is the integrated continuum flux density (with error), FWHM and PA are respectively the size and the position angle (measured East of North) of the synthesized beam, $F_{\textrm{ff}}$ is the estimated free-free contamination upper limit.}
\tablebib{Observations presented in: (1)~\citet{2012ApJ...760L..17P} (2)~\citet{Kwon:2015lr} (3) this work.}

\end{table*}

\subsection{CARMA observations of DR Tau and FT Tau}

Observations of DR Tau at 1.3mm were obtained using CARMA between 2007 October 2007 and 2011 December 2011. Multiple array configurations (A, B,
and C) provide a uv-coverage spanning 15-1290 \klambda. Double-sideband single-polarization receivers were tuned to a LO frequency of 230 GHz in A configuration, 227.75 GHz in B configuration, and 228.60 GHz in C configuration. For optimal continuum sensitivity, the spectral windows in the correlator were configured to the maximum possible bandwidth per spectral window (0.47 GHz). The number of continuum spectral windows varied for different configurations: a total bandwidth of 1.9 GHz was available for B and C configuration observations, while during the A configuration observations the total bandwidth was 8 GHz. Observations of complex gain calibrators (0530+135 and 0449+113) were interleaved with science target observations. Additionally, a strong quasar was observed to calibrate the complex bandpass. The absolute flux density scale was derived from observations of a secondary flux density calibrator (3C84 or 3C273), whose flux density was monitored by the CARMA observatory, resulting in a fractional uncertainty of $\sim15\%$ in the absolute flux density calibration. Calibration was performed with the MIRIAD software package, with each dataset being calibrated separately.
The 1 and 3 mm CARMA observations of FT~Tau are presented in \citet{Kwon:2015lr}, to which refer for the observational details.  Here we give a brief description. The data were obtained over a period of about 2 years, between 2008 Oct 15 and 2011 Jan 5.  For good \textit{uv} coverage, multiple array configurations were employed: 1 mm observations in B and C configurations provide a \textit{uv} coverage of 17.0--620 \klambda, and 3 mm observations in A and C configurations gives a \textit{uv} coverage from 4.1--727.6 \klambda.  MIRIAD \citep{sault1995} was used for the data calibration. The absolute flux density was obtained through observations of a reliable flux calibrator (Uranus) and resulted in fractional uncertainties respectively of 10\% and 8\% for observations at 1.3 and 2.6\,mm. The complex gains were obtained through observations of a nearby bright point source (3C111). In the case of the A configuration data at 3 mm the C-PACS \citep{2010ApJ...724..493P} was employed for removing short period atmospheric turbulence. {Note that while the images of \citet{Kwon:2015lr} are produced with a Briggs robust parameter of 0 (which produces images with lower signal-to-noise ratio but better beam resolution), the images in this paper use the natural weighting (which does not apply any density weighting function to the observed uv-points thus producing images with the best signal-to-noise ratio). Through the analysis procedure described in Section~\ref{sec:fitting.method} we fit the flux measured at each uv-point, therefore we adopt the natural weighting scheme since is the most suitable to perform a direct comparison between the observed and the model data.}

\subsection{VLA observations of DR Tau and FT Tau}
Observations of DR Tau and FT Tau using the Karl G. Jansky Very Large
Array (VLA) of the National Radio Astronomy Observatory\footnote{The
National Radio Astronomy Observatory is a facility of the National
Science Foundation operated under cooperative agreement by Associated
Universities, Inc.} were made as part of the Disks@EVLA project (AC982)
between 2010 November and 2012 August.  DR Tau was observed using the
Q-band ($\lambda\sim$7mm) receivers with two 1GHz basebands centered
at 41.5 and 42.5GHz in the C and B configurations, providing projected
{\it uv}-spacings from 5 to 1500~\klambda.  FT Tau was observed using the
Ka-band ($\lambda\sim$1cm) receivers with two 1GHz basebands centered at
30.5 and 37.5GHz in the C and B configurations, providing projected {\it
uv}-spacings from 8 to 1300~\klambda.  For both targets the complex gain
was tracked using frequent observations of J0431+2037 (in C configuration)
or J0431+1731 (in B configuration), and the complex bandpass was
measured using 3C84.  The absolute flux density scale was derived from
observations of 3C147 (e.g., \citealt{Perley:2013rf}), and its overall
accuracy is estimated to be 10\%.  The data were calibrated, flagged,
and imaged using a modified version of the VLA Calibration Pipeline (see
\href{https://science.nrao.edu/facilities/vla/data-processing/pipeline/scripted-pipeline}{https://science.nrao.edu/facilities/vla/data-processing/pipeline/scripted-pipeline}).
At Ka-band the calibrator source J0431+2037 turned out to have multiple
components that required the source to be modeled before being used to
derive calibration solutions.  In addition, because of the substantial
time period covering the observations, corrections for source proper
motion and/or other systematic position offsets between datasets
(caused, e.g., by the structure of J0431+2037) also had to be applied.
The astrometry reported here corresponds to that derived from the B
configuration data. The VLA observations shown in this paper have been imaged using natural {weighting}.

Both sources were also observed with the C-band ($\lambda\sim$6cm)
receivers in the VLA's most compact, D configuration in 2010 July,
in order to evaluate any potential contamination from ionized gas at
shorter wavelengths.  Two 1GHz basebands were centered at 5.3 and 6.3GHz.
Complex gain variations were tracked through observations of J0431+2037,
the bandpass was measured using 3C84, and the absolute flux density scale
was obtained through observations of 3C147.  DR Tau was detected with an
integrated flux density $F_{\rm 6cm} = 99\pm31\mu$Jy, while for FT Tau
a 3$\sigma$ upper limit on the 6cm flux density of 72$\mu$Jy was obtained.

\section{Results}
\label{sec:results}
For each disk, we report a set of statistical and physical results that we now describe in turn. In this Section we discuss in detail the results for FT~Tau and in Appendix \ref{app:fits.results} we report detailed plots for the other disks.

\begin{figure}
\centering
\resizebox{\hsize}{!}{\includegraphics{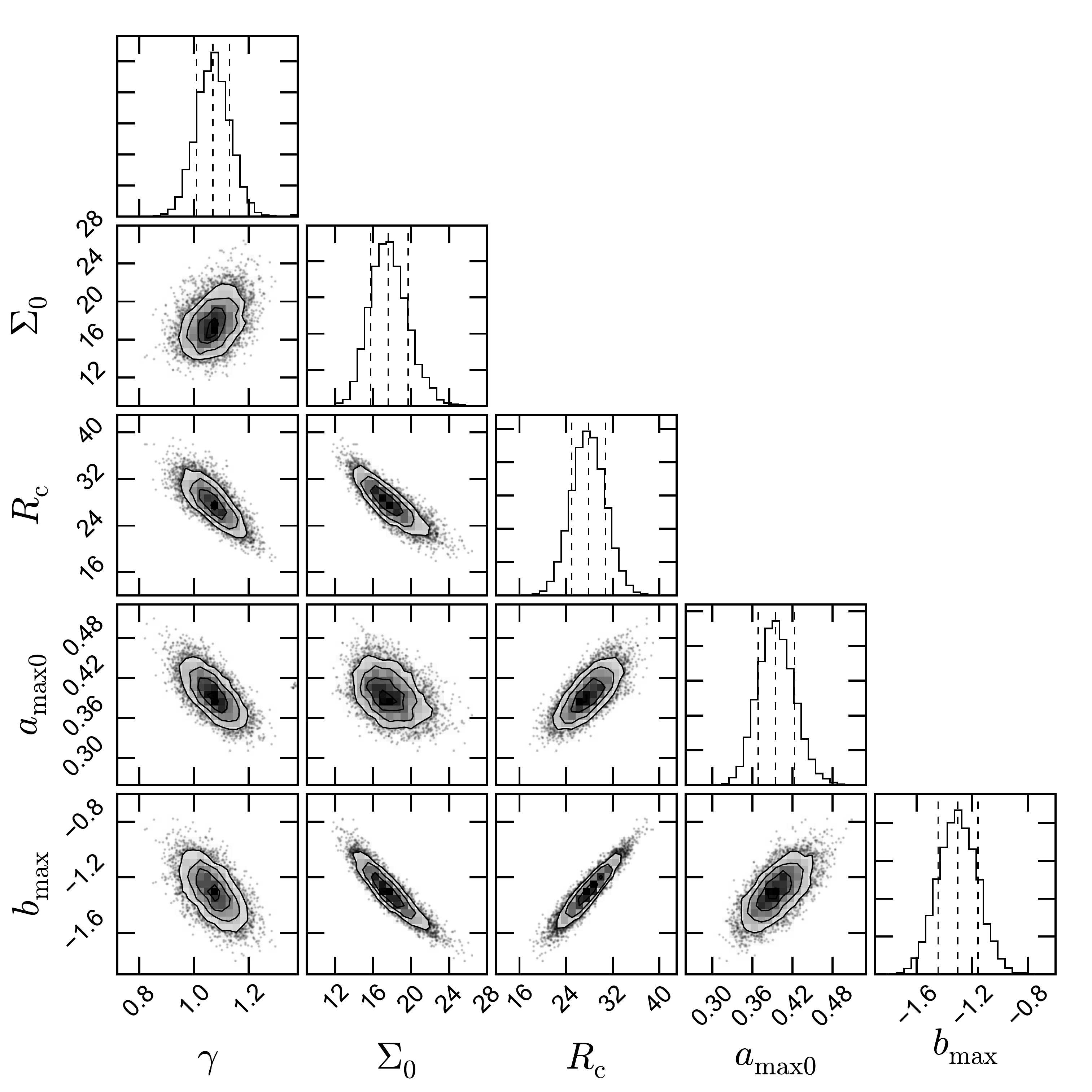}}
\caption{Representation of the MCMC results for FT Tau. On the top diagonal, the 1D histograms are the marginalized distributions of the fitted parameters, with the vertical dashed lines representing the 16\%, the 50\% and the 84\% percentiles, respectively from left to right. The 2D density plots represent the bi-variate distributions for each couple of parameters, with one dot representing one sample. The plot shows the posterior sampling provided by 500 steps of the 1000-walkers chain (800 burn-in steps were performed to achieve convergence). Note that in order to obtain an independent set of samples, the chain has been thinned by a factor equal to the autocorrelation time ($\sim 80$ steps in this case).}
\label{fig:fttau.triangle}
\end{figure}

In Figure \ref{fig:fttau.triangle} we present a staircase plot showing the posterior PDF computed from the chain, after proper thinning (the fit needed 800 burn-in steps, and 500 further steps to sample the posterior). On the diagonal of Figure \ref{fig:fttau.triangle} we show the marginalized 1D distributions for each parameter, which display a Gaussian-like shape, while off the diagonal we show the 2D marginalized distributions which give an overview of the correlations between the parameters. For FT Tau we obtain the following parameter values: 
\begin{eqnarray*}
\gamma =& &1.07\pm 0.06 \\
\Sigma_0 =& &18\pm 2\,\mathrm{g/cm}^2\\
R_c =& &28\pm 3\,\mathrm{AU}\\
a_\mathrm{max0} =& &0.40\pm 0.03\,\mathrm{cm}\\
b_\mathrm{max} =& &-1.3\pm 0.1
\end{eqnarray*}
where the errors are given by the central credibility interval of their marginalized distribution (i.e. the 16\% and 84\% percentiles).
Note that the fit has more parameters, namely four pairs of directional offsets [one pair $(\Delta \alpha_0,\ \Delta \delta_0)$ for each wavelength], but for the clarity of the plot we do not show them. In Table \ref{table:fits.disk.center} we report the derived position of the disk centroid $(\alpha,\ \delta)$ determined by the fit, where $\alpha=\alpha_0+\Delta \alpha_0$ and $\delta=\delta_0+\Delta \delta_0$. For each disk, we also report from the SIMBAD database the star reference position at Epoch 2000.0 (from Hipparcos or 2MASS measurements), the proper motion estimates (from Hipparcos or US Naval Observatory Catalogs) and, for each interferometric dataset, the fitted disk center position with derived uncertainties. Within the uncertainties, the derived positions are consistent with the expected stellar positions based on the astrometric and proper motion measurements. 

\begin{table*}
\caption{Fitted disk centroid positions.}
\centering
\begin{tabular}{lccccccc}
\hline
\hline
Object 	& Telescope & Epoch & $\alpha$ & $\delta$ & $\Delta \alpha$, $\Delta \delta $ & p.m. ($\alpha$) & p.m. ($\delta$)\vspace{3pt} \\
        & 			& (UT)  & (J2000.0) & (J2000.0) & mas & mas/yr & mas/yr \vspace{3pt} \\
\hline 
AS 209	& -- 		& 2000.0	     &	16 49 15.30324 			& $-$14 22 08.6346  			& -- & $-$7.69 & $-$22.84 \\
		& SMA 	& 2006-05-12 & 16 49 15.282\phantom{00} 	& $-$14 22 08.77\phantom{00} 	& 0.005 & &   \\
		& CARMA	& 2009-12-10 & 16 49 15.303\phantom{00} 	& $-$14 22 08.853\phantom{0} 	& 0.005 & &   \\
		& VLA 	& 2011-05-20 & 16 49 15.298\phantom{00} 	& $$-14 22 08.914\phantom{0} 	& 0.005 & &   \\
DR Tau	& -- 		& 2000.0 	     & 04 47 06.209\phantom{00}  	& +16 58 42.81\phantom{00}  	& -- & 12.6 & $-$17.1\\
		& CARMA 	& 2011-12-06 & 04 47 06.219\phantom{00} 	& +16 58 42.711\phantom{0}  	& 0.002 & & \\
       		& VLA   	& 2012-08-08 & 04 47 06.217\phantom{00} 	& +16 58 42.652\phantom{0} 	& 0.006 &  & \\
FT Tau	& -- 		& 2000.0 	     & 04 23 39.193\phantom{00}	& +24 56 14.11\phantom{00}  	& -- &  10.3 & $-$21.4 \\
		& CARMA	& 2008-10-15 & 04 23 39.193\phantom{00} 	& +24 56 14.003\phantom{0}  	& 0.002 & & \\
		& VLA 	& 2011-03-28 & 04 23 39.196\phantom{00} 	& +24 56 13.977\phantom{0}  	& 0.004 & & \\
\hline
\end{tabular}
\tablefoot{For each object we report the astrometric coordinates at the reference Epoch 2000.0 with proper motion estimates from SIMBAD\@. For each interferometric dataset we report the coordinate of the disk center derived from the fit, with the estimated errors ($\Delta \alpha$ and $\Delta \delta $ are found to be approximately equal). These are the formal uncertainties of the fitting process and do not account for the uncertainties introduced by the phase calibration (position accuracy of the calibrator and potentially uncorrected phase offsets), we estimate these to contribute at the 0.1$^{\prime\prime}$ level for our observations.}
\label{table:fits.disk.center}
\end{table*}

\begin{figure}
\centering
\resizebox{\hsize}{!}{\includegraphics{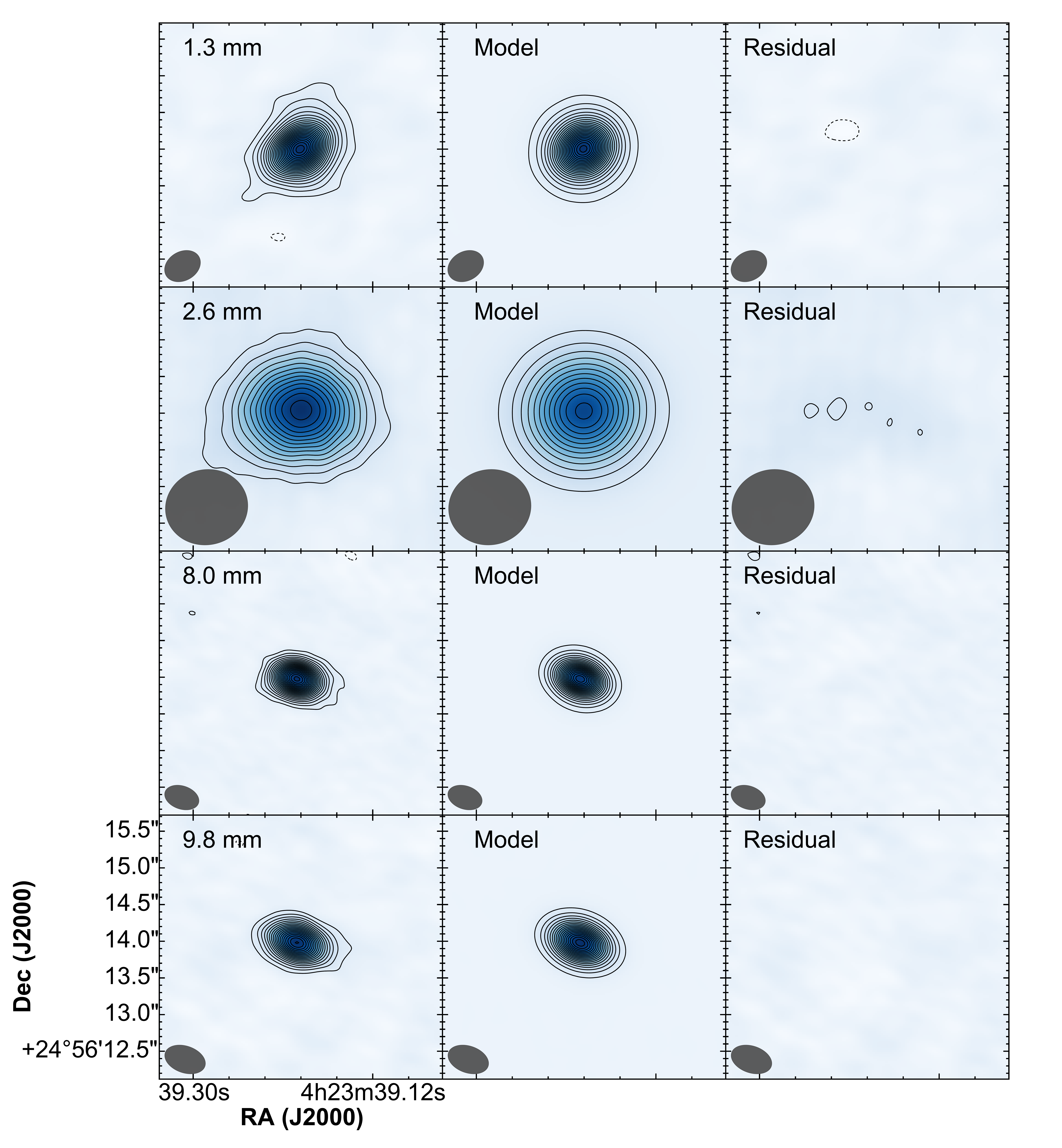}}
\caption{Comparison between the observed and the best-fit model images at different wavelengths of the FT Tau protoplanetary disk. The best-fit model is defined by the following parameters: $\gamma=1.07$, $\Sigma_0=18\,$g/cm$^2$, $R_c=28\,$AU, $a_\mathrm{max0}=0.4\,$cm, $b_\mathrm{max}=-1.3$. The observed images are shown in the left panels, the model images in the center panels, the residuals in the right panels. The positive and negative contour levels are spaced by 3$\sigma$ (starting from -3$\sigma$) and are the same in all the panels. The synthesized beam FWHM is represented as a grey ellipse in the bottom-left corner of each map.}
\label{fig:fttau.residuals}
\end{figure}

In Figure \ref{fig:fttau.residuals} we compare the observed and the model images at each wavelength, showing the residuals obtained by imaging the residual visibilities (obtained by subtracting the noise-free model visibilities from the the observed visibilities). The best fit model represented in Figure \ref{fig:fttau.residuals} corresponds to the model with median values of the marginalized distributions ($\gamma=1.07$, $\Sigma_0=18\,$g/cm$^2$, $R_c=28\,$AU, $a_\mathrm{max0}=0.4\,$cm, $b_\mathrm{max}=-1.3$) and we have verified that it is among the models with lowest reduced $\tilde\chi^2\simeq 1.01$. To produce the maps we have applied the CLEAN algorithm \citep{Clark:1980lr} with natural weighting. The residuals are small at all the wavelengths, with one negative 3$\sigma$ residual left at 1.3mm, a few 3$\sigma$ residuals left at 3mm and no residuals within $\pm 3\sigma$ left at 8.0 and 9.8 mm.

\begin{figure*}
\centering
\resizebox{0.4\hsize}{!}{\includegraphics{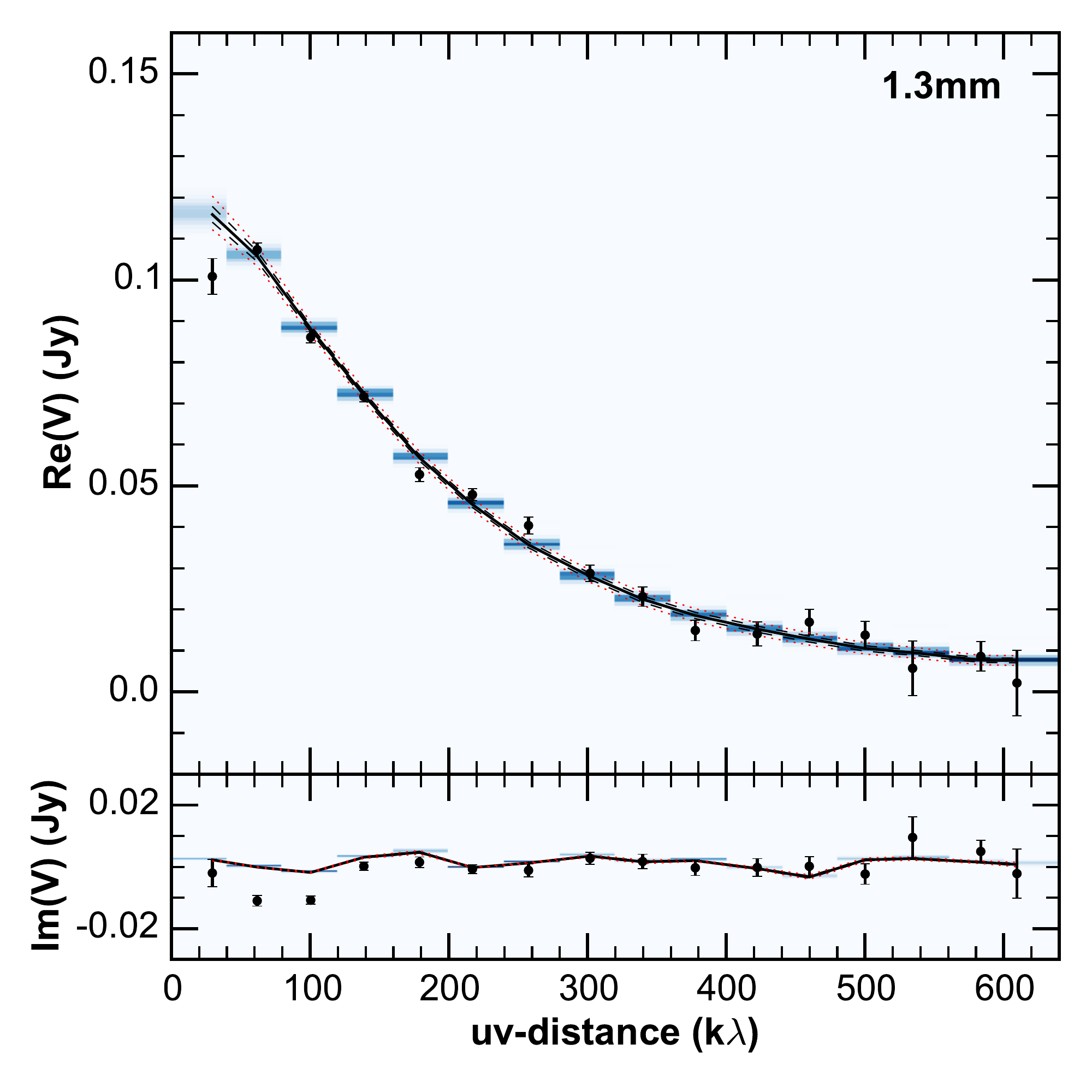}}
\resizebox{0.4\hsize}{!}{\includegraphics{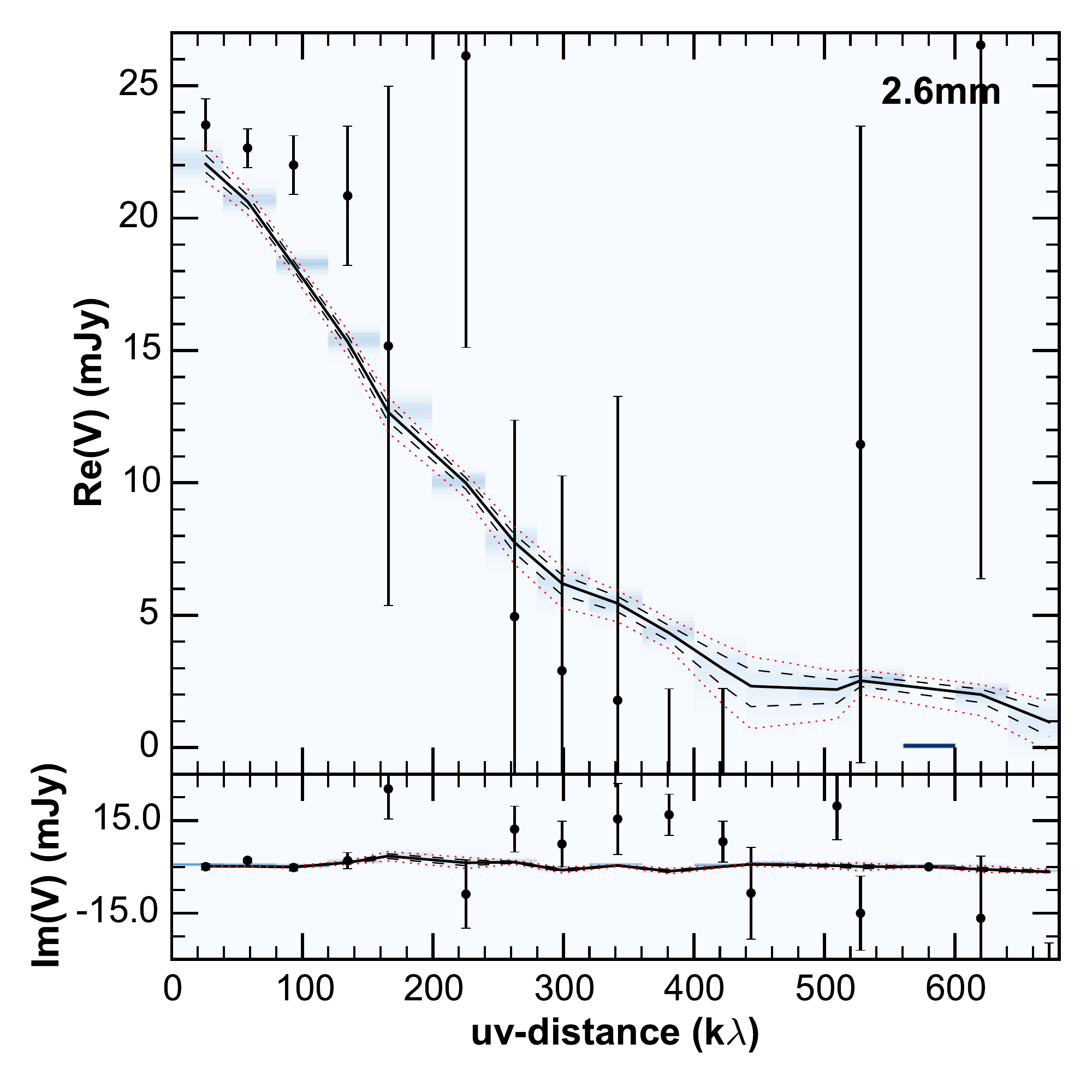}}\\
\resizebox{0.4\hsize}{!}{\includegraphics{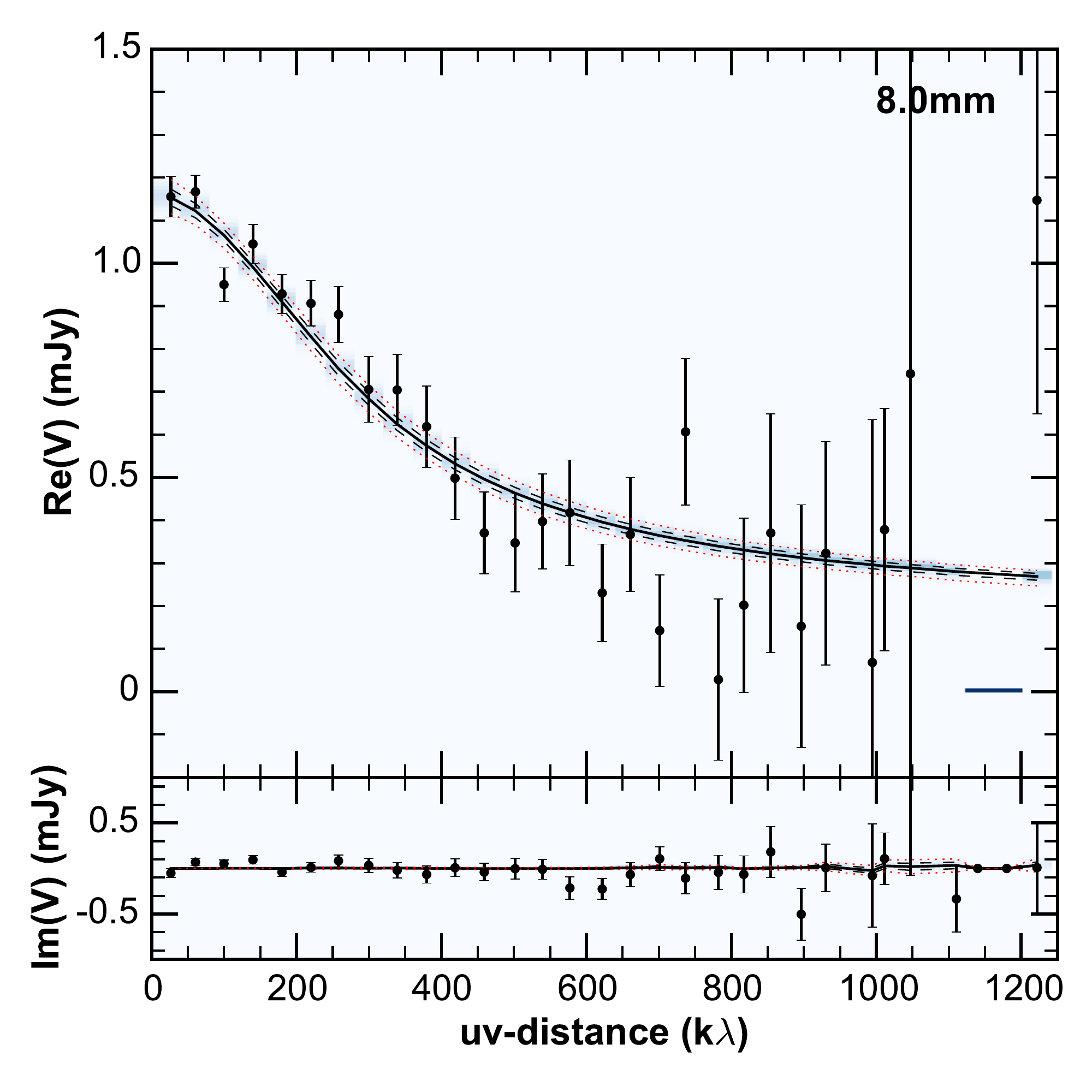}}
\resizebox{0.4\hsize}{!}{\includegraphics{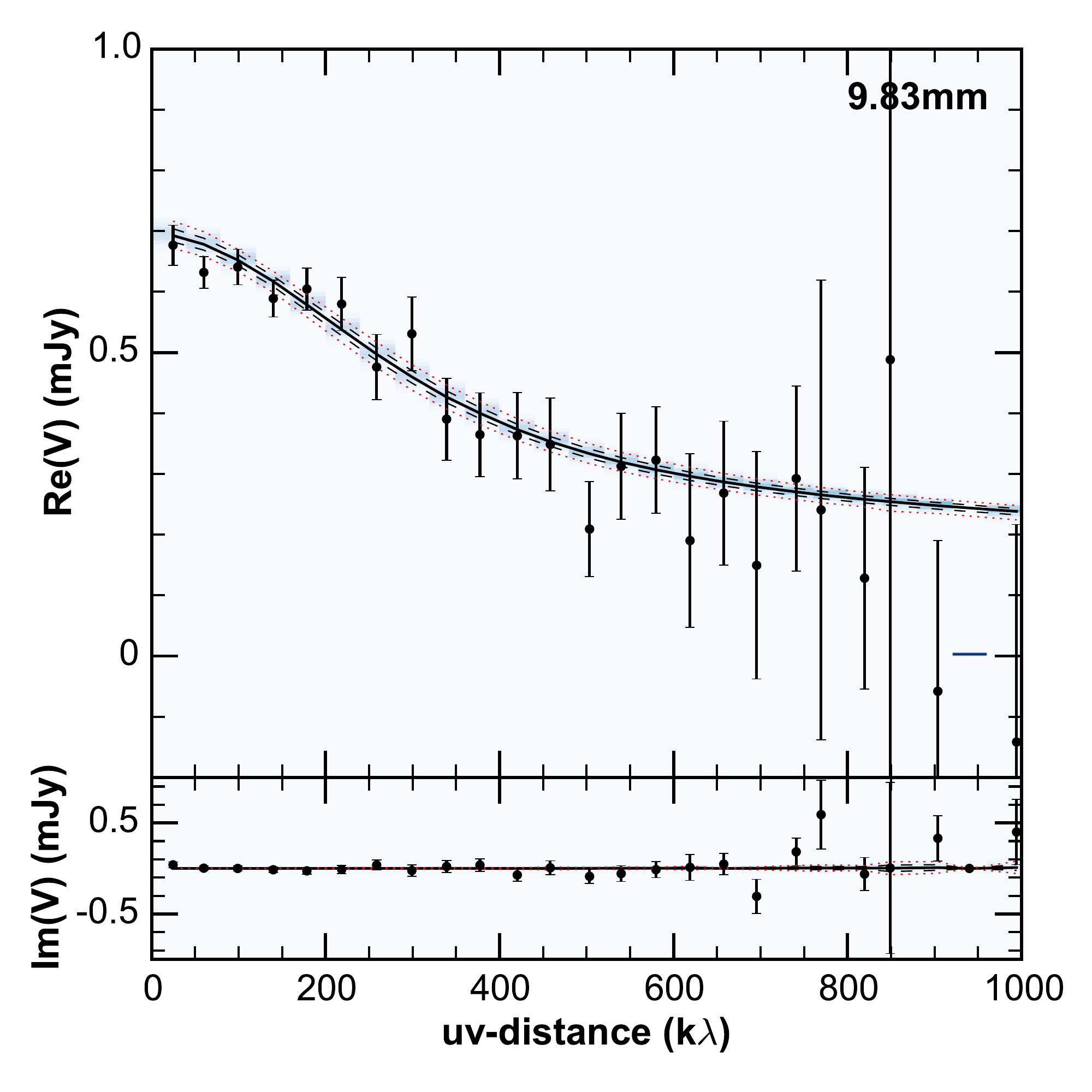}}
\caption{Comparison between the model and the observed visibilities as a function of deprojected baseline length (uv-distance) for FT Tau. The data are binned in 40k$\lambda$ bins. Black dots represent the observed data, the colored boxes represent the probability distribution of model visibilities for each uv-distance bin (the $x$-axis extent of each box is the bin size, while the $y$-axis extent is not fixed as it depends on the probability distribution of the model at that particular uv-distance bin; note that in some cases they are very close to each other). The color scale represents the density of the distribution. The solid black curve is the median, the dashed black lines are the 16\% and 84\% percentiles, and the red dotted lines are the 2.3\% and 97.7\% percentiles.}
\label{fig:fttau.uvplots}
\end{figure*}

Figure \ref{fig:fttau.uvplots} shows the comparison between the probability distribution of the model visibilities and the observations, as a function of the deprojected baseline length (uv-distance). Each panel corresponds to one wavelength, with the upper frame showing the real part ${\mathrm{Re}(V)}$ and the lower frame showing the imaginary part ${\mathrm{Im}(V)}$. %Both the observations and the models are represented in bins of 40${k\lambda}$. 
The declining profile of ${\mathrm{Re}(V)}$ with increasing uv-distance shows that the disk is spatially resolved at all the wavelengths. Furthermore, we note that the visibility profile at longer wavelengths has a steeper decline with increasing baseline than the visibility profile at shorter wavelengths, thus confirming the rather general observational feature that the size of the (sub-)mm emitting region is anti-correlated with the observing wavelength \citep{2012ApJ...760L..17P, Testi:2014kx}.

The models fit the observations with a very good agreement at the shortest wavelength (1.3 mm), and a lower degree of agreement at longer wavelengths. This should not be a surprise for two main reasons: first, the observations at the shorter wavelength also have higher signal-to-noise ratio and therefore have more weight in the fit; secondly, we are modeling all the wavelengths simultaneously, therefore the resulting best-fit models are not necessarily the models that best fit each wavelength separately. We note, indeed, that although the VLA observations at 8.0 and 9.83 mm have a worse signal-to-noise ratio, the models are still able to reproduce the observed total flux density (short uv-distances) and the average flux density at the longer spatial frequencies extremely well. In the case of FT Tau, the observations at 2.6 mm have a very low signal to noise ratio compared to those at 1.3, 8.0 and 9.8 mm.

\begin{figure*}
\centering
\resizebox{0.45\hsize}{!}{\includegraphics{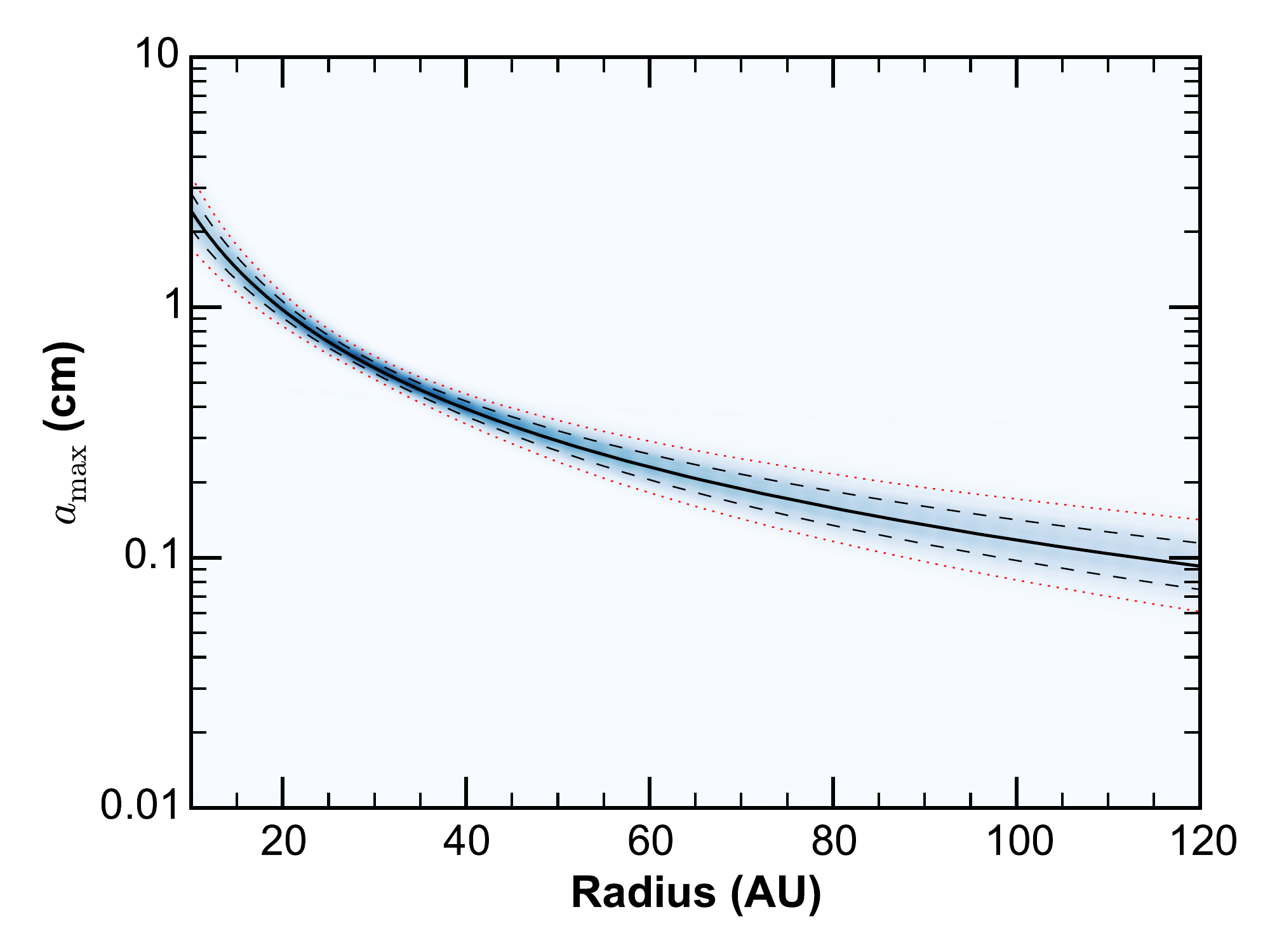}}
\resizebox{0.45\hsize}{!}{\includegraphics{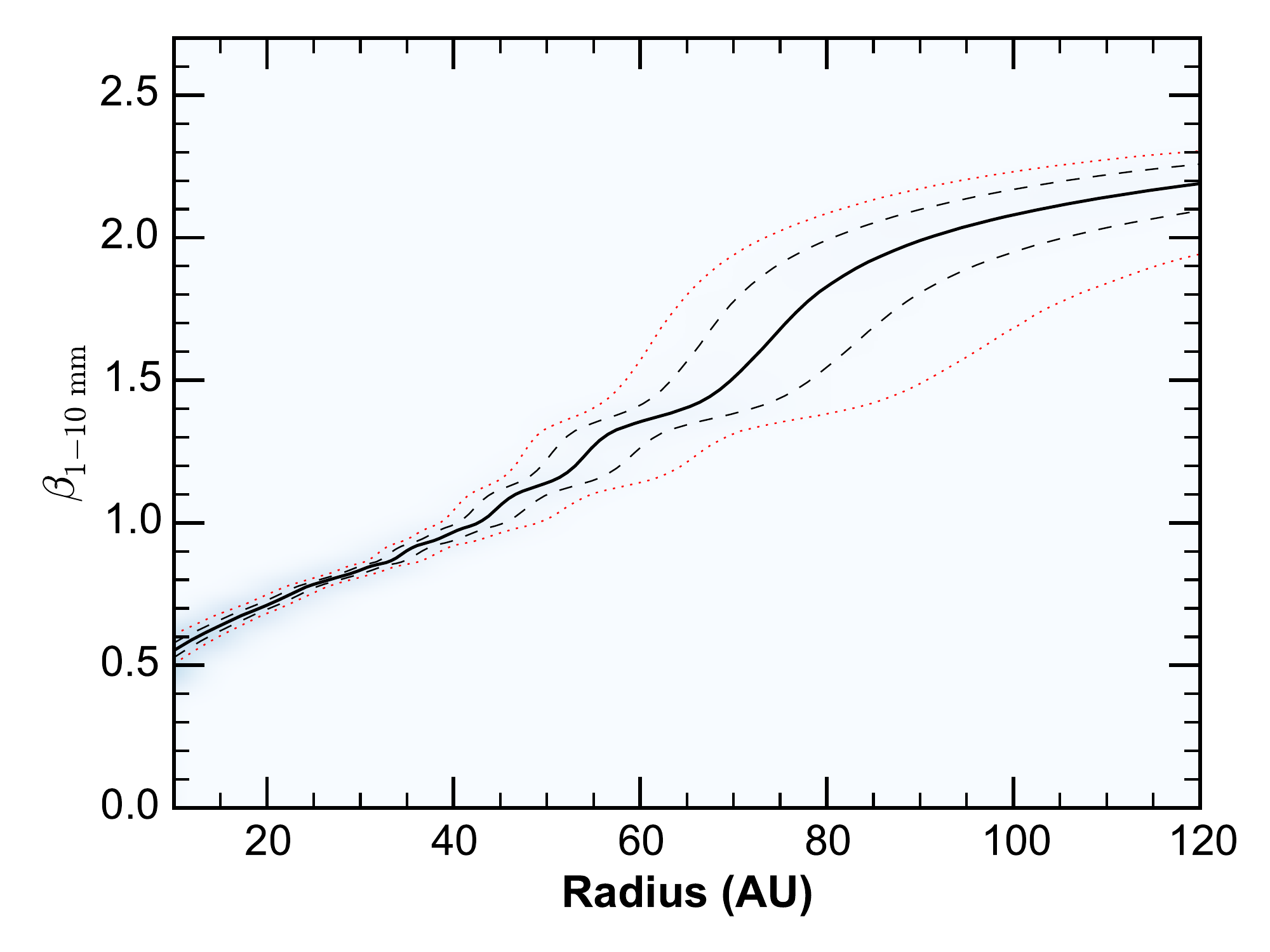}}\\
\resizebox{0.45\hsize}{!}{\includegraphics{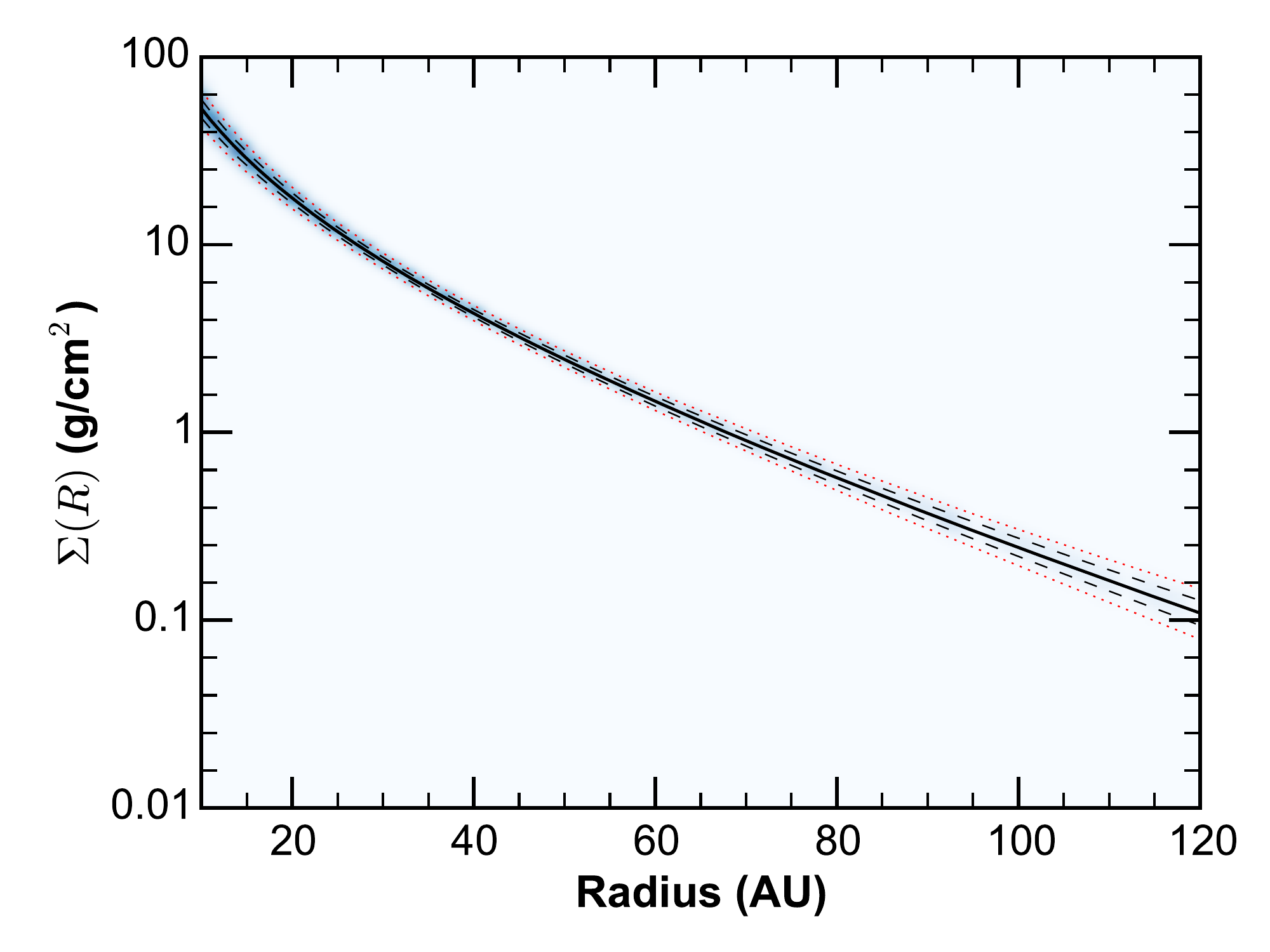}}
\resizebox{0.45\hsize}{!}{\includegraphics{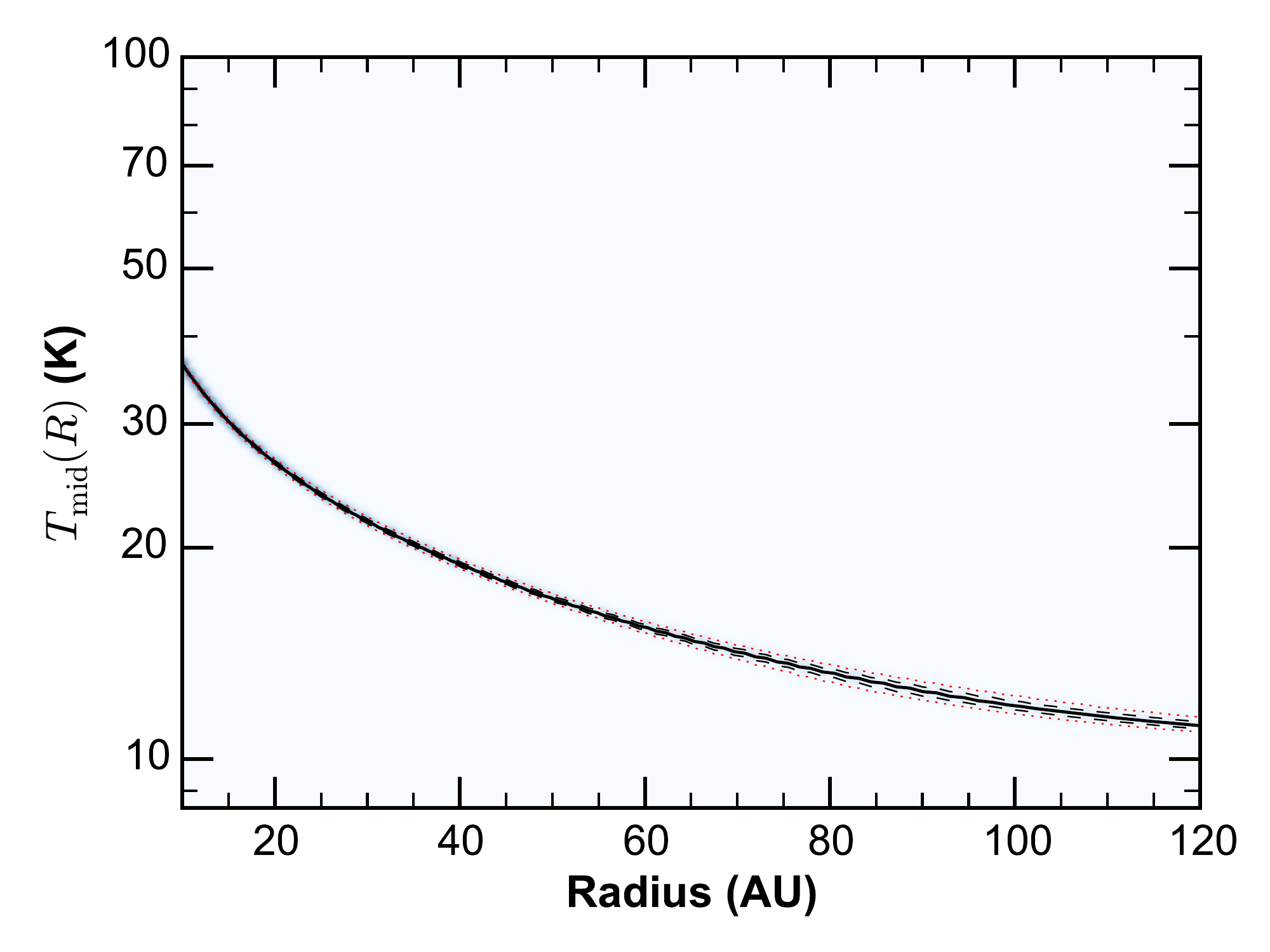}}
\caption{Results of the FT Tau fit. \textit{Top:} in the left panel, the posterior PDF of the maximum dust grain size $a_\mathrm{max}$ as a function of the disk radius; in the right panel, the posterior PDF  of the dust spectral radial profile $\beta(R) $ between 1 and 10 mm. {\textit{Bottom:} in the left panel, the posterior PDF of the gas surface density; in the right panel, the posterior PDF of the midplane temperature.} Line conventions are the same as those in Fig. \ref{fig:fttau.uvplots}.}
\label{fig:fttau.amaxbeta}
\end{figure*}

The top-left panel of Figure~\ref{fig:fttau.amaxbeta} shows the posterior PDF of the maximum dust grain size ${a_{\max}}$ as a function of the disk radius. The maximum dust grain size is larger in the inner disk with respect to the outer disk, changing by one order of magnitude from 1\,cm at 20\,AU to 1\,mm at 120\,AU. Note that the smoothness of the ${a_{\max}(R)}$ profile follows from the power-law parametrization in Eq. \ref{eq:amax.parametrization}. The 2$\sigma$ error bars (given in terms of the 2.3-97.7\% credibility interval) are smaller than 10\% at all radii and allow us to conclude that there is a clear signature of a radial gradient in the maximum dust grain size throughout the disk.
In the top-right panel of Figure \ref{fig:fttau.amaxbeta} we report the posterior PDF of the dust spectral index radial profile ${\beta(R)}$ between 1 and 10\,mm, computed given the $a_{\max}(R)$ posterior PDF according to:
\begin{equation}
\beta(R) = \frac{\partial \log\kappa_\nu(R)}{\partial \log\nu}\,.
\end{equation}
The spectral index ${\beta}$ increases with radius: the small values ${\beta< 1}$ for ${R<50\,}$AU signal the presence of dust grains that have reached sizes comparable to $1\,$mm or more \citep{Natta:2004yu}, whereas in the outer disk the spectral index approaches $\beta\gtrsim \beta_{\mathrm{ISM}}=1.7$, a signature of the presence of smaller grains. The fact that in the outer disk we obtain $\beta\gtrsim \beta_{\mathrm{ISM}}$ for grains somewhat larger ($a_{\max}\approx 1\,$mm) than the usual ISM dust grains ($a_{\max}\approx 1-10\,\mu$m) 
%should not surprise because it 
is consistent with the observations at different wavelengths: these observations give us information on different spatial scales in the disk and they are all fitted well by a model with the $a_{\max}$ profile shown in Figure~\ref{fig:fttau.amaxbeta}. {The bottom plots in Figure~\ref{fig:fttau.amaxbeta} present the physical structure derived for FT~Tau: the gas surface density (bottom-left panel) and the midplane temperature (bottom-right) profiles. The surface density profile is monothonically decreasing, with a power-law index ${\gamma=1.07\pm0.06}$, a normalization value ${\Sigma_{0}=18\pm2\,}$g/cm${^{2}}$ at 40\,AU, and a cut-off radius ${R_{c}=28}$AU. The midplane temperature profile decreases from 40K in the inner disk to 11K in the outer region.}

The AS 209 and DR Tau protoplanetary disks have been fitted with the same analysis presented here and the results are shown in Appendix \ref{app:fits.results}. For both these disks the fit performed well, as can be seen from the maps of the residuals and the comparison between the model and the observed visibilities at all the wavelengths. We note that the observations of DR Tau at 1.3 mm display an asymmetry (in the NE region) that an axisymmetric disk model like the one we are using here is not able to account for. 
In Table \ref{table:fits.results} we summarize the results of the fits for the FT Tau, AS 209 and DR Tau protoplanetary disks. 
In all disks we find sharply decreasing dust grain sizes, with ${a_{\max}\approx 0.5\,}$cm at ${R\lesssim 40}$AU, with a radial power law slope ${-1.8\leq b_{\max} \leq -1.17}$. 
We also notice that AS 209, DR Tau and FT Tau are fitted with ${\gamma>0}$ and ${b_{\max}<0}$. A degeneracy between ${\gamma}$ and ${b_{\max}}$ is also apparent from the bi-variate distributions in Figure~\ref{fig:fttau.triangle} (bottom-left panel) and was already noticed by \citet{2013A&amp;A...558A..64T}.

\begin{table}
\caption{Parameters derived from the fits.}
\centering
\resizebox{\hsize}{!}{
\begin{tabular}{lccccc}
\hline
\hline
Object & ${\gamma}$ & ${\Sigma_{0}}$ & ${R_{c}}$ & ${a_{\max 0}}$ & ${b_{\max}}$ \vspace{1pt}\\
 & & (g/cm$^2$) & (AU) & (cm) & \vspace{3pt}\\
\hline 
AS 209  		& $0.91^{+0.03}_{-0.03}$ & $7.0^{+0.4}_{-0.4}$ & $78^{+3}_{-3}$ & $0.62^{+0.02}_{-0.02}$ & $-1.17^{+0.07}_{-0.07}$ \vspace{4pt} \\
DR Tau 		& $1.10^{+0.08}_{-0.1}$ & $20^{+3}_{-3}$ & $21^{+3}_{-3}$ & $0.24^{+0.03}_{-0.02}$ & $-1.8^{+0.2}_{-0.2}$ \vspace{4pt}\\
FT Tau 		& $1.07^{+0.06}_{-0.06}$ & $18^{+2}_{-2}$ & $28^{+3}_{-3}$ & $0.40^{+0.03}_{-0.03}$ & $-1.3^{+0.1}_{-0.1}$ \vspace{4pt}\\
\hline
\end{tabular}
}
\tablefoot{For each parameter of the fit, we report the median value, with error bars given by the 16\% and 84\% percentiles.}
\label{table:fits.results}
\end{table}

In Table \ref{table:fits.models.physical} we report physical quantities derived from the models: the total mass of the disk $M_\mathrm{disk}$ (computed as the sum of the dust and the gas mass), the radius $R_{90}$ containing 90\% of the disk mass, and the radius $\Rbar$ within which the temperature is computed using the two layer approximation (see Section \ref{sec:disk.model}). 
It is reassuring that for all the disks $\Rbar$ is larger than or comparable to the radius containing the 90\% of the mass, thus implying that the assumption of a power law temperature profile in the region $R>\Rbar$ has minimal influence on the computation of the total flux density. 
We observe that the disk masses that we obtain with our multi-wavelength analysis are comparable within a factor of 2 (or 4 in the worst case) with those derived by previous single-wavelength studies. The \citet{Andrews:2009zr} analysis of AS~209 found $M_{\mathrm{disk}}=28\times 10^{-3}\Msun$ ($\gamma=0.4$, $R_c=126\,$AU), the \citet{2009ApJ...701..260I} analysis of DR~Tau found $M_{\mathrm{disk}}=63\times 10^{-3}\Msun$ ($\gamma=-0.3$, $R_c=41\,$AU), and the \cite{2011A&amp;A...529A.105G} analysis of FT~Tau found $M_{\mathrm{disk}}=7.7\times 10^{-3}\Msun$ ($\gamma=-0.17$, $R_c=43\,$AU). We note that in two cases (DR~Tau and FT~Tau) we obtain $\gamma>0$, whereas past analyses obtained $\gamma<0$. Furthermore, in all the cases we obtain $\gamma$ values larger than previous single-wavelength studies. As noticed by \cite{2013A&amp;A...558A..64T}, this can be understood looking at the anti-correlation between $\gamma$ and $b_{\max}$, clearly visible in the bottom-left frame in the staircase plot in Figure \ref{fig:fttau.triangle}: single-wavelength studies (that adopt an opacity constant with radius and therefore $b_{\max}=0$) obtain smaller $\gamma$ values than a multi-wavelength analysis, where $b_{\max}$ and $\gamma$ are constrained simultaneously.

\begin{table}
\caption{Models: physical quantities}
\centering
\begin{tabular}{lccc}
\hline
\hline
Object & $M_\mathrm{disk}$ & $R_{90}$ & $\Rbar$\vspace{5pt}\\
 & ($10^{-3}\Msun$) & (AU) & (AU) \\
\hline 
AS 209	& $14.9^{+0.4}_{-0.4}$ 	& $161^{+5}_{-3}$ 	& $155^{+5}_{-4}$ \vspace{4pt} \\
DR Tau	& $14^{+4}_{-1}$ 	& $52^{+3}_{-3}$ 		& $65^{+4}_{-2}$  \vspace{4pt} \\
FT Tau	& $15^{+1}_{-1}$ 	& $69^{+6}_{-4}$ 		& $99^{+7}_{-6}$   \vspace{4pt} \\
\hline
\end{tabular}
\tablefoot{For each disk, we report the total disk mass  $M_\mathrm{disk}$, the radius $R_{90}$ within which the 90\% of the disk mass resides and the radius $\Rbar$ within which the disk temperature is computed with the two layer disk model.}
\label{table:fits.models.physical}
\end{table}

\section{Discussion}
\label{sec:discussion}
In the last section we showed how our new multi-wavelength fitting technique allows us to constrain simultaneously the disk structure and the radial variation of the maximum dust grain size. 
This is the most important difference between our method and previous attempts to use multi-wavelengths observations to constrain the dust properties. {We derive a unique, yet simplified, disk physical structure which describes the emission at all the observing wavelengths and at the same time we obtain a self-consistent distribution} of particle sizes which we assume to be a continuous function. Previous analyses have either assumed a non-self-consistently derived temperature distribution across the disk \citep{2011A&amp;A...529A.105G} or have used different disk physical structure fits at different wavelengths to infer from their differences a constraint on the dust properties \citep{Banzatti:2011ff,2012ApJ...760L..17P}. The method we developed, extending the work of \citet{2013A&amp;A...558A..64T}, improves on previous results by attempting a self-consistent modeling of the disk structure and dust radial stratification. Our models produce results that are in qualitative agreement with previous studies (larger grains in the inner disk than in the outer disk), but do show some quantitative difference, even when using the same assumptions about the dust composition.

We used the AS~209 protoplanetary disk to perform a detailed comparison of the results from our multi-wavelength analysis with those of \cite{2012ApJ...760L..17P}, who constrained the disk structure and the dust radial distribution fitting each wavelength separately. The details of the comparison are reported in Appendix \ref{app:code.benchmark}, while here we briefly summarize the main results. Adopting the same disk model and the same dust properties as \cite{2012ApJ...760L..17P} we fitted the AS 209 protoplanetary disk separately at each wavelength and found an extremely good agreement with the disk structure obtained by \cite{2012ApJ...760L..17P}. Then, we performed a multi-wavelength fit again using the same disk model and dust prescriptions used by \cite{2012ApJ...760L..17P} and we compared the resulting $a_{\max}(R)$ and $\beta(R)$ profiles.
The two techniques provide $a_{\max}(R)$ profiles in good agreement almost throughout the disk, with some differences in the inner region where the emission is not spatially resolved. 
The main differences between the two approaches arise from the different derivation of the dust temperature profile. The modeling by 
\citet{2012ApJ...760L..17P} produces independent temperature profiles at different wavelengths, whereas our multi-wavelength fit derives a unique temperature profile for the disk midplane that holds at all wavelengths (this is made possible by imposing a fixed parametrization of the maximum grain size with radius, in our case a power law\footnote{This choice is justified by the outer disk maximum grain size distribution derived by other authors and by simple fits to the predictions of global dust evolution models, e.g., \citet{Birnstiel:2012yg}.}). Obtaining an accurate estimate of the temperature in the outer disk is difficult, due to the low optical depth to the stellar radiation and the low temperature reached by the disk midplane (moreover, other external heating effects possibly start to play a role). Given the expectation that within the range of wavelengths at which we observe the major contribution of the emission always comes from the midplane dust, modeling such emission with a unique temperature profile is more physically founded than using several different temperature profiles that apply in different disk regions. This consideration constitutes the main motivation for the development of our joint multi-wavelength analysis. 

\begin{figure*}
\centering
\resizebox{0.49\hsize}{!}{\includegraphics{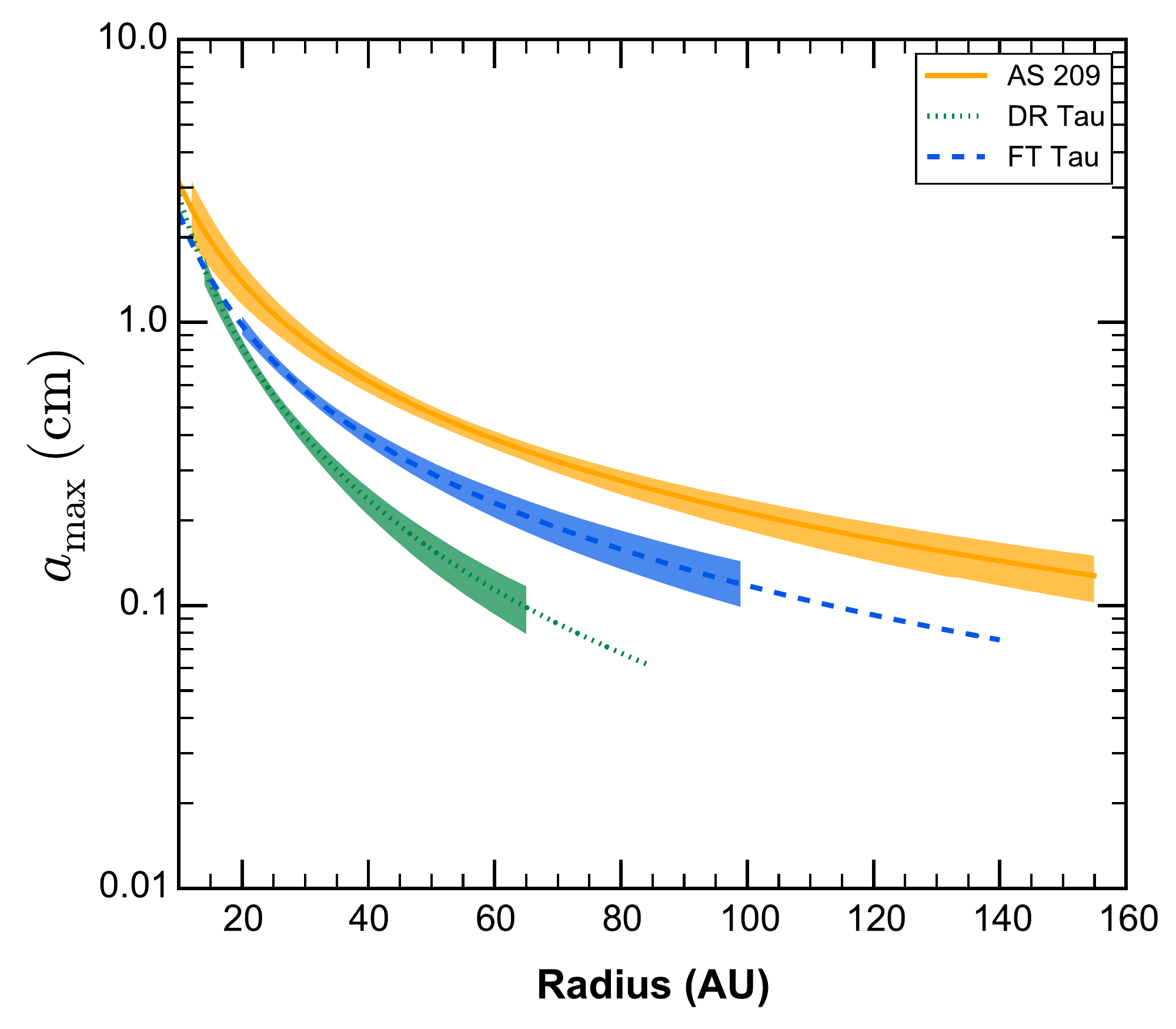}}
\resizebox{0.49\hsize}{!}{\includegraphics{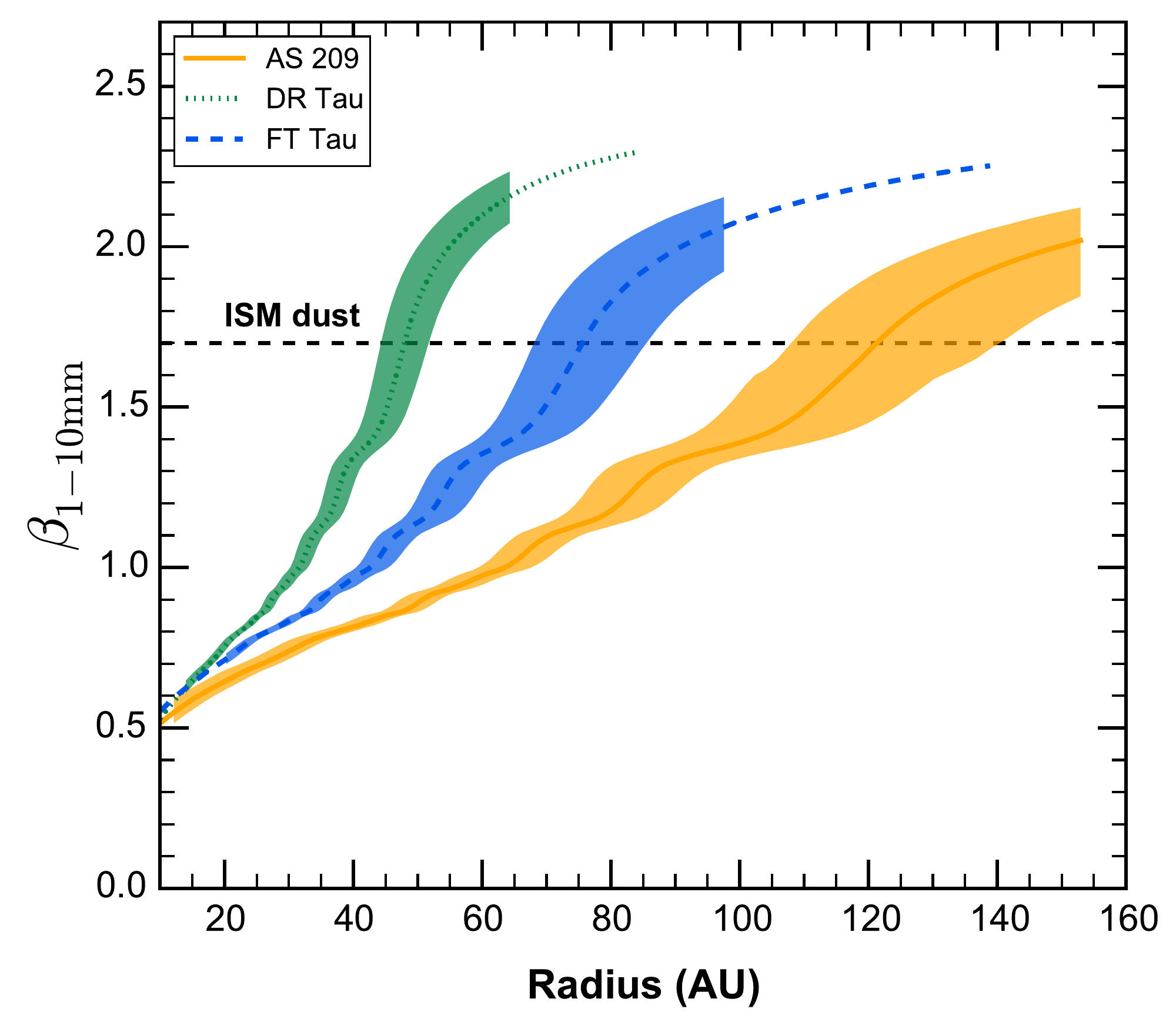}}
\caption{\textit{Left panel:} radial profile of the maximum dust grain size ${a_{\mathrm{max}}}$ constrained from the multi-wavelength fits. \textit{Right panel:} radial profile of the dust opacity spectral slope ${\beta(R)}$ between 1\,mm and 10\,mm. The dashed black horizontal line at ${\beta_{\mathrm{ISM}}=1.7}$ represents the typical value of $\beta$ for small ISM dust grains. \textit{In both panels:} the thick lines represent the median (i.e., the best-fit) model, and the shaded areas represent the 1$\sigma$ credibility intervals. The best-fit model lines are plotted wherever the signal-to-noise ratio is larger than 3 (computed for the observation displaying the most extended disk emission); the shaded areas are truncated at half the average beam size (inner regions) and at $R=\Rbar$ (outer regions).}
\label{fig:comparison.amaxbeta}
\end{figure*}

In the left panel of Figure \ref{fig:comparison.amaxbeta} we compare the radial profiles ${a_{\max}(R)}$ obtained for the disks in the sample. 
The observed declining profile of ${a_{\max}(R)}$ with radius is in line with the expected outcome of the viscous evolution of disks according to which the smaller dust particles (more coupled with the gas) are brought to large stellocentric distances, whereas the larger dust grains (less coupled with the gas and more sensitive to the gas drag) drift inwards  \citep{Weidenschilling:1977lr,Brauer:2008kq,Birnstiel:2010jk, Armitage:2010fk}.
In both figures, we plot the best-fit models as lines and the 1$\sigma$ credibility intervals as shaded areas. The best-fit models are truncated at the radius where the signal-to-noise ratio of the observations becomes smaller than 3, computed for the observation displaying the most extended disk emission. The shaded areas are truncated in the inner region at half the synthesized beam size to give a visual representation of the average angular resolution of the observations, and in the outer regions at $R=\Rbar$ . All the objects support evidence of large grains in the inner disk ($a_\mathrm{max}\approx 1\,$cm) and smaller grains in the outer disk ($a_\mathrm{max}\lesssim 3\,$mm), with changes in size by at least one order of magnitude. 
Given the constrained $a_{\max}(R)$ profiles, we compute the dust spectral index $\beta(R)$ profiles (shown in the right panel of Figure \ref{fig:comparison.amaxbeta}), which grow accordingly from $\beta\approx 0.5$ in the inner disk to $\beta\gtrsim 1.7$ in the outer disk. These findings confirm the earlier evidence of  $\beta(R)$ increasing with radius obtained by \cite{2012ApJ...760L..17P}, \cite{Banzatti:2011ff} and \citet{2011A&amp;A...529A.105G} with different techniques, and by \citet{2013A&amp;A...558A..64T} with an initial implementation of this joint multi-wavelength analysis. 

The ${a_{\max}(R)}$ radial profiles derived for the three disks tend towards similar values ${a_{\max}(R)\simeq}2\,$cm in the innermost spatially resolved region $10\,\mathrm{AU}<R<20\,$AU but display apparent differences in the slope with AS~209, FT~Tau and DR~Tau showing respectively an increasing steepness. The differences we observe in the slopes may be due to several factors: different disk ages (grain growth processes can lead to time-dependent grain size distributions), different initial grain size distributions of the primordial material out of which the disks formed, different dust compositions, and/or different disk morphologies. The limited sample of disks analyzed here clearly does not enable us to investigate in detail these effects on the dust size distribution, but it is remarkable to note the overall similarity between the profiles. The three disks AS~209, FT~Tau and DR~Tau appear to have progressively more concentrated large grains. Extension of our analysis to a larger sample of objects will possibly allow us to understand what drives these differences in the overall distribution of grain sizes in disks.
Our work here lays down the methodology for this type of study.

\section{Conclusions and outlook}
This paper presents the architecture and the capabilities of a new multi-wavelength analysis designed to constrain the structure and the dust properties of protoplanetary disks through a simultaneous fit of interferometric (sub-)mm observations at several wavelengths. The analysis adopts a Bayesian approach and performs a fit in the uv-plane. It requires models for the disk thermal emission and the dust opacity. The architecture of the analysis is highly modular (the disk and the dust models can be changed independently of each other) and therefore is particularly suitable to test other models for the dust opacity or the disk structure (e.g., disks with holes, or with non-axisymmetric morphology). 

For this study, we have modeled the disk with a two layer disk approximation \citep{1997ApJ...490..368C,2001ApJ...560..957D} and the dust opacity with Mie theory. We have applied the fit technique to three protoplanetary disks (AS 209, DR Tau, FT Tau) for which sub-mm, mm and cm observations were available. We have combined observations from CARMA, SMA, and VLA interferometers, with different angular resolution and signal-to-noise ratios. Despite the heterogeneity of the observations, the analysis technique has proven to be effective in fitting simultaneously all the datasets available for each object, as the visibility comparisons and the residual maps show. Furthermore, the convergence of the Markov Chain Monte Carlo was assessed through careful statistical checks. 
The strength of our method lies in the fact that it allows us to derive a unique and self-consistent disk structure ($\Sigma(r)$, T(r)) that is applied to all wavelengths to derive the overall variation of the maximum grain size with radius under the simplifying assumption that the radial profile $a_{\max}(R)$ can be approximated with a smooth power law (which is a realistic assumption given the angular resolution of the observations we are analyzing here).

In the three disks analyzed here, we have found a common trend of larger (cm-size) grains in the inner disks (R<30-40\,AU) and smaller (mm-size) grains in the outer disks, but different slopes of ${a_{\max}(R)}$ for different disks. A natural question that arises is whether this is an evolutionary trend (caused by dust growth processes) or an intrinsic variability of disk properties. It is not possible to answer this question with the very limited sample we are analyzing here, {but the analysis method is ready to be performed} on larger samples that will have the potential of giving us some insight in possible correlations and intrinsic variability.

The highly modular architecture of the analysis makes it suitable to test other dust opacity and disk models with relatively little effort from the coding point of view. From this perspective, it will become important to develop new models that are, on the one hand more computationally efficient, and on the other hand refined enough to describe the complex structures now seen in the dust and gas distribution of protoplanetary disks (e.g., \citealt{2041-8205-808-1-L3}). The versatility of the method makes this kind of multi-wavelength analysis suitable for tackling many interesting questions about the dust and the gas evolution in protoplanetary disks.

\begin{acknowledgements}
MT and LT acknowledge support by the DFG cluster of excellence
Origin and Structure of the Universe (\href{http://www.universe-cluster.de}{www.universe-cluster.de}). A.I. acknowledge support from the NSF award AST-1109334/1535809 and from the NASA Origins of Solar Systems program through the award number NNX14AD26G. The fits have been carried out on the computing facilities of the Computational Center for Particle and Astrophysics (C2PAP) as part of the approved project ``Dust evolution in protoplanetary disks''. MT and LT are grateful for the experienced support by F. Beaujean (C2PAP). Figures have been generated using the Python-based \texttt{matplotlib} package \citep{Hunter:2007fk}. Staircase plots of PDFs have been generated with a user-modified version of the Python-based \texttt{triangle} package \citep{dan_foreman_mackey_2014_11020}. Plots of the residuals have been generated with APLpy, an open-source plotting package for Python hosted at http://aplpy.github.com.
This research has made use of the SIMBAD database,
operated at CDS, Strasbourg, France.
This work was partly supported by the Italian Ministero dell\'\,Istruzione, Universit\`a e Ricerca through the grant Progetti Premiali 2012 -- iALMA (CUP C52I13000140001).
\end{acknowledgements}

\appendix
%%%%%%%%%%%%%%%%%%%%%%%%%%%%%%%%%%
%%%% APPENDIX BAYESIAN DETAILS %%%
%%%%%%%%%%%%%%%%%%%%%%%%%%%%%%%%%%
\section{Bayesian analysis: MCMC details and implementation}
\label{app:mcmc.details}
{
To efficiently compute the posterior distributions for all the free parameters (the 5 physical parameters that define the disk model plus 2 offset parameters per each fitted wavelength) we use a variant of the Markov Chain Monte Carlo (MCMC) algorithm \citep{Mackay:2003bs,Press:2007nr} which proved to be effective in finding global maxima for a wide range of posteriors. 
We adopt an affine-invariant ensemble sampler for MCMC proposed by \citet{Goodman:2010} which is designed to run simultaneously several Markov chains (also called \textit{walkers}) interacting with each other to converge to the maximum of the posterior. For unimodal posteriors (i.e., posteriors that exhibit only one global maximum) this algorithm is rather efficient in avoiding getting stuck in local maxima and allows the computation to be massively parallelized over the walkers. According to our experience, we have always observed unimodal posteriors, thus making this algorithm particularly suitable for our purpose.

In this study, we let the MCMC (usually of 1000 walkers) evolve for an initial \textit{burn-in} phase, which is needed to allow the MCMC to perform a reasonable sampling of the parameter space and to find the posterior maximum (usually this is achieved after 10 \textit{autocorrelation times}\footnote{The autocorrelation time of a MCMC is an estimate of the number of posterior PDF evaluations needed to produce a large number of independent samples of the target density.}). After the burn-in phase, we let the MCMC run for other several (3-4) autocorrelation times to get a sufficient number of independent posterior samples. Since two consecutive steps in the MCMC are correlated \citep{Goodman:2010}, in order to extract a set of independent posterior samples out of the whole MCMC, we need: (1) to discard the samples of the burn-in phase and (2) to thin the remaining chain, i.e., to consider only steps separated by one autocorrelation time discarding all the steps between them\footnote{In our case, since the autocorrelation time is usually observed to be smaller than 100 steps, the thinning does not reduce the effective samples size, but allows us to save a lot of computational time during post-processing, e.g., when producing the uv-plots that show the comparison between the observed visbilities and the density of synthetic visibilities.}. For completeness, we note that to estimate the chain convergence we have also analyzed the \textit{acceptance ratio}, i.e., the ratio of accepted over proposed moves, verifying it to be within the acceptable range between 0.2 and 0.5 in all cases. The number of steps needed to achieve convergence varies from disk to disk and depends on several factors, thus making it not predictable {\it a priori}; for the disks analyzed here, we needed at most 2000 steps including burn-in. Note that with 1000 walkers, 2000 steps and an acceptance fraction of $\sim 0.2-0.5$, the method requires the computation of several million models and likelihoods, hence it is necessary to exploit the efficient Message Passing Interface (MPI) parallelization of the computation by advancing several hundred walkers in parallel.

As explained in Section~\ref{sec:fitting.method}, in addition to the disk structure and the dust size distribution, we also fit the position of the disk centroid by adding two \textit{nuisance} parameters for each wavelength, namely $\Delta \alpha_0$  and $\Delta \delta_0$. To implement these offset parameters we exploit the fact that a translation in the real space corresponds to a phase shift in the conjugate (Fourier) space. Therefore, to shift the disk emission computed by the model by $(\Delta \alpha_0,\ \Delta \delta_0)$ on the sky, we multiply the model visibilities $V^{\mathrm{mod}}_{j}(u,v)$ by the phase-shift $\exp\left[2\pi i(u\Delta \alpha_0/\lambda+v\Delta \delta_0/\lambda)\right]$, with $\Delta \alpha_0$ and $\Delta \delta_{0}$ given in radian units and $\lambda$ in meters. 

From the computational point of view, the main architecture of the {analysis} is written in Python and delegates to C-- and Fortran--compiled external libraries the most demanding tasks, like the disk model evaluation or the visibility sampling. Writing the main architecture of the {analysis } in Python allows us to use the affine-invariant MCMC algorithm implemented in the Python-based \textit{emcee} package\footnote{The code can be found at  \href{https://github.com/dfm/emcee}{https://github.com/dfm/emcee}.} which enables a massive parallelization of the overall computation. By far the most costly part of the sampling is the evaluation of the posterior, which \textit{emcee} allows to do simultaneously for half the walkers (it exploits the Message Passing Interface (MPI) protocol do distribute the computation to several cores). In our case, we ran the fits on hundreds of cores hosted at the Computational Center for Particle and Astrophysics (C2PAP), decreasing the overall computational time to the order of one or two days. After a careful profiling of the {analysis method}, we noted that the bottleneck of the posterior evaluation (i.e., the single walker computation) is given by the several Fourier Transform computations\footnote{To compute the Fourier Transforms we use the numpy implementation.} (one for each wavelength that is being fitted) and by their sampling at the discrete locations where the antennas sampled the sky. 
}

%%%%%%%%%%%%%%%%%%%%%%%%%%%%%%%%%%
%%%% APPENDIX BENCHMARK AS209 %%%%
%%%%%%%%%%%%%%%%%%%%%%%%%%%%%%%%%%
\section{AS 209: Comparison with previous study}
\label{app:code.benchmark}
We report the results of the comparison between our analysis and the previous one by \citet{2012ApJ...760L..17P}. As described below, the comparison is performed using the same observational datasets, the same disk model and the same dust opacity used by \citet{2012ApJ...760L..17P}. 

As presented in Section \ref{sec:fitting.method}, our analysis consists of a self-consistent modeling of the disk structure and the radial distribution of the dust grains that provides us with a unique model that is capable of reproducing the multi-wavelength observations simultaneously. In this framework, 
the radial profile of $a_{\max}$ is constrained from observations at several wavelengths.
Then, from the resulting $a_{\max}(R)$ profile we derive the corresponding $\beta(R)$ profile through the dust opacity model.

The analysis of \citet{2012ApJ...760L..17P} consists of two steps: first, assuming a constant $a_{\max}$ throughout the disk they separately fit the observations at different wavelengths and obtain radial profiles of the disk temperature and the optical depth $\tau_\lambda(R)$; then, assuming that the disk surface density $\Sigma$ is unique, they interpret the wavelength dependence of $\tau_\lambda$ in terms of radial variations of $\beta$. In the end, they provide constraints on $a_{\max}(R)$ by fitting the $\beta(R)$ profile with a dust opacity model. In order to check whether we recover the results of \cite{2012ApJ...760L..17P} we first fit each wavelength separately. Subsequently, we perform a multi-wavelength fit with the same disk and dust model and compare the results.

\begin{figure}
\centering
\resizebox{\hsize}{!}{\includegraphics{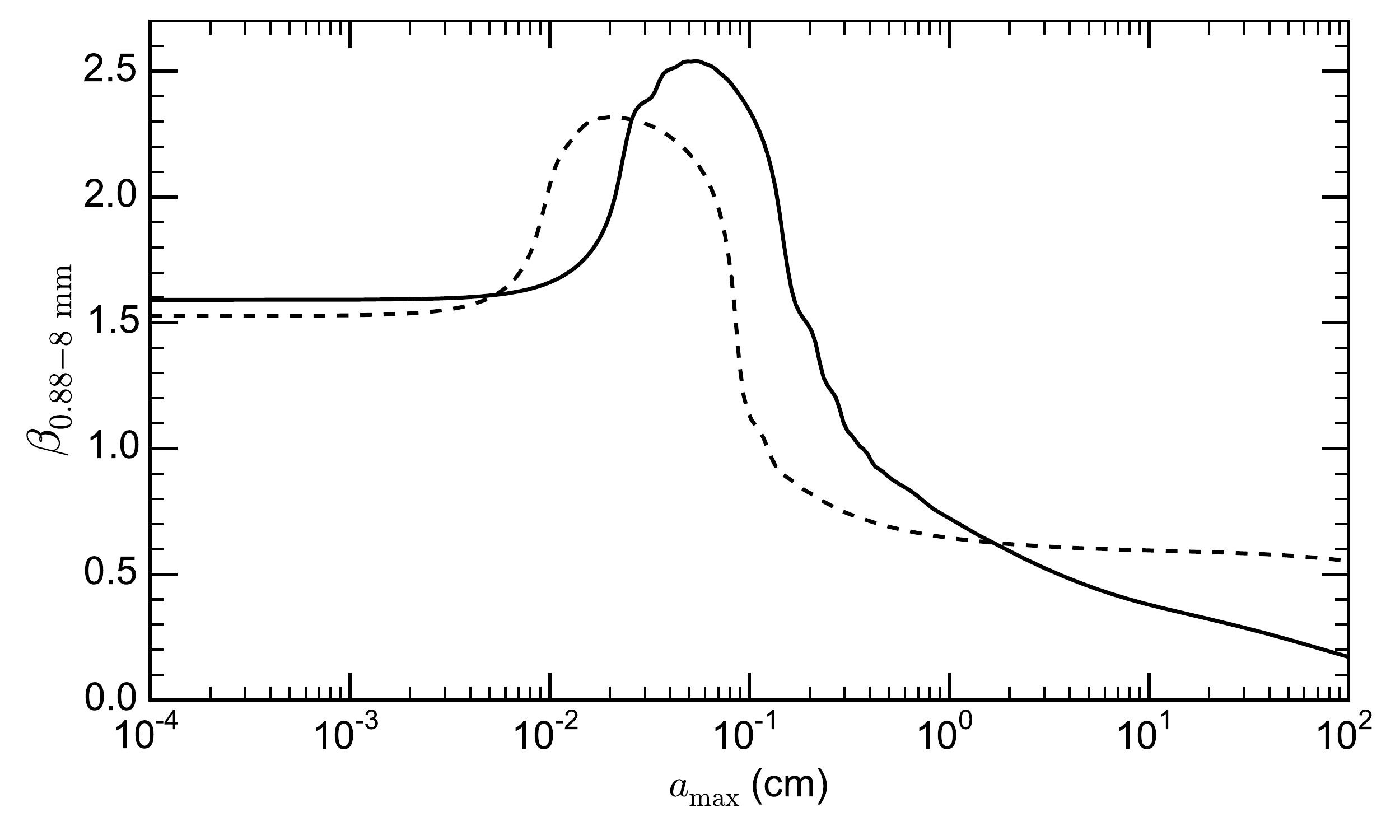}}
\caption{Dust opacity spectral index $\beta$ between 0.88 and 8 mm  as a function of the maximum dust grain size $a_{\max}$. The solid line refers to our dust composition ($q=3$, 5.4\% astronomical silicates, 20.6\% carbonaceaous material, 44\% water ice and 30\% vacuum), whereas the dashed line refers to the dust composition used by \cite{2012ApJ...760L..17P} ($q=3.5$, 7.7\% astronomical silicates, 29.5\% carbonaceaous material, 62.8\% water ice).}
\label{fig:beta.vs.amax.Perez.composition}
\end{figure}

We perform single-wavelength fits of each observation of the AS 209 protoplanetary disk, using the same setup for the disk model and the dust properties used by \citet{2012ApJ...760L..17P}. For the disk model, we use the two-layer model described by \citet{2009ApJ...701..260I} which assumes the following surface density profile \citep{1998ApJ...495..385H}:
\begin{equation}
\Sigma(R) = \Sigma_{\mathrm{T}}\left(\frac{R}{R_{\mathrm{T}}} \right)^{-\gamma}\exp\left\{-\frac{1}{2\left(2-\gamma \right)}\left[\left(\frac{R}{R_{\mathrm{T}}} \right)^{2-\gamma}-1 \right] \right\}\,,
\end{equation}
where ${R_{\mathrm{T}}}$ is the radius at which the radial component of the gas velocity changes sign (gaseous material at ${R<R_{\mathrm{T}}}$ moves inwards, at ${R>R_{\mathrm{T}}}$ moves outwards). The dust size distribution is defined with the parametrization in Eq. \eqref{eq:amax.parametrization}, with ${a_{\min}^{\mid}=a_{min}^{sur}=10\,}$nm, ${a_{\max}^{\mid}=1.3\,}$mm constant throughout the disk (${b_{\max}^{\mid}=0}$), and ${q_{\mid}=q_{\sur}=3.5}$. The dust grains are assumed to be compact spherical grains made of astronomical silicates (7.7\%), carbonaceaous material (29.5\%) and water ice (62.8\%) with an average dust grain density 0.9\,g/cm$^3$ (the correct value would be 1.3\,g/cm$^3$, but we adopt 0.9\,g/cm$^3$ in order to have the same setup used by \citealt{2012ApJ...760L..17P}). The dust opacity is computed through Mie theory as described in Section \ref{sec:dust.model} using the same optical constants. In Figure \ref{fig:beta.vs.amax.Perez.composition} we show a comparison between the $\beta(a_{\max})$ profile for the dust we used in our joint multi-wavelength fits (presented in Section \ref{sec:results}) and the dust used by \cite{2012ApJ...760L..17P}.

In Table \ref{tab:as209.comparison.perez} we report the comparison of the single wavelength fits. 
\begin{table}
\centering
\caption{Comparison of the single wavelength fits of AS 209 performed with our {analysis} and the results reported by \citet{2012ApJ...760L..17P}.}
\begin{tabular}{l c c c c}
\hline
\hline
${\lambda}$ & ${\gamma}$ & ${\Sigma_{\mathrm{T}}}$ & ${R_{\mathrm{T}}}$ \vspace{3pt} & Ref.\\
\hline 
0.88 	& $0.25^{+0.04}_{-0.05}$ & $0.44^{+0.02}_{-0.02}$ & $60^{+2}_{-2}$ & 1\vspace{5pt} \\
 		& ${0.20^{+0.03}_{-0.05}}$ & ${0.43^{+0.02}_{-0.01}}$ & ${61^{+1}_{-2}}$ & 2 \vspace{5pt} \\
2.80 	& $0.76^{+0.09}_{-0.09}$ & $0.38^{+0.07}_{-0.07}$ & $69^{+6}_{-9}$ & 1\vspace{4pt}\\
		& ${0.60^{+0.10}_{-0.05}}$ & ${0.47^{+0.03}_{-0.09}}$ & ${60^{+7}_{-3}}$ & 2 \vspace{5pt} \\
8.00 	& $0.364^{+0.14}_{-0.17}$ & $1.5^{+0.3}_{-0.3}$ & $24^{+2}_{-4}$ & 1\vspace{4pt}\\
		& ${0.36^{+0.09}_{-0.18}}$ & ${1.75^{+0.39}_{-0.21}}$ & ${24^{+1}_{-3}}$ & 2 \vspace{5pt} \\
9.83 	& $0.375^{+0.16}_{-0.17}$ & $1.748^{+0.37}_{-0.34}$ & $27^{+3}_{-4}$ & 1 \\
		& ${0.31^{+0.15}_{-0.18}}$ & ${1.97^{+0.47}_{-0.31}}$ & ${26^{+2}_{-4}}$ & 2 \vspace{5pt} \\
\hline
\end{tabular}
\label{tab:as209.comparison.perez}
\tablefoot{For each parameter, we report the median value, with uncertainties given by the 16\% and 84\% percentiles of its marginalized distribution.}
\tablebib{(1) This work; (2)  \citet{perez:phdthesis}.}
\end{table}
The agreement between our results and those by \cite{2012ApJ...760L..17P} is extremely good, with all the values compatible within 1$\sigma$ and only in few cases within 2$\sigma$. Similarly to \cite{2012ApJ...760L..17P} we derive larger disks ($R_{\mathrm{T}}\gtrsim 60\,$AU) at the shorter wavelengths and smaller disks ($R_{\mathrm{T}}\approx 25\,$AU) at the longer wavelengths, thus confirming the observational result that the size of emitting region is anticorrelated with the observing wavelength. As a further check, we note that the estimates of the uncertainty we obtain from our MCMC fits are similar to those obtained by \cite{2012ApJ...760L..17P}.

\begin{figure*}
\resizebox{0.5\hsize}{!}{\includegraphics{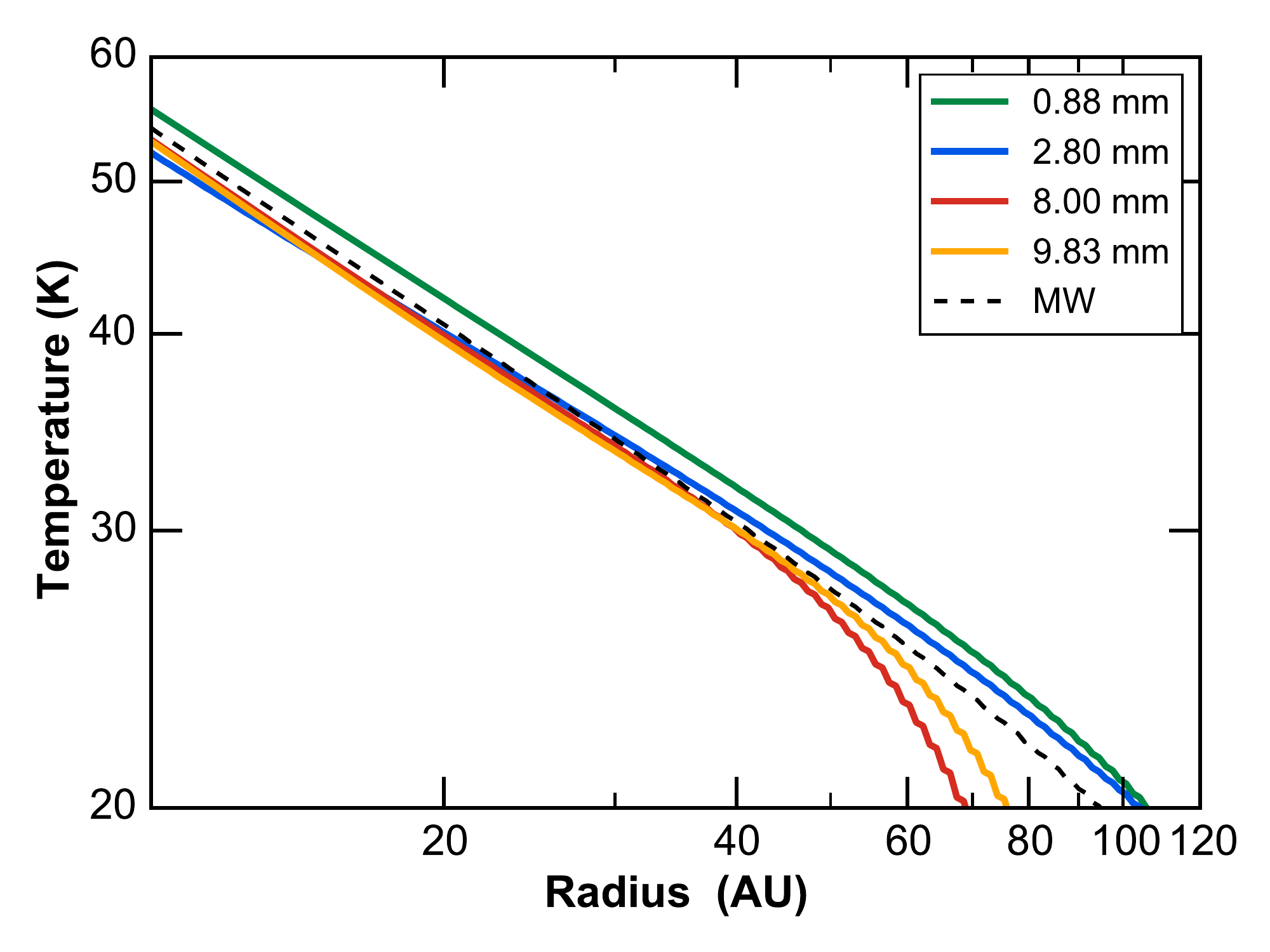}}
\resizebox{0.5\hsize}{!}{\includegraphics{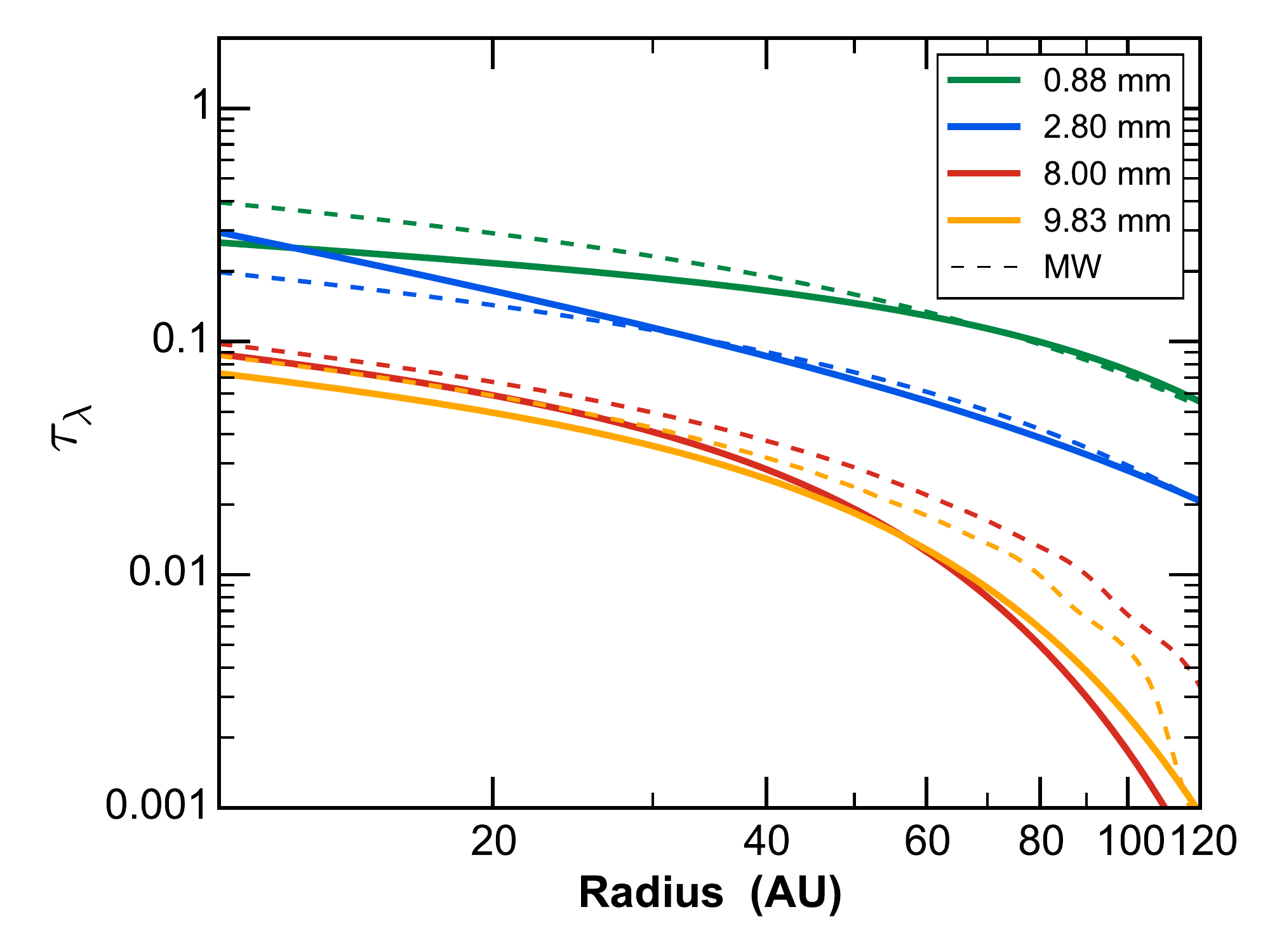}}
\caption{
\textit{Left panel:} best-fit model midplane temperature obtained from fitting separately each wavelength (solid lines, one line per each wavelength) or from our multi-wavelength fit (dashed line). \textit{Right panel:} for the same best-fit models, the optical depth $\tau_{\nu}=\kappa_\nu \Sigma$ of the disk midplane to its own thermal radiation. \textit{Both panels:} the single-wavelength fits have been performed assuming a constant $a_{\max} = 1.3\,$
mm (i.e. constant dust opacity) throughout the disk. The dashed lines refer to the best-fit model obtained through multi-wavelength modeling.}
\label{fig.comparison.perez.structure}
\end{figure*}

In Figure \ref{fig.comparison.perez.structure} we show the midplane temperature profiles (left panel) and the optical depth profiles $\tau_{\nu}=\kappa_\nu \Sigma$ (right panel) obtained with the single wavelength fits. We remark that both the temperature and the optical depth profiles obtained with our single wavelength modeling are found to be in complete agreement with those computed by \cite{2012ApJ...760L..17P}. The agreement occurs at all the fitted wavelengths.

In Table \ref{table:as209.perez.mw.results} we report the results of the multi-wavelength fit that has been performed with the same disk model and dust assumptions as the single wavelength fits, with the only difference that in the multi-wavelength fit $a_{\max}$ is not constant throughout the disk and its radial profile is constrained from the observations at several wavelengths.
\begin{table}
\caption{Parameters derived from the multi-wavelength fit of AS 209.}
\centering
\begin{tabular}{ccccc}
\hline
\hline
${\gamma}$ & ${\Sigma_{\mathrm{T}}}$ & ${R_{\mathrm{T}}}$ & ${a_{\max0}}$ & ${b_{\max}}$ \vspace{1pt}\\
 & (g/cm$^2$) & (AU) & (cm) & \vspace{2pt}\\
\hline 
$0.91^{+0.08}_{-0.04}$ & 
$0.81^{+0.07}_{-0.08}$ &
$48^{+3}_{-4}$ &
$0.37^{+0.04}_{-0.05}$ &
$-1.3^{+0.1}_{-0.1}$ \vspace{2pt}\\
\hline
\end{tabular}
\tablefoot{For each parameter of the fit, we report the median value, with error bars given by the 16\% and 84\% percentiles.}
\label{table:as209.perez.mw.results}
\end{table}
In Figure \ref{fig.comparison.perez.structure} the multi-wavelength results are represented with a dashed black line. 

The midplane temperature profile obtained with the multi-wavelength fit is an average temperature between the four single wavelength profiles. The agreement between the multi-wavelength temperature profile and those derived at 0.88 and 2.80 mm is good throughout the disk, whereas at 8.00 and 9.83 mm some discrepancies arise at $R>40\,$AU. The comparison of the optical depth profiles displays a similar behaviour, with a good agreement at 0.88 and 2.80 mm and larger discrepancies at 8.00 and 9.83 mm in the outer disk. The discrepancies in the optical depth can be understood by considering that the $\Sigma(R)$ profiles (and therefore the $\tau_\lambda(R)$ profiles) obtained with the single wavelength fits are totally independent of each other, whereas the multi-wavelength fit is defined by a unique $\Sigma(R)$ and produces different $\tau_\lambda(R)$ slopes at different wavelengths through radial variation of $a_{\max}$. In other words, the ability of the multi-wavelength fit to produce different $\tau_\lambda(R)$ profiles at different wavelengths depends on the degrees of freedom of the $a_{\max}$ parametrization\footnote{Note that this limitation is related to the dust parametrization, not to the multi-wavelength approach of the fit: the implementation of more sophisticated $a_{\max}(R)$ parametrizations is not only possible but is one of the advantages of having a fit architecture that is highly modular.}. That said, the net advantage of using the multi-wavelength fit lies in the fact that it provides a unique, self-consistent disk model with a dust radial distribution, as opposed to several single wavelength fits that provide as many different disk structures.

\begin{figure*}
\centering
\resizebox{0.49\hsize}{!}{\includegraphics{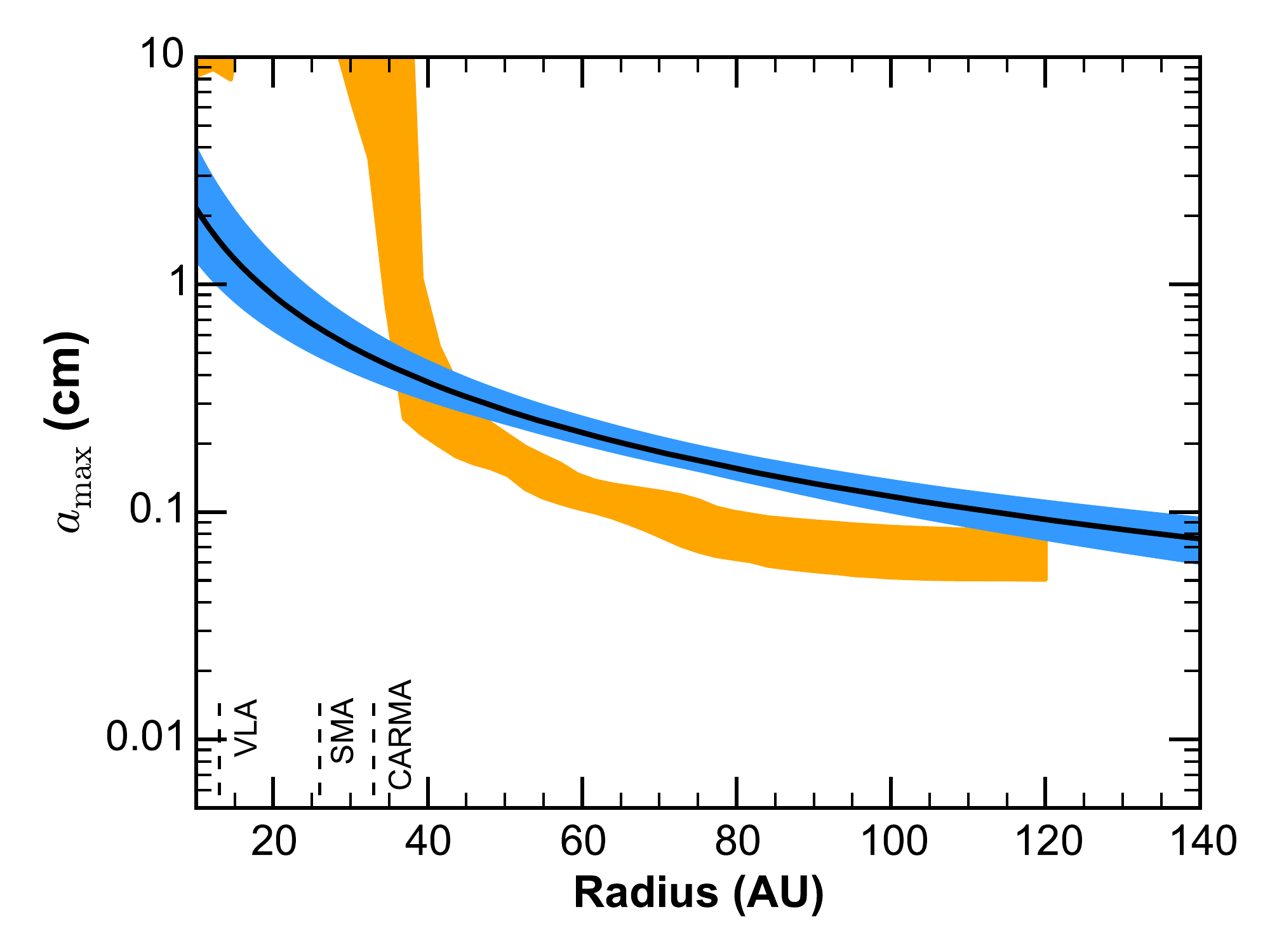}}
\resizebox{0.49\hsize}{!}{\includegraphics{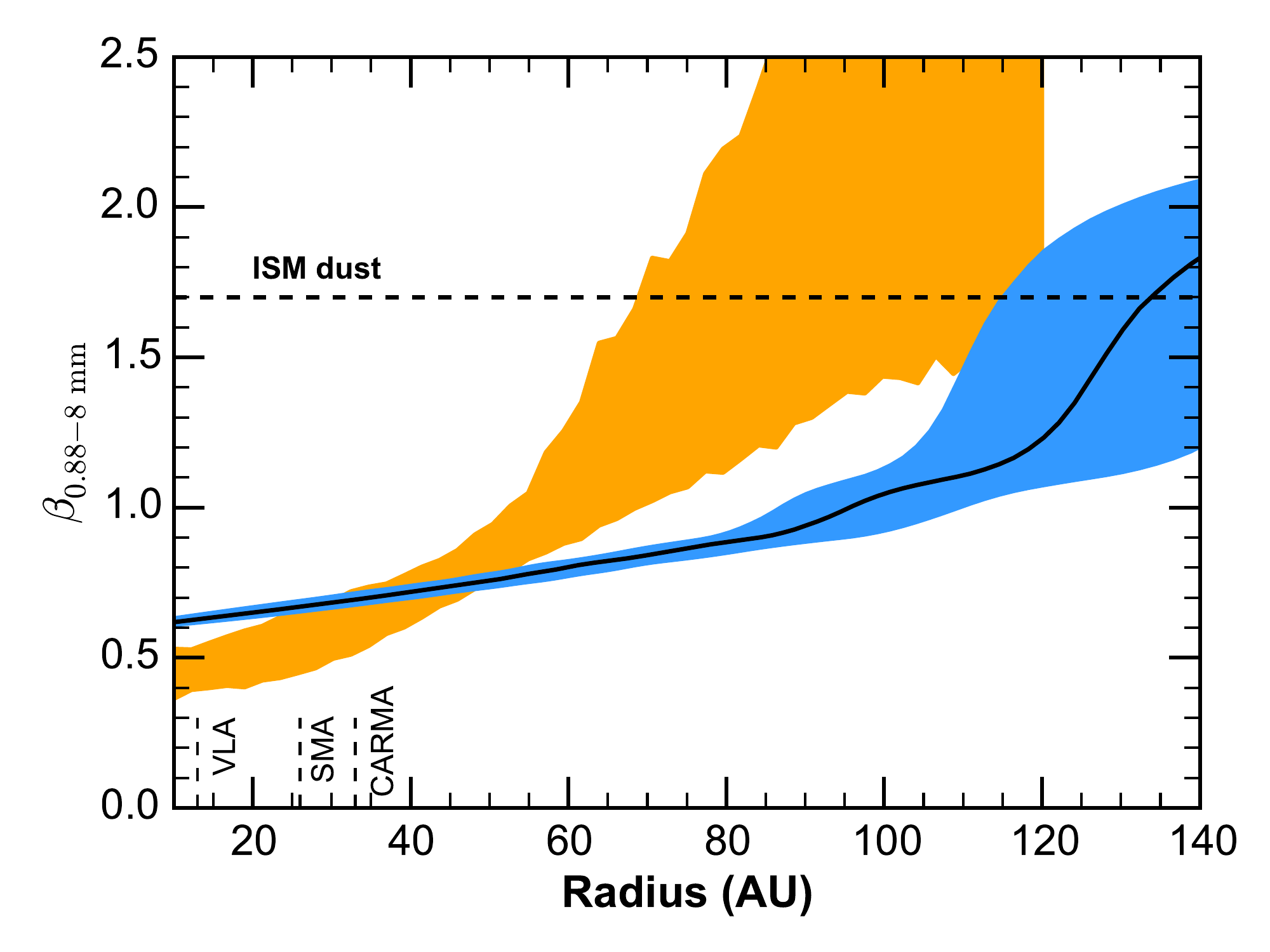}}
\caption{\textit{Left panel:} maximum dust grain size $a_{\max}$ as a function of the disk radius. \textit{Right panel:} dust opacity spectral index $\beta$ between 0.88 and 8 mm as a function of the disk radius. 
\textit{Both panels:}
the solid black line with the shadowed blue area represent the best-fit and the 3$\sigma$ region constrained by our multi-wavelength analysis. The yellow shaded area represent the 3$\sigma$ region obtained by \cite{2012ApJ...760L..17P}. The vertical dashed lines represent the spatial resolution of the observations.}
\label{fig:amaxbeta.Perez.comparison}
\end{figure*}

We now compare the $a_{\max}(R)$ and $\beta(R)$ profiles obtained with our multi-wavelength fit and those obtained by \cite{2012ApJ...760L..17P}. In the left panel of Figure \ref{fig:amaxbeta.Perez.comparison} we compare the $a_{\max}(R)$ profiles, which agree to within a factor less than 2 in the region where most of the signal comes from (between 40 and 140 AU the disk emission is spatially resolved and with signal-to-noise ratio larger than 3). It is reassuring that we both derive the same absolute dust grain size and the radial slope throughout the disk. The discrepancy visible at $R<$40\,AU our $a_{\max}(R)$ should not be a source of concern for two reasons: first, at $R<$40\,AU the disk is not spatially resolved at any wavelength. Secondly, $a_{\max}(R)$ is computed differently: our $a_{\max}(R)$ profile is by definition a power law, therefore it cannot become arbitrarily steep since it has to accommodate both the inner and the outer disk simultaneously; conversely, the $a_{\max}(R)$ profile derived by \cite{2012ApJ...760L..17P} is independent at each radius, but goes to extremely large values $a_{\max}\gtrsim 10\,$cm due to the high degeneracy in the $\beta(a_{\max})$ curve (cf.\ Figure \ref{fig:beta.vs.amax.Perez.composition}). 

In the right panel of Figure \ref{fig:amaxbeta.Perez.comparison} we compare the radial profile of $\beta$ between 0.88 and 8.0 mm. The two profiles agree in that they find common evidence of $\beta(R)$ increasing with radius from small values $\beta\sim 0.5$ in the inner disk and $\beta\gtrsim \beta_{\mathrm{ISM}}=1.7$ in the outer disk. Nevertheless, they also display some important  differences that indeed can be understood recalling the method used to derive them. In our multi-wavelength analysis, the constraint is posed on the $a_{\max}(R)$ profile, while $\beta(R)$ is  calculated as a post-processing result of the analysis through Mie theory as explained in Section \ref{sec:dust.model}. It is then natural that a slowly decreasing $a_{\max}(R)$ profile results in a slowly increasing $\beta(R)$ profile. The analysis by  \cite{2012ApJ...760L..17P}, on the other hand, poses a direct constraint on $\beta(R)$, that is computed substantially as the ratio of the optical depth profiles $\tau_\lambda(R)$ at 0.88 and 8.00 mm that has been obtained from the single-wavelength fits (the actual procedure they use is more refined since they employ a MCMC to compute a PDF for $\beta(R)$ given the $\tau_\lambda(R)$ profiles and and average temperature ${\overline{T}(R)}$). For these reasons, the $\beta(R)$ profiles obtained by these two methods are not directly comparable point by point in radius, nevertheless they show common evidence of an increasing $\beta$ with radius.

%%%%%%%%%%%%%%%%%%%%%%%%%%%%%%%%%%
%%%% APPENDIX FITS RESULTS %%%%%%%
%%%%%%%%%%%%%%%%%%%%%%%%%%%%%%%%%%
\pagebreak
\section{Fits results}
\label{app:fits.results}
\begin{figure*}
\centering
\resizebox{0.35\hsize}{!}{\includegraphics{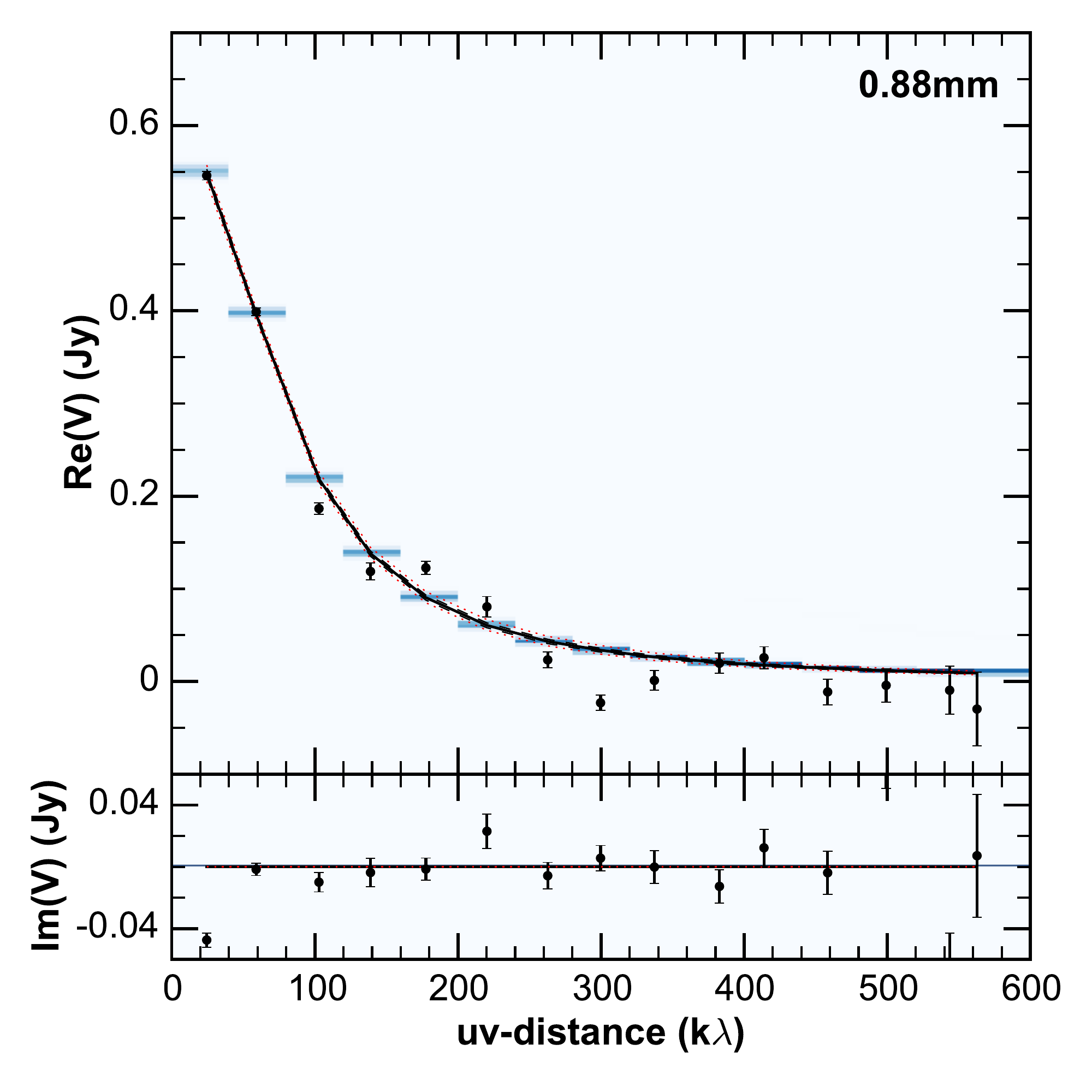}}
\resizebox{0.35\hsize}{!}{\includegraphics{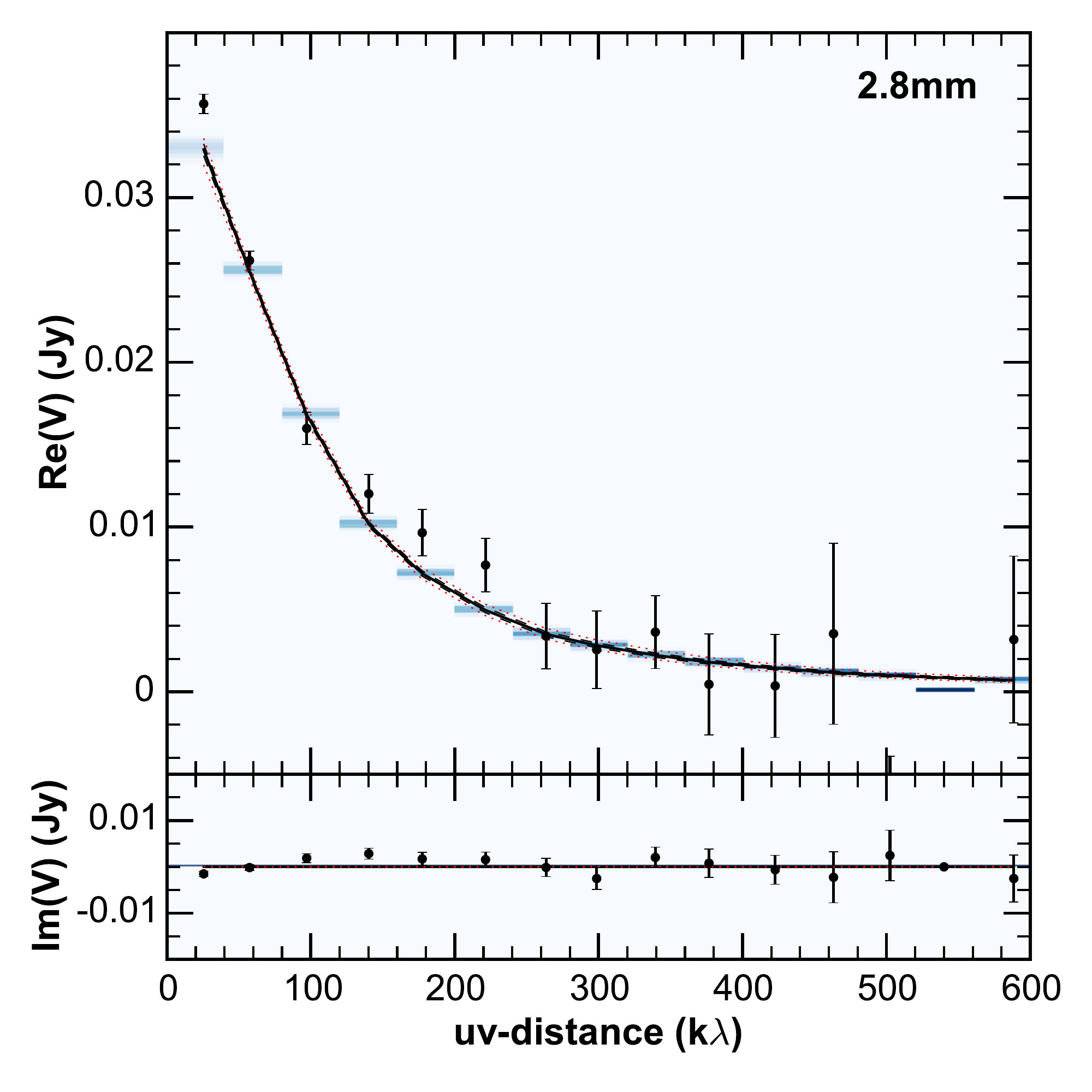}}\\
\resizebox{0.35\hsize}{!}{\includegraphics{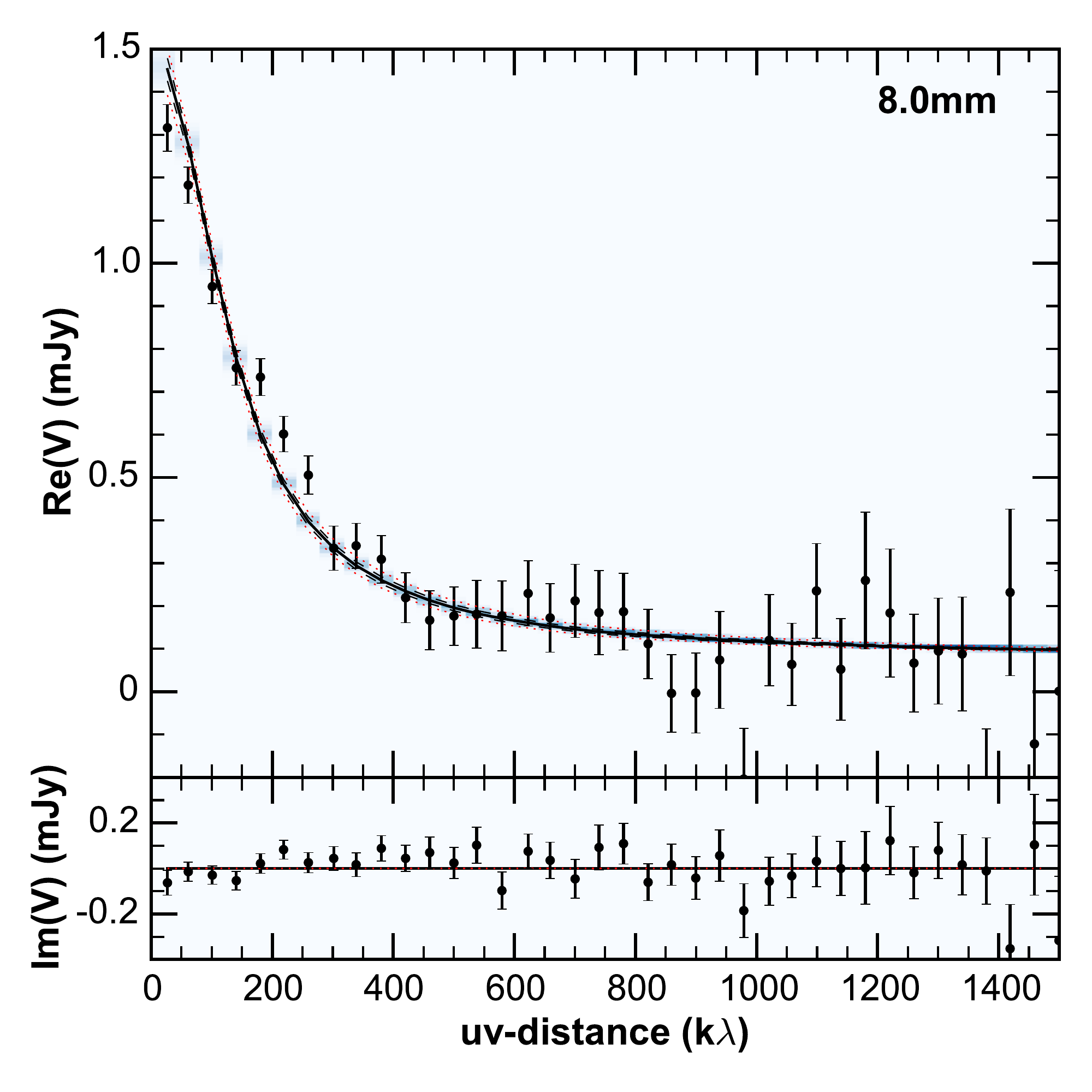}}
\resizebox{0.35\hsize}{!}{\includegraphics{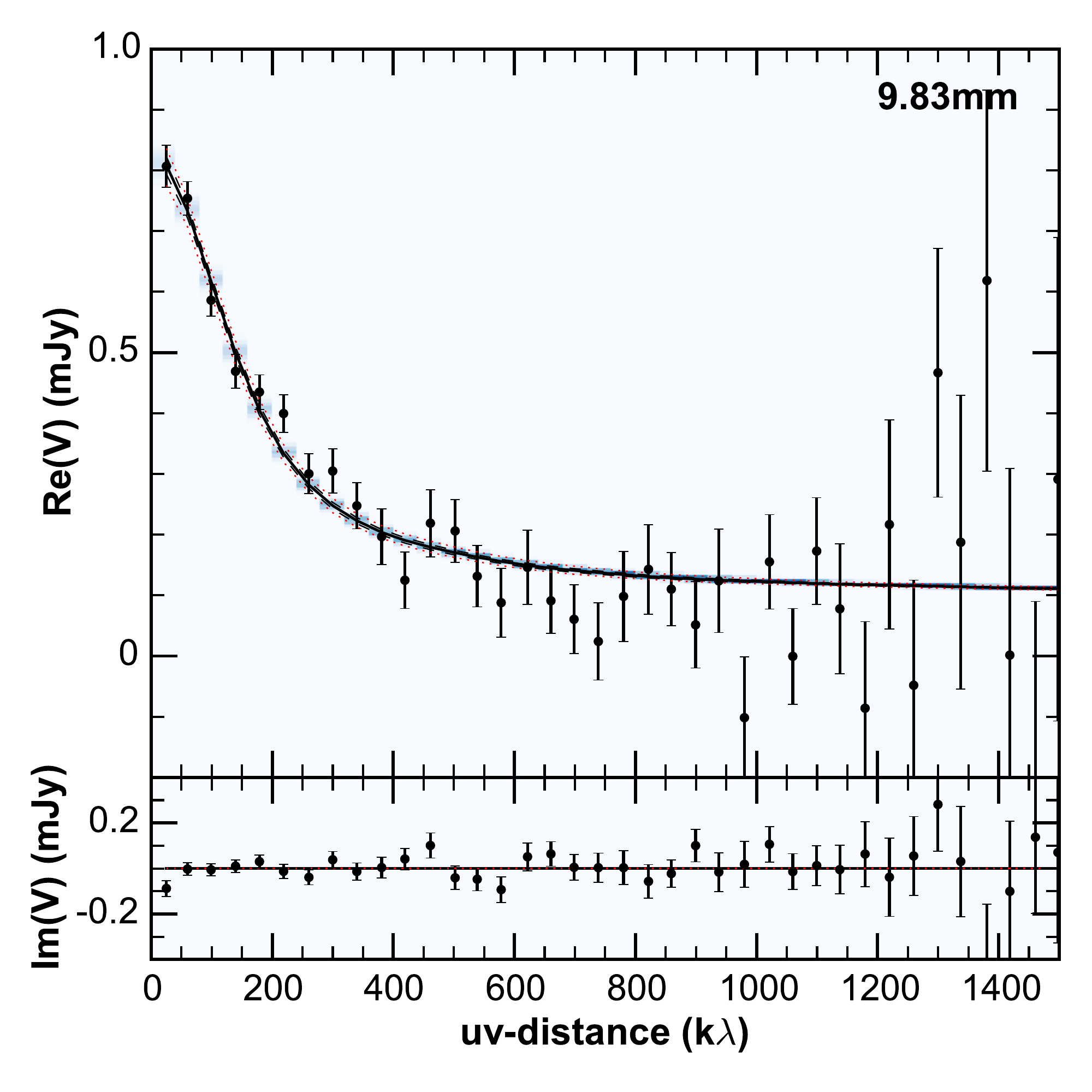}}
\caption{AS 209 bin-averaged visibilities as a function of deprojected baseline length (uv-distance). Black dots represent the observed data, the colored area represents the density of models for each uv-distance bin and the lines are defined as in Figure \ref{fig:fttau.amaxbeta}.}
\label{fig:as209.uvplots}
\end{figure*}

\begin{figure*}
\centering
\resizebox{0.46\hsize}{!}{\includegraphics{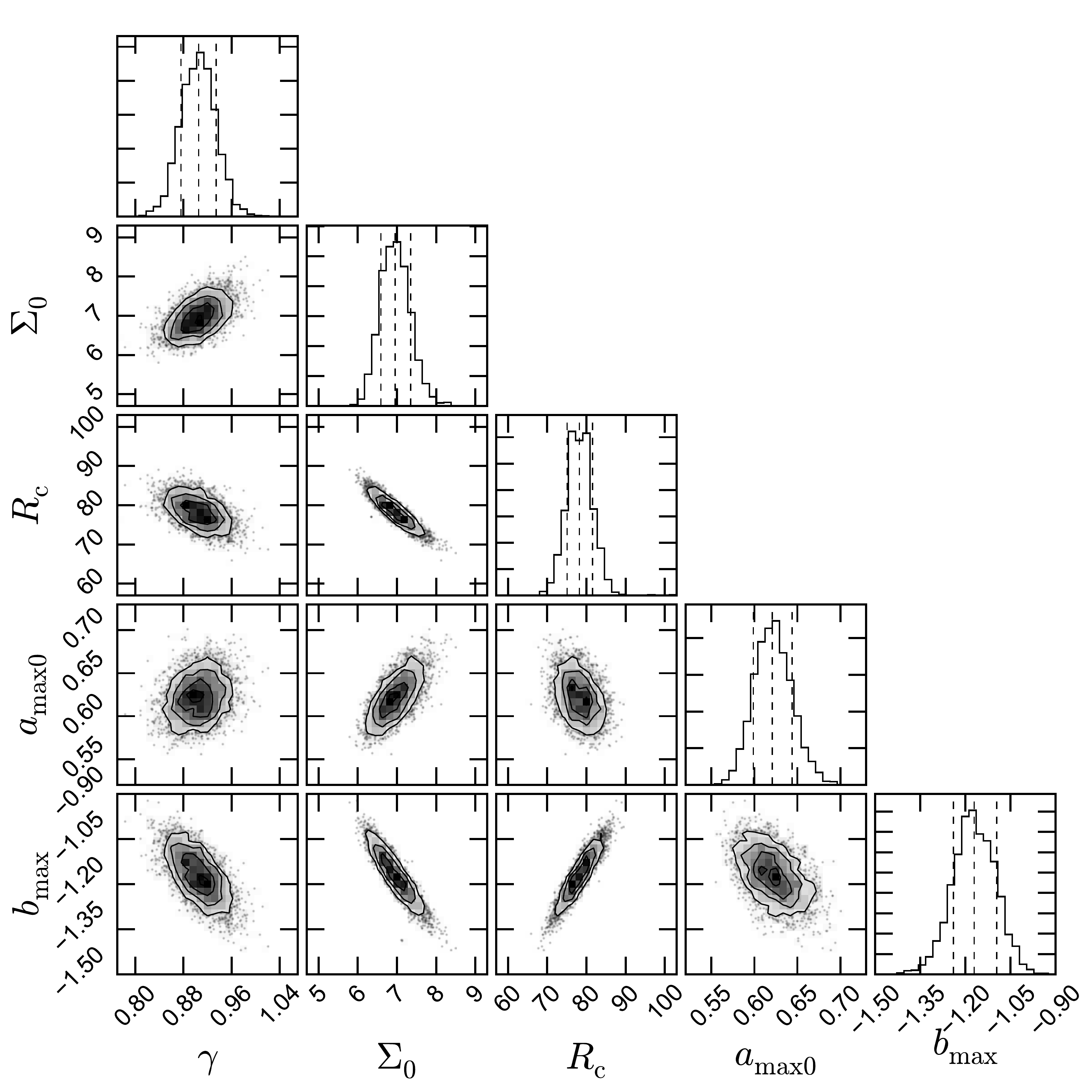}}
\resizebox{0.45\hsize}{!}{\includegraphics{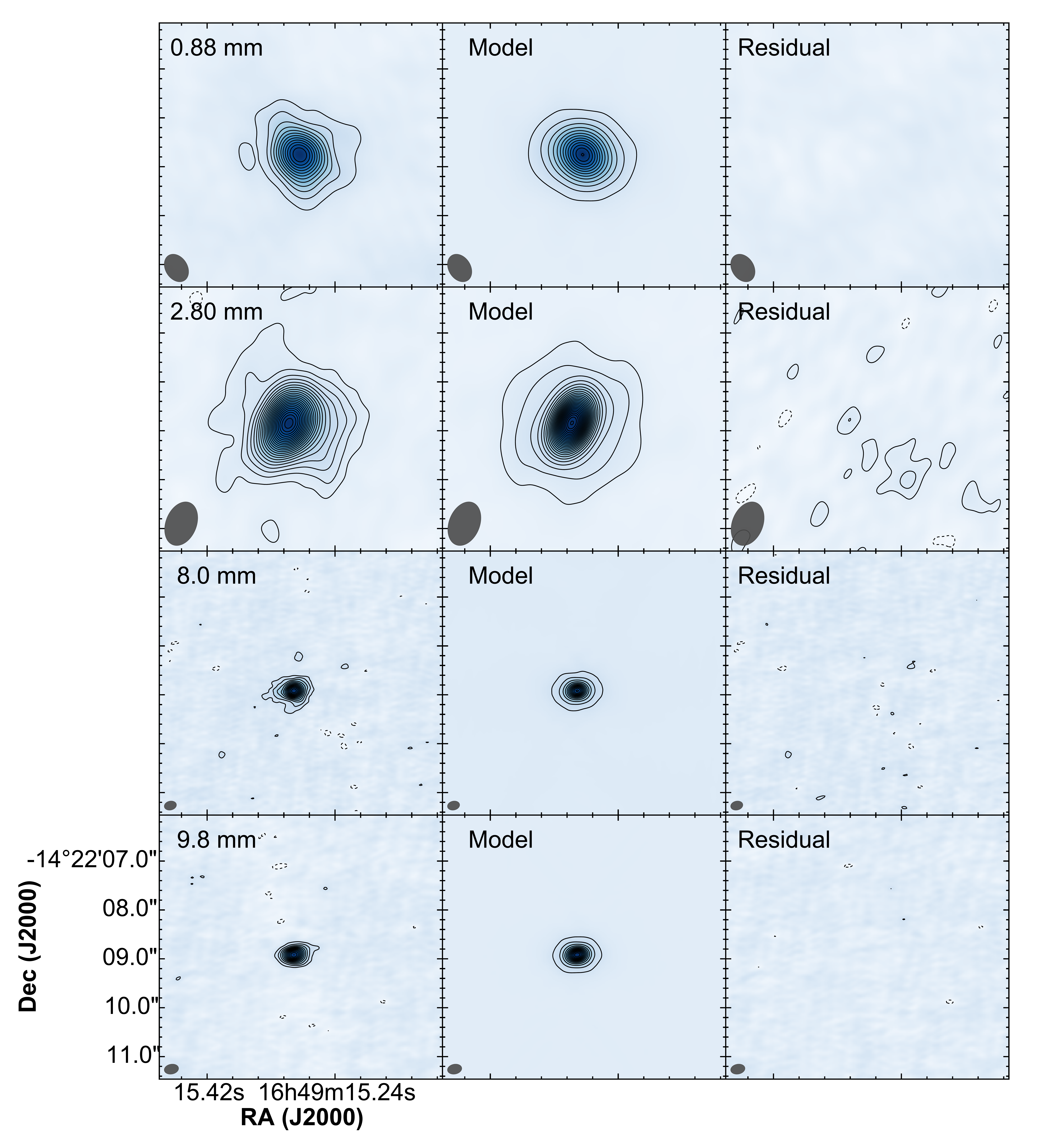}}
\caption{\textit{Left panel: }Staircase plot showing the marginalized and bi-variate probability distributions resulting from the fit for AS 209. \textit{Right panel: }AS 209 maps of the residuals at the fitted wavelengths.}
\label{fig:as209.residuals}
\end{figure*}

\begin{figure*}
\centering
\resizebox{0.45\hsize}{!}{\includegraphics{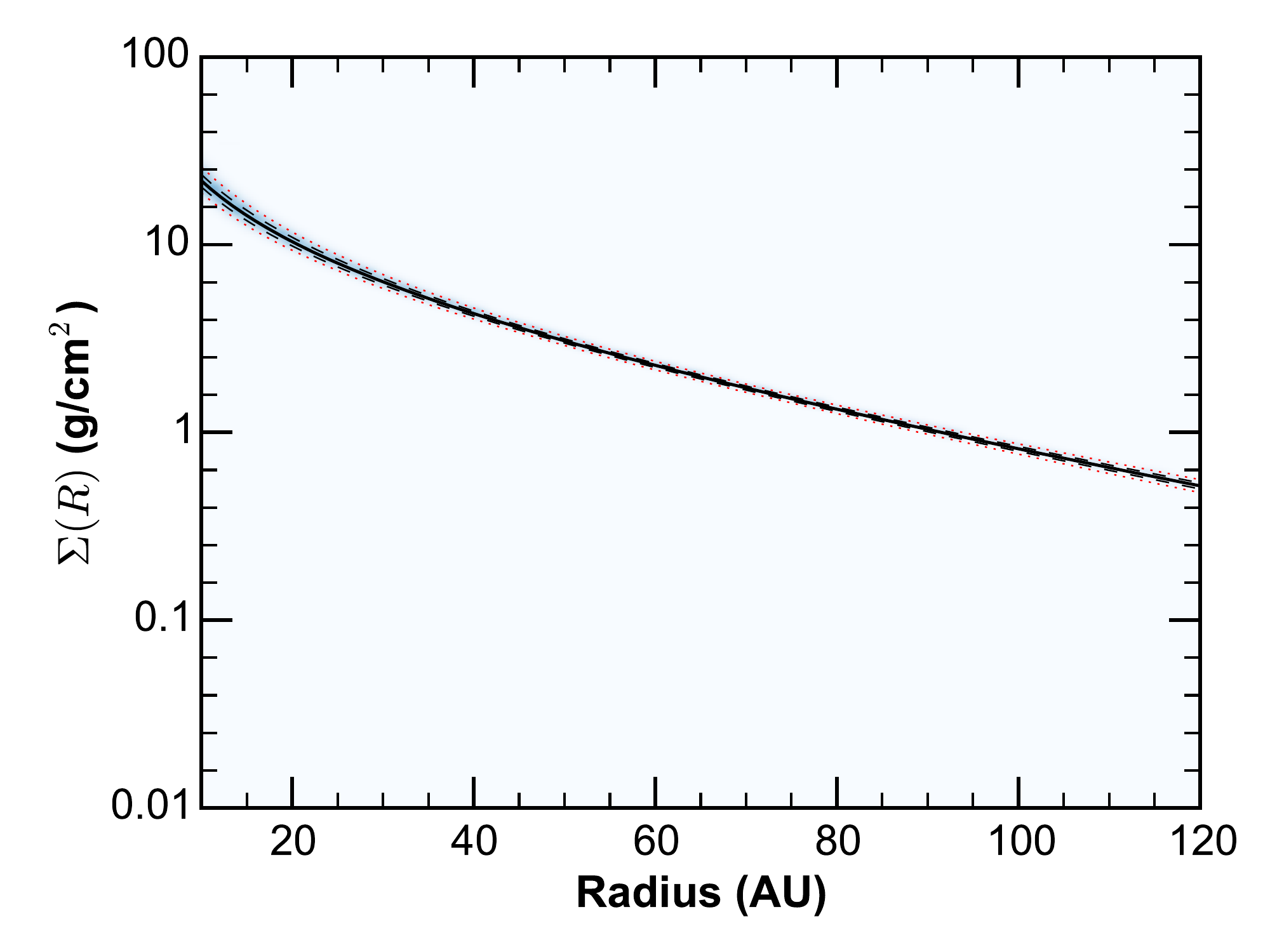}}
\resizebox{0.45\hsize}{!}{\includegraphics{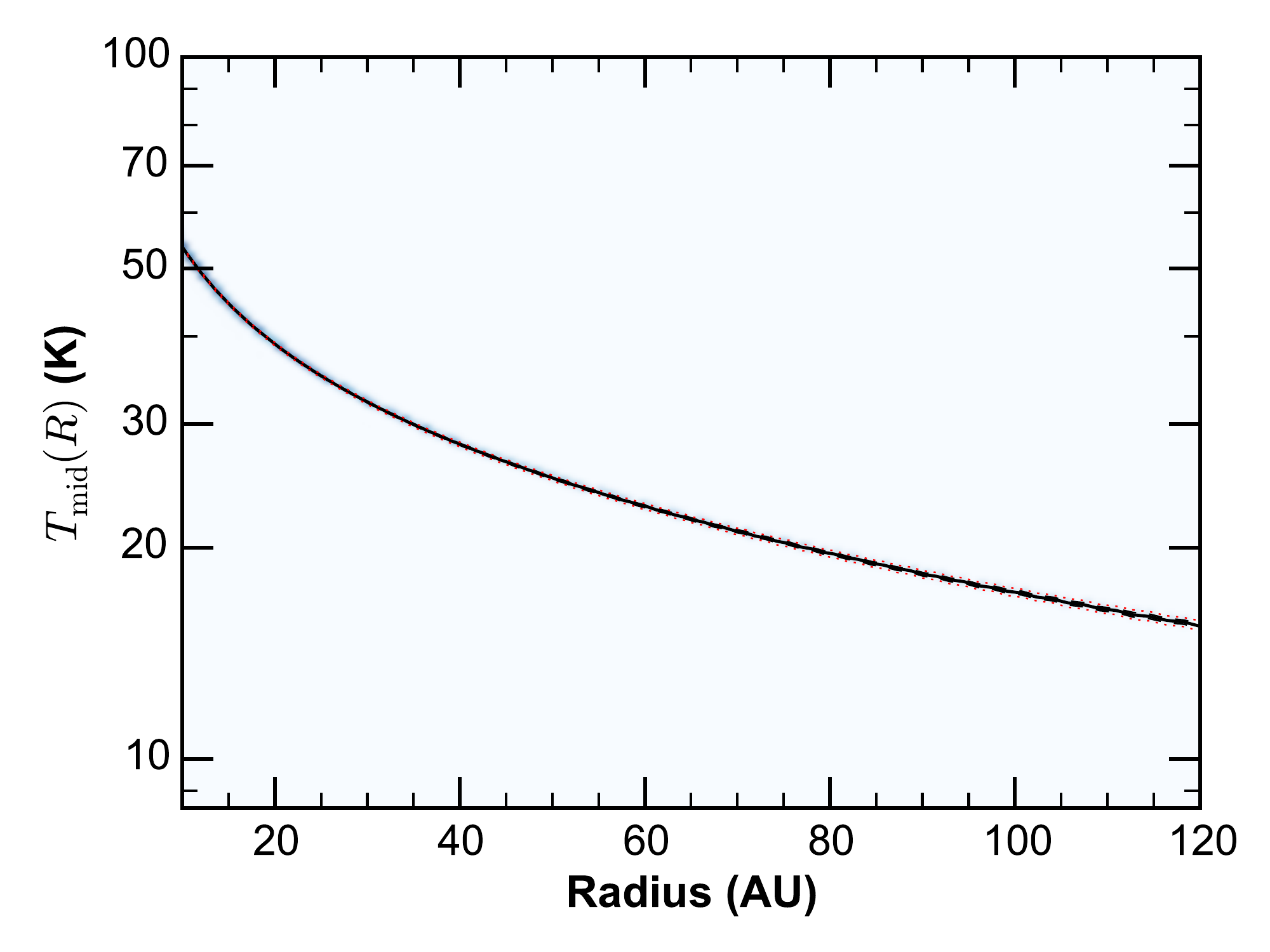}}
\caption{
{Results of the AS 209 fit.
\textit{Left panel: }posterior PDF of the gas surface density. 
\textit{Right panel:} posterior PDF of the midplane temperature. Line conventions are the same as those in Fig. \ref{fig:fttau.uvplots}.}}
\label{fig:as209.structure}
\end{figure*}

\begin{figure*}
\centering
\resizebox{0.32\hsize}{!}{\includegraphics{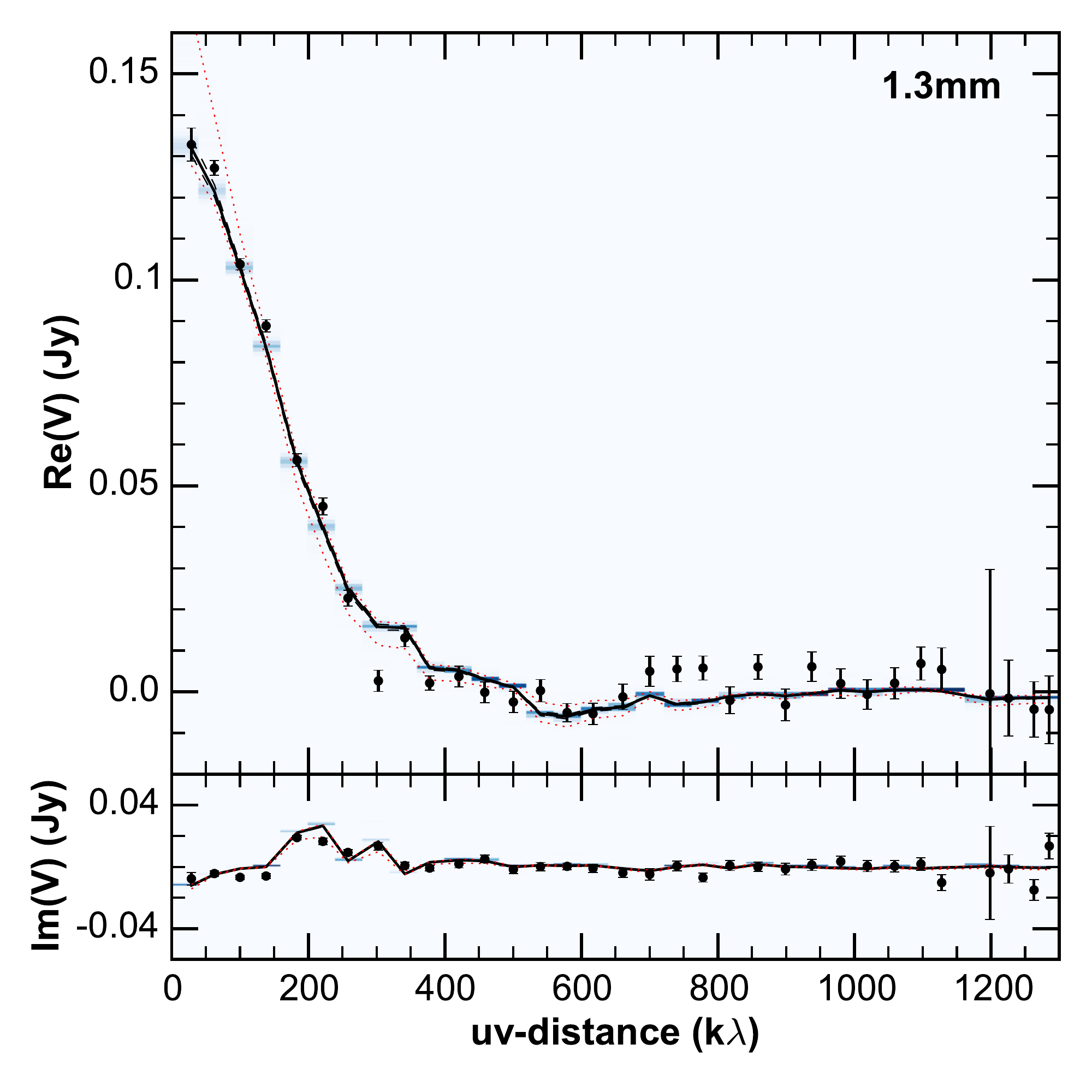}}
\resizebox{0.32\hsize}{!}{\includegraphics{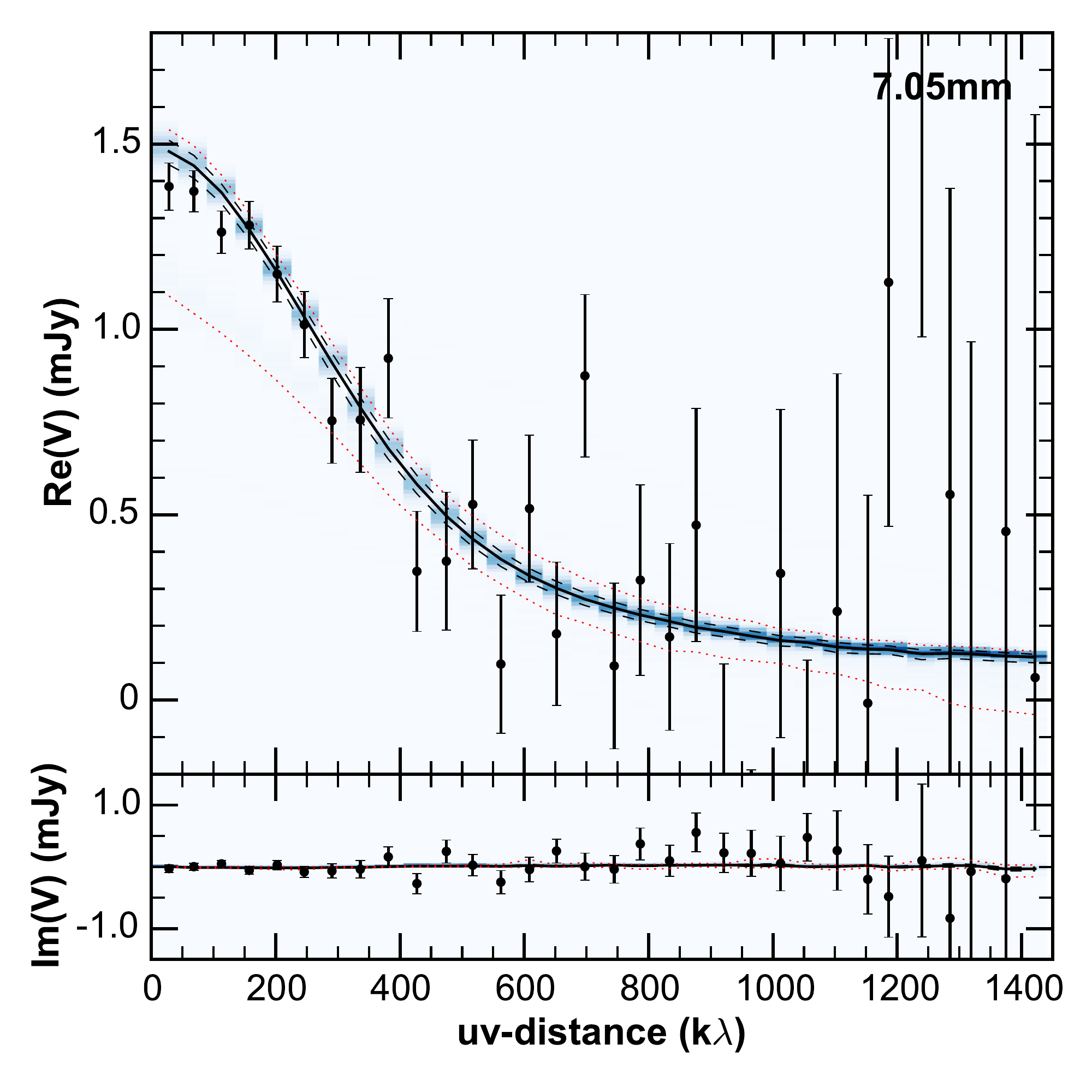}}
\resizebox{0.32\hsize}{!}{\includegraphics{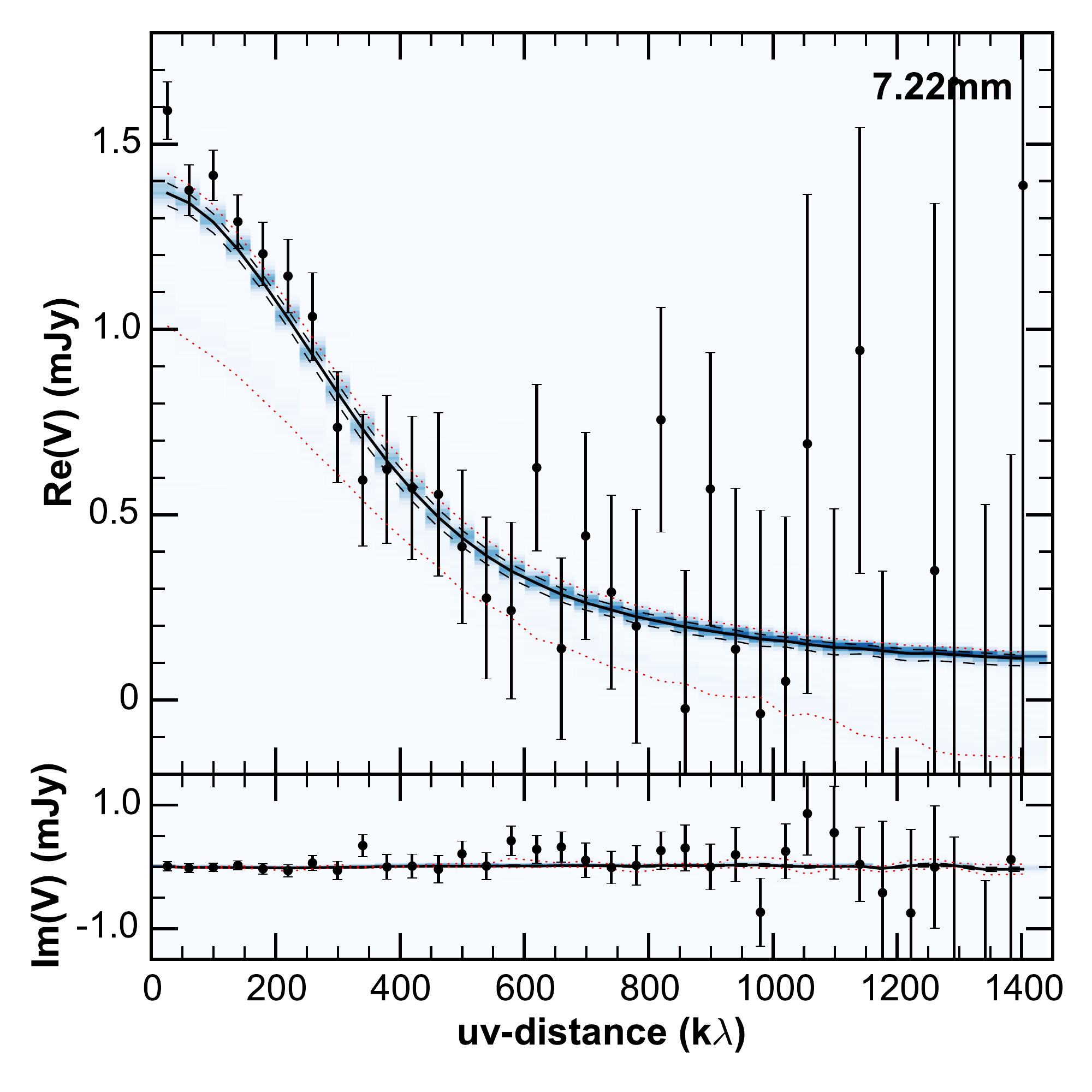}}
\caption{DR Tau bin-averaged visibilities as a function of deprojected baseline length (uv-distance). Color and line conventions are defined in Figure \ref{fig:fttau.amaxbeta}.}
\label{fig:drtau.uvplots}
\end{figure*}

\begin{figure*}
\centering
\resizebox{0.45\hsize}{!}{\includegraphics{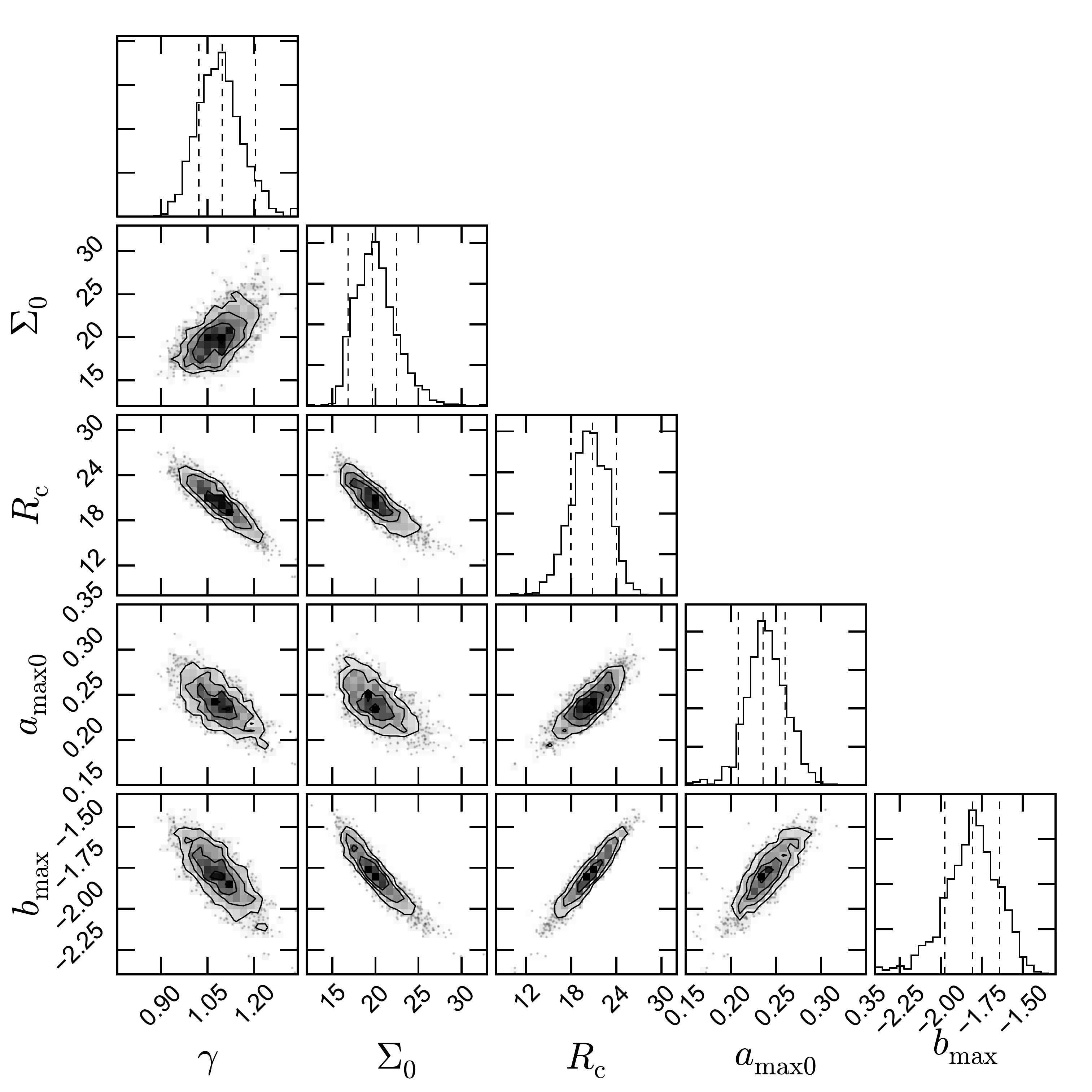}}
\resizebox{0.52\hsize}{!}{\includegraphics{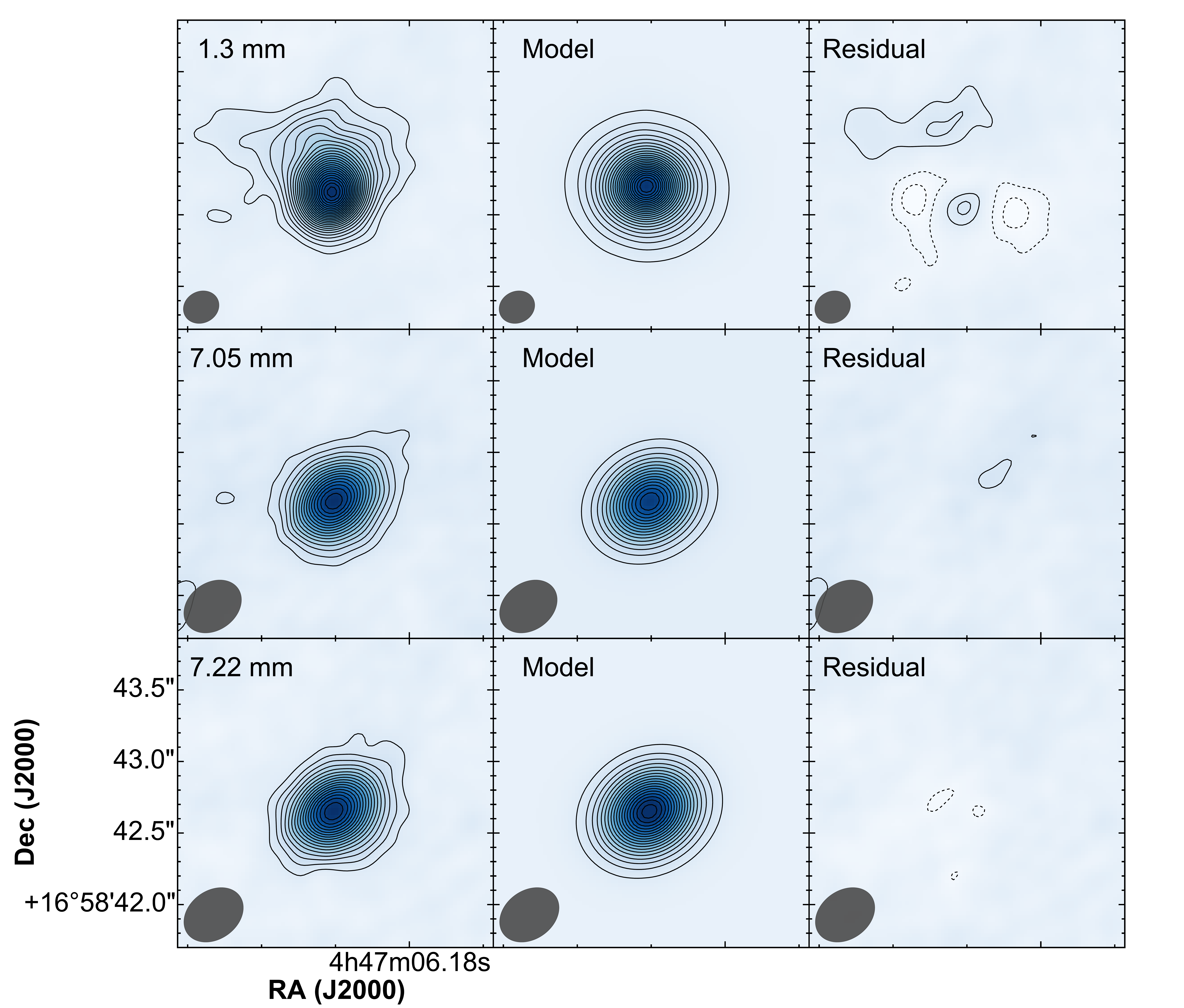}}
\caption{\textit{Left panel: }Staircase plot showing the marginalized and bi-variate probability distributions resulting from the fit for DR Tau. \textit{Right panel: }DR Tau maps of the residuals at the fitted wavelengths.}
\label{fig:drtau.residuals}
\end{figure*}

\begin{figure*}
\centering
\resizebox{0.45\hsize}{!}{\includegraphics{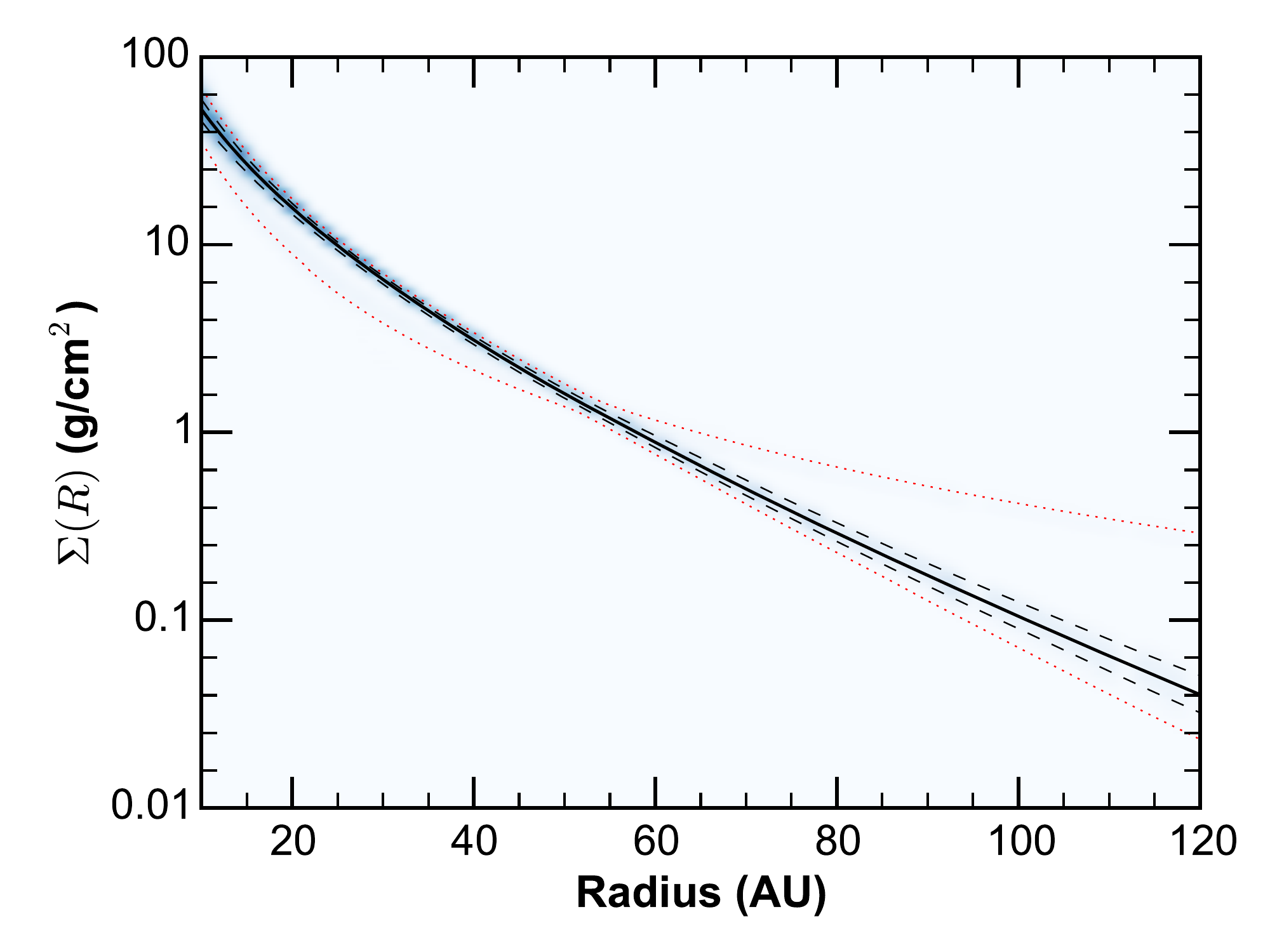}}
\resizebox{0.45\hsize}{!}{\includegraphics{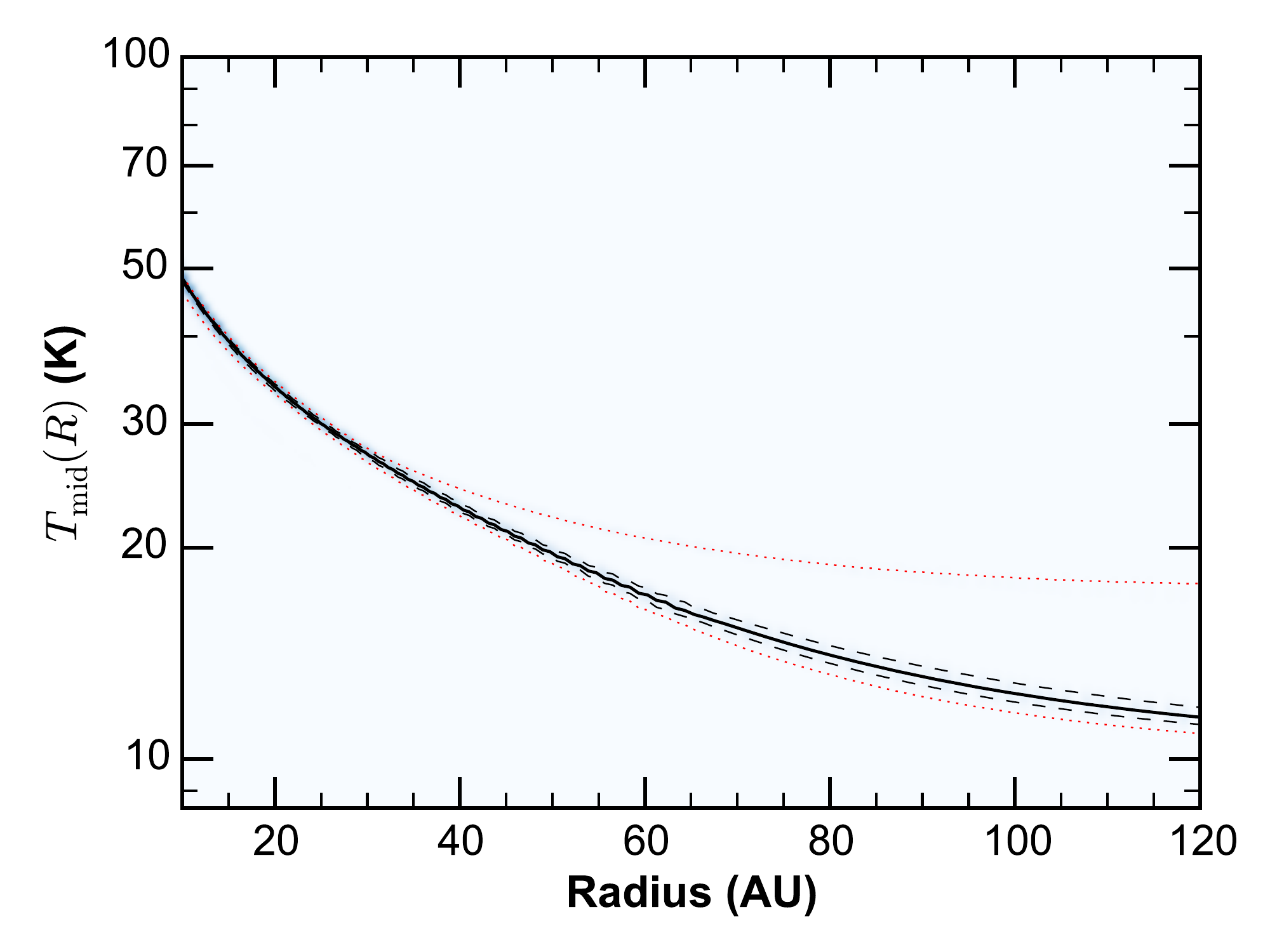}}
\caption{
{
Results of the DR Tau fit.
\textit{Left panel: }posterior PDF of the gas surface density. 
\textit{Right panel:} posterior PDF of the midplane temperature. Line conventions are the same as those in Fig. \ref{fig:fttau.uvplots}}.}
\label{fig:drtau.structure}
\end{figure*}

As anticipated in Section \ref{sec:results}, here we report the results of the multi-wavelength fits for AS 209 and DR Tau. For each disk we present staircase plots with the {1D and 2D marginalized posterior PDFs}, maps of the residuals at each wavelength (obtained subtracting the best-fit model from the observations) and a comparison between the observed and the model visibilities at each wavelength.{ We also present the physical structure derived for each disk: the gas surface density and the midplane temperature profile. In Figures \ref{fig:as209.uvplots}, \ref{fig:as209.residuals} and \ref{fig:as209.structure} we present the results of the fit for AS~209, showing respectively: the comparison of model and observed visibilities, the residual maps and the posterior PDFs, the derived disk structure. In Figures \ref{fig:drtau.uvplots}, \ref{fig:drtau.residuals} and \ref{fig:drtau.structure} we present the same plots for DR~Tau.}

\bibliographystyle{aa}
\bibliography{mt_disks}
\end{document}